\documentclass[reprint, superscriptaddress, aps, prx, floatfix, longbibliography]{revtex4-2}

\usepackage[normalem]{ulem}
\usepackage{graphicx}
\usepackage{color}
\usepackage{verbatim}
\usepackage{subfigure}
\usepackage{amssymb}
\usepackage{amsmath}
\usepackage{enumitem}
\usepackage{comment}
\usepackage{csquotes}
\MakeOuterQuote{"}
\usepackage{graphicx}
\usepackage{xcolor} 
\usepackage{dcolumn}
\usepackage{bm}

\usepackage[pangram]{blindtext}
\usepackage{subfigure}
\newcommand*\diff{\mathop{}\!\mathrm{d}}
\def\*#1{\mathbf{#1}}

\newcommand{\ed}[1]{{\color{black}#1}}

\begin{document}

\title{Theory of Crystallization versus Vitrification}

\author{Muhammad R. Hasyim}
\email{muhammad$_$hasyim@berkeley.edu}
\affiliation{Department of Chemical and Biomolecular Engineering, University of California, Berkeley, CA, USA}

\author{Kranthi K. Mandadapu}
\email{kranthi@berkeley.edu}
\affiliation{Department of Chemical and Biomolecular Engineering, University of California, Berkeley, CA, USA}
\affiliation{Chemical Sciences Division, Lawrence Berkeley National Laboratory, Berkeley, CA, USA}


\begin{abstract}
The competition between crystallization and vitrification in glass-forming materials manifests as a non-monotonic behavior in the time-temperature transformation (TTT) diagrams, which quantify the time scales for crystallization as a function of temperature. 
We develop a coarse-grained lattice model, the Arrow-Potts model, to explore the physics behind this competition. 
Using Monte Carlo simulations, the model showcases non-monotonic TTT diagrams resulting in polycrystalline structures, with two distinct regimes limited by either crystal nucleation or growth.
At high temperatures, crystallization is limited by nucleation and results in the growth of compact crystal grains. 
At low temperatures, crystal growth is influenced by glassy dynamics, and proceeds through dynamically heterogeneous and hierarchical relaxation pathways producing fractal and ramified crystals. 
To explain these phenomena, we combine the Kolmogorov-Johnson-Mehl-Avrami theory with the field theory of nucleation, a random walk theory for crystal growth, and the dynamical facilitation theory for glassy dynamics.
The unified theory yields an analytical formula relating crystallization timescale to the nucleation and growth rates through universal exponents governing glassy dynamics of the model. We show that the formula with the universal exponents yields excellent agreement with the Monte Carlo simulation data and thus, it also accounts for the non-monotonic TTT diagrams produced by the model. Both the model and theory can be used to understand structural ordering in various glassy systems including bulk metallic glass alloys, organic molecules, and colloidal suspensions.  
\end{abstract}

\maketitle

\section{Introduction}
When cooled down slowly from its molten state, a material forms a polycrystalline solid whose microstructure consists of compacted and randomly oriented crystal grains. When quenched very rapidly, the molten state vitrifies into a glass where the molecular structure is indistinguishable from a liquid. For any given cooling protocol, both crystallization and vitrification are simultaneously present and the competition between them can radically alter the final microstructure and properties of a material. Understanding this competition is crucial for a wide range of materials science problems including the formation of bulk metallic glasses \citep{Masuhr1999,Schroers1999,Hays1999} and their subsequent recrystallization \citep{Kim2002}, the crystallization of organic molecules and pharmaceuticals \citep{Huang2018, Karmwar2011, Blaabjerg2016}, and the solidification of geological melts \citep{Iezzi2011}. 
\begin{figure}
    \centering
    \includegraphics[width=0.85\linewidth]{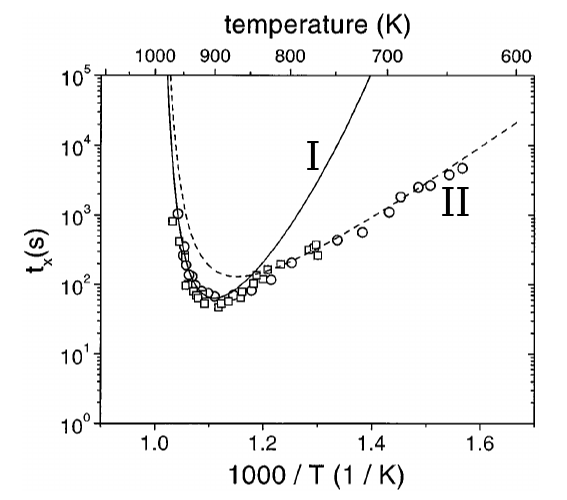}
    \caption{TTT diagram for $\mathrm{Zr}_{41.2}\mathrm{Ti}_{13.8}\mathrm{Cu}_{12.5}\mathrm{Ni}_{10.0}\mathrm{Be}_{22.5}$ showcasing non-monotonic and cross-over behavior. The crystallization time scale $t_\mathrm{x}$ is measured from the time in which the release of enthalpic heat is first observed. The liquidus temperature is $1024\,$K. Adapted from Ref.~\citep{Masuhr1999}.}
    \label{fig:experimental}
\end{figure}

The competition between crystallization and vitrification can be observed in the time-temperature-transformation (TTT) diagrams, which plot a measure of the crystallization timescale $t_\mathrm{x}$ as a function of temperature. These TTT diagrams universally exhibit non-monotonic or "nose-like" shape \citep{Masuhr1999,Schroers1999,Hays1999,Karmwar2011,Blaabjerg2016}, which is shown in Fig.~\ref{fig:experimental} for a metallic glassy system \cite{Masuhr1999}. The intuition behind this phenomenon is well-known. Near the melting temperature $T_m$, crystallization is slow because the nucleation free-energy barrier is high despite the rapid diffusion of atoms/molecules. As the temperature decreases, this barrier also decreases allowing crystallization to proceed faster; see Fig.~\ref{fig:experimental}. Eventually, dynamics dramatically slows down crystallization and promotes the kinetic arrest of the liquid until its full vitrification at the glass transition temperature $T_g$. While this intuition is roughly correct \citep{Limmer2011,limmer2013corresponding}, no theory has been developed yet to explain the trends in TTT diagrams, especially the apparent cross-over of Trend I and Trend II in Fig.~\ref{fig:experimental}, and which is also consistent with kinetic microscopic mechanisms found in molecular dynamics (MD) simulations \citep{Donati1998,Gebremichael2004,hurley1995kinetic,valeriani2012compact, Puser2009,Sanz2011,Sanz2014}. 

On a related note, it is well-known that the viscosity of supercooled liquids increases dramatically in a \textit{super-Arrhenius manner} 
when the temperature is decreased \citep{Angell2000}. This trend is also accompanied by a key microscopic phenomenon known as dynamical heterogeneity \citep{Donati1998,Gebremichael2004,hurley1995kinetic}. Simply put, microscopic reorganization of atoms and molecules are clustered into localized mobile regions and extended immobile/frozen regions. The mobile regions can be detected by computing the particles' relative displacements from MD simulations \citep{Donati1998,Gebremichael2004}. In addition, dynamical heterogeneity can also be observed in colloidal suspensions using confocal microscopy \citep{Kegel2000,Weeks2000,Gao2007,Fris2011}.  At a closer look, these mobile regions form chain-like patterns, which exist because the displacement of one particle cascades into the displacement of the next neighboring particle. 

The cooperative motions in the liquid have a dramatic effect on crystallization under supercooled conditions. For instance, a molecular dynamics study \citep{valeriani2012compact} has shown that crystallization of single-component hard spheres produce fractal and ramified crystals under super-compressed conditions. In this regime, crystal nucleation is no longer rare, and growth proceeds through a cooperative and percolation-like process that requires \textit{no particle self-diffusion} \citep{Puser2009}. In fact, the growth mechanism resembles more of re-crystallization in the corresponding glass where growth proceeds through the chain-like motion of mobile particles triggering an immediate "avalanche" of crystal ordering \citep{Sanz2011,Sanz2014}.

Guided by these findings, we seek to construct a theory of how crystallization proceeds in glass-forming liquids, with specific consideration to the microscopic mechanisms of dynamical heterogeneity integral to glass formers, and the mechanisms of crystalline ordering found in super-compressed hard spheres. We address this by developing a coarse-grained lattice model, which contains minimal ingredients to reproduce the TTT diagrams found in experiments, as well as the microstructure and kinetic mechanisms found from MD simulations. The model will also guide us in the construction of a unified theory, yielding an analytical formula for the crystalliation time scales that encompasses the competition between crystallization and vitrification. Furthermore, we will discuss the influence of cooling protocols towards the TTT diagrams, and argue how protocols play a fundamental role in interpreting the trends found in TTT diagrams and impact the resulting microstructure of the polycrystalline solid. 

The article is organized as follows: In Section~\ref{sec:arrow-potts} we introduce the Arrow-Potts model to understand the competiting effects of crystallization and vitrification. Section~\ref{sec:arrowpotts_thermodynamics} describes the thermodynamics and phase diagrams of the Arrow-Potts model. Section~\ref{sec:arrowpotts_results} presents the TTT diagrams of the Arrow-Potts model and the resulting microstructures from two different cooling protocols. Section~\ref{sec:arrowpotts_theory} derives an analytical formula for crystallization time scales from the TTT diagrams by combining the field theory of nucleation, dynamical facilitation theory for glassy dynamics, random walk theory of crystal growth, with the  Kolmogorov-Johnson-Mehl-Avrami theory. We finally test the analytical formulas and corresponding scaling laws in Section~\ref{sec:arrowpotts_theoryvssim}.  

\section{A Model for Crystallization in Glass-Formers} \label{sec:arrow-potts}
\begin{figure}
\centering
\includegraphics[width=\linewidth]{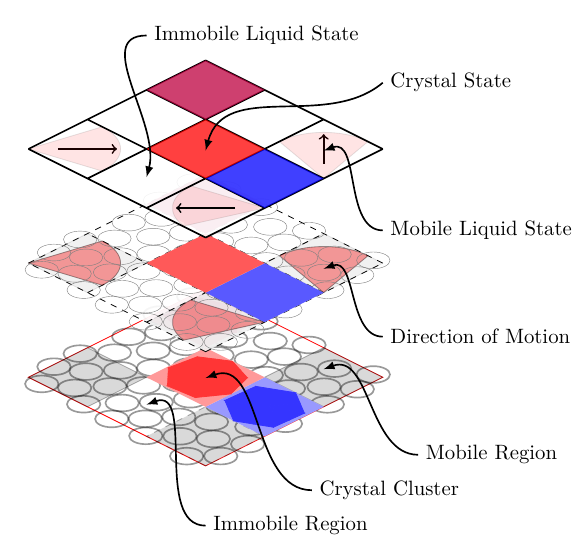}
\caption{An illustration of the coarse-graining concept leading to the Arrow-Potts model. A supercooled liquid consists of regions with low/high mobility and crystalline order, which are identified on the first layer. The grey and white colors indicate mobile liquid and immobile liquid regions, respectively, while other colors indicate crystal clusters in different orientations.  On the second layer, we associate each mobile region with its direction of motion, which dictates its direction of facilitation. On the third layer, we finally obtain a coarse-grained lattice model, which consists of immobile/mobile liquid states and crystal states with arrows representing direction of facilitation.}
\label{fig:arrowpotts_illus}
\end{figure}

At the microscopic level, a supercooled liquid exhibits both dynamical heterogeneity in terms of mobile and immobile regions, and crystallization. We use this fact to construct our model, herein referred to as \textit{the Arrow-Potts model}, to understand the competition between crystallization and vitrification. Conceptually, the Arrow-Potts model is built as a coarse-grained description of a glass-former, as illustrated in Fig.~\ref{fig:arrowpotts_illus}. It is a lattice model, which consists of a $d$-dimensional cubic lattice. Each lattice site represents one coarse-grained region of the glass former, and contains two degrees of freedom: (1) a spin variable $\* n_i$ representing mobility of particles, and (2) another spin variable $s_i$ representing crystalline order, where $i$ is the index of the lattice site. 

To model glassy dynamics and study its effect on crystallization, we use the perspective of dynamical facilitation theory \citep{Garrahan2002, Chandler2010, Biroli2013, Keys2011}. Dynamical facilitation theory is motivated by observations in MD simulations \citep{Keys2011,Donati1998,Gebremichael2004} where mobile liquid regions facilitate the motion of neighboring regions in a \textit{hierarchical} manner. The theory takes into account the dynamical heterogeneity and hierarchical motion, and explains the emergence of super-Arrhenius relaxation behaviors. The idea of dynamical facilitation is inspired from lattice-based kinetically constrained models \citep{Ritort2003,palmer1984models,Garrahan2003} where the facilitation mechanism is achieved by imposing constraints on the kinetics of spin variables leading to hierarchical relaxation. In what follows, we describe the Arrow model \cite{Garrahan2003} in Section~\ref{sec:arrow_describe}, a kinetically constrained model in $d$-dimensions that captures glassy dynamics in the absence of crystallization. We then extend the model to include the possibility of crystallization. In this case, since we are interested in polycrystal formation consisting of single crystalline regions with multiple orientations, we take inspiration from Potts models \citep{Potts1952,Wu1982} to allow for the degeneracy of crystal states. Taken together, they form the Arrow-Potts model presented in Section~\ref{sec:arrowpotts_describe}.

\subsection{The Arrow Model} \label{sec:arrow_describe}
As mentioned before, a supercooled liquid exhibits coexisting mobile and immobile regions, which are represented by the mobility spin variable $\* n_i$ in the Arrow model \citep{Garrahan2003}. The variable $\mathbf{n}_i$ is a unit vector that assigns a lattice site with either an immobile ($|\mathbf{n}_i|=0$) state or a mobile ($|\mathbf{n}_i | = 1$) state. Furthermore, in a mobile state, the unit vector points at the corners of the $d$-dimensional cubic lattice allowing for mobility in different directions, and hence is $2^d$-degenerate.

\begin{figure}[t]
\centering
\includegraphics[width=\linewidth]{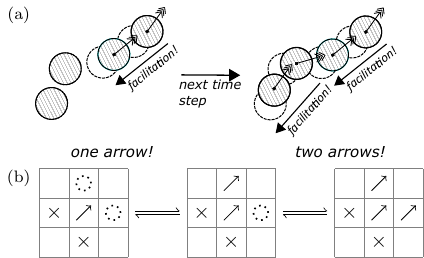}
\caption{A schematic of facilitated dynamics. (Top) Motion leads to more motion, and so the facilitation arrow indicates where the next set of motion is going to occur. (Bottom) An illustration of possible moves for mobile-immobile liquid transitions elucidating the kinetic constraint on a lattice. Dashed circle and
$\times$ indicate sites where a transition is allowed or forbidden, respectively.}
\label{fig:arrowpotts_liqfacilitate}
\end{figure}

The unit vector of a mobile state represents the \textit{direction of facilitation} and points in the opposite direction of collective motion \citep{Garrahan2003}, since motion facilitates more cooperative motion in a hierarchical and directional manner. This directionality is empirically observed in MD simulations \citep{Donati1998,Gebremichael2004, Keys2011} and illustrated in Fig.~\ref{fig:arrowpotts_liqfacilitate}a. On a lattice, the Arrow model implements the facilitation mechanism by kinetically constraining the reversible transition between an immobile and mobile state of some direction, with a nearest-neighboring mobile state of the same direction (Fig.~\ref{fig:arrowpotts_liqfacilitate}b). 

The equilibrium statistics of liquid mobility obey that of non-interacting spins 
where each mobile region of the liquid or an \textit{excitation} costs a characteristic energy $J_0$. In the atomistic picture, this implies that the rate of non-trivial particle-displacement events is Arrhenius as a function of temperature, which is observed below the onset temperature for glassy dynamics $T_\mathrm{o}$ \citep{Keys2011}. This is reflected in  the Arrow model with a non-interacting Hamiltonian
\begin{equation}
\mathcal{H}_\text{liq} = J_0\sum_i(1-\delta_{|\mathbf{n}_i|,0})\,, \label{eq:arrowpotts_arrowpart}
\end{equation}
which yields the equilibrium concentration of mobile states $c_\mathrm{eq} = \langle |\* n_i | \rangle = 2^d/\left( 2^d+e^{\beta J_0} \right) \to e^{-\beta J_0}$ as $T\to 0$, where $\beta = 1/k_\mathrm{B} T$. 

\begin{figure}
\includegraphics[width=\linewidth]{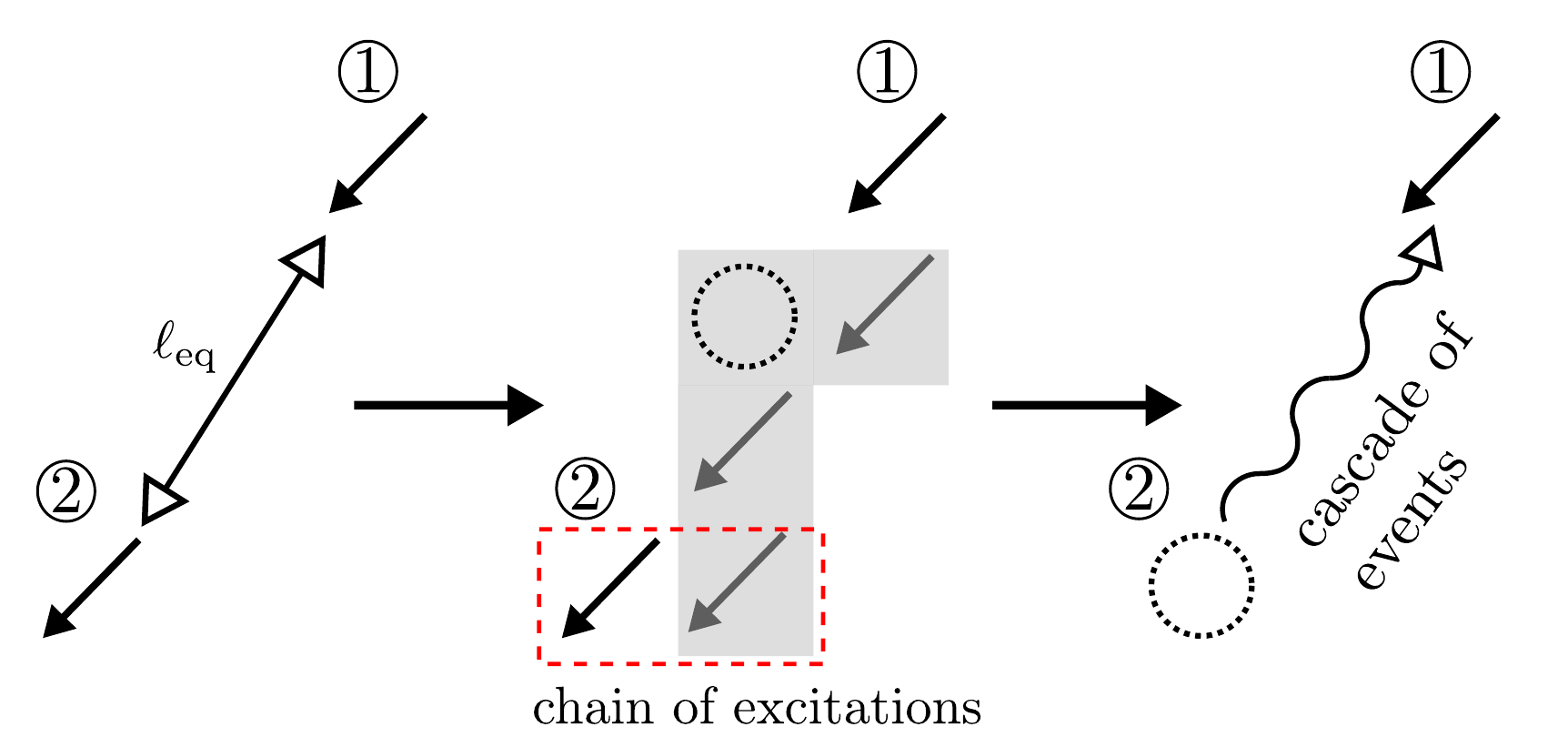}
\caption{An illustration of the emergent relaxation behavior due to facilitation mechanism. The system relaxes when a mobile state (labeled as \raisebox{.5pt}{\textcircled{\raisebox{-.9pt} {2}}}) changes its spin using another mobile state (labeled as \raisebox{.5pt}{\textcircled{\raisebox{-.9pt} {1}}}) $\ell_\mathrm{eq}$ away. The minimal energy pathway corresponds to \textit{a chain of excitations} of length $\ell_\mathrm{eq}$, which starts from excitation \raisebox{.5pt}{\textcircled{\raisebox{-.9pt} {1}}} to excitation \raisebox{.5pt}{\textcircled{\raisebox{-.9pt} {2}}}.  
Note that the pathway may generate holes, relaxing a few excitations, to achieve a lower energy barrier at the transition state. 
Once this transition state is achieved, a cascade of events can be initiated, starting from the excitation pair boxed in red, where the chain retracts back to excitation \raisebox{.5pt}{\textcircled{\raisebox{-.9pt} {1}}}.}
\label{fig:arrowpotts_liqchain}
\end{figure}

\begin{figure}
    \centering
    \includegraphics[width=\linewidth]{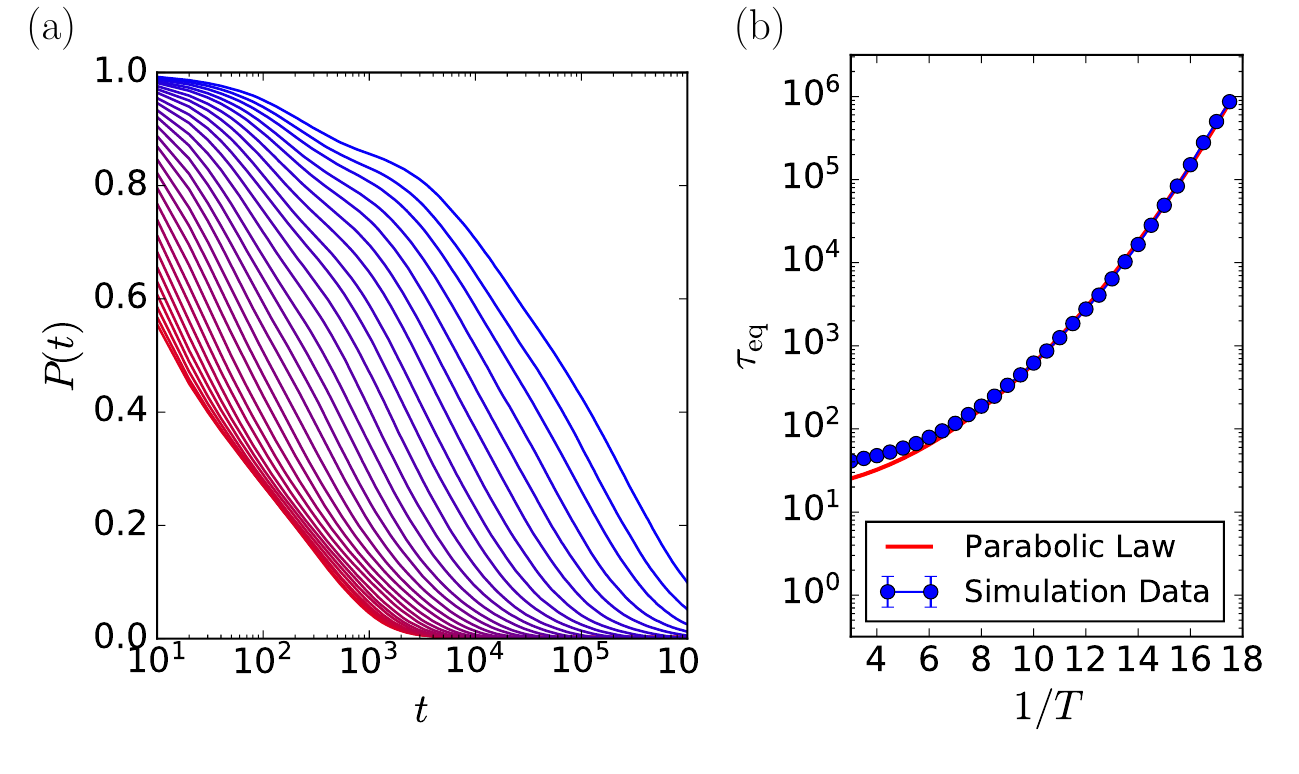}
    \caption{(Left) The decay of the persistence function $P(t)$ as a function of time, where red to blue indicates higher to lower equilibrium temperatures. (Right)  Equilibrium relaxation time $\tau_\mathrm{eq}$, measured in Monte Carlo sweeps (MCS). 
    Parameters: $J_0=0.25$ and $k_\mathrm{B}=1$. }
    \label{fig:arrow_eqdata}
\end{figure}

As the temperature $T\to 0$,  $c_\mathrm{eq}$ is low such that excitations become sparse and are separated by a distance $\ell_\mathrm{eq}$. 
The relaxation time of the system corresponds to a time scale for all the spins to change their state at least once, which in the atomistic picture corresponds to particles moving at least a particle diameter.
However, a lone excitation, i.e., one without any nearest-neighboring excitations, cannot relax by itself due to the kinetic constraint. In fact, relaxation is achieved when the lone excitation relaxes with the help of another excitation at distance $\ell_\mathrm{eq}$ creating a hierarchical chain of excitations, as illustrated in Fig.~\ref{fig:arrowpotts_liqchain}. The result of this chain-like mechanism is two-fold:
\begin{itemize} \setlength{\itemsep}{0pt}
    \item The overall energy barrier $E$ for relaxation scales logarithmically with $\ell_\mathrm{eq}$. In other words, $E = \gamma J_0 \ln \ell_\mathrm{eq}$ where $\gamma$ is a proportionality constant that sets the hierarchical nature of relaxation at different length-scales. In the Arrow model, $1/(2\ln 2) < \gamma < 1/(\ln 2)$ in any dimension \citep{Aldous2002,chleboun2014influence}. 
    \item On a lattice, $\ell_\mathrm{eq} = (2^d/c_\mathrm{eq})^{1/d}\to e^{\beta J_0/d}$ as $T\to 0$. This implies the energy barrier $E \to \gamma \beta (J_0)^2/d$. The relaxation time of a supercooled system is then given by 
    \begin{equation}
    \tau_\mathrm{eq} \sim \exp(\beta E) \sim \exp(\gamma \beta^2 J_0^2/d)\, \label{eq:parablaw}
    \end{equation}
    resulting in the ``parabolic law'' corresponding to the super-Arrhenius behaviors of relaxation times \citep{Sollich1999,Sollich2003}. 
\end{itemize}
The equilibrium relaxation time $\tau_\mathrm{eq}$ of the Arrow model can be calculated from the decay of the persistence function $P(t)$ given by \citep{Buhot2001}
\begin{equation}
P(t) = \left\langle \frac{1}{L^d} \sum_{i} P_i(t) \label{eq:arrow_persistfuncdef} \right\rangle \,,
\end{equation}
where $\langle \ldots \rangle$ is an ensemble average, $L$ is the linear size of the $d$-dimensional cubic lattice, and $P_i(t)=1$ if a spin at the indexed site $i$ has not flipped once at time $t$ and $P_i(t)=0$ otherwise. Using Eq.~\eqref{eq:arrow_persistfuncdef}, $\tau_\mathrm{eq}$ is defined as the instantaneous relaxation timescale corresponding to $P(t)$, i.e., $P(\tau_\mathrm{eq})=1/e$ \citep{Buhot2001,Ritort2003}.
The result of this calculation in 2D is shown in Fig.~\ref{fig:arrow_eqdata}, alongside the persistence function $P(t)$, in agreement with the parabolic law. Note that the parabolic law is obeyed not only in the Arrow model, but also in other kinetically constrained models, notably the East model \citep{Jackle1991,Sollich1999,Sollich2003,Faggionato2012,chleboun2014influence}. Furthermore, the parabolic law Eq.~\eqref{eq:parablaw} collapses the experimental viscosity data for a wide range of single-component \citep{Elmatad2009} and multi-component \citep{Katira2019} glass formers onto a single universal curve.

\subsection{The Arrow-Potts Model} \label{sec:arrowpotts_describe}
In addition to glassy dynamics, a supercooled liquid is also thermodynamically driven to crystallize. We model the nucleation and growth of crystals under the influence of glassy dynamics by adding a new degree freedom to the Arrow model, which finally transforms it to the Arrow-Potts model.  This new degree freedom is the scalar spin variable $s_i=\{0,1,\ldots,q\}$, which assigns a lattice site to be either a liquid ($s_i=0$) state or crystal ($s_i > 0$) states (see Fig.~\ref{fig:arrowpotts_illus}). The parameter $q$ denotes the degeneracy of crystal states, which originates from the discretization of crystal grain orientation relevant for a polycrystalline system. 

To capture the melting transition, we need the energetic cost to have a liquid-solid interface as a model parameter. Furthermore, to capture the formation of polycrystals during the freezing transition, we need the grain boundary energy, which provides an energetic penalty for having two adjacent crystal clusters with distinct orientations. A Hamiltonian $\mathcal{H}_\text{xtl}$ that combines these two energetic parameters can be written as
\begin{eqnarray}
\mathcal{H}_\text{xtl} =&& -\frac{\Delta \epsilon}{2} \sum_{\langle ij \rangle}\left((1-\delta_{s_i,0})\delta_{s_j,0}+(1-\delta_{s_j,0})\delta_{s_i,0}\right) \nonumber
\\
&& -\epsilon \sum_{\langle ij \rangle}\delta_{s_i,s_j}+h(T)\sum_i\delta_{s_i,0} \label{eq:arrowpotts_pottspart} \,,
\end{eqnarray}
where $h(T)$ is the temperature-dependent chemical potential that drives melting/freezing. From Eq.~\eqref{eq:arrowpotts_pottspart}, it can be seen that adjacent crystal-liquid sites cost $-\Delta \epsilon/2$, while both adjacent crystal-crystal sites of the same orientation and liquid-liquid sites cost $-\epsilon$.  Therefore, the energetic cost to have a crystal-liquid interface relative to either pure liquid or crystal states is $\epsilon - \Delta \epsilon/2$.  Thus, we need $\epsilon > \Delta \epsilon/2$ to obtain a positive interfacial tension. In addition, adjacent crystal-crystal sites with different orientations costs no energy, which implies that the energetic cost for a grain boundary relative to pure crystals is $\epsilon$.

The total Hamiltonian including the energetics of both liquid and crystalline states can be constructed by summing over Eq.~\eqref{eq:arrowpotts_arrowpart} and Eq.~\eqref{eq:arrowpotts_pottspart}
\begin{equation}
\mathcal{H}=\mathcal{H}_\text{liq}+\mathcal{H}_\text{xtl}+\sum_i C[\mathbf{n}_i,s_i]. \label{eq:arrowpotts_fullhamilton}
\end{equation}
The constraint function $C[\mathbf{n}_i,s_i]=\infty$ if $s_i \neq 0$ and $|\mathbf{n}_i| \neq 0$, and zero otherwise. As a result, the spin variables only create \textit{three} different types of states: (i) an immobile liquid state $(|\mathbf{n}_i|=0,s_i=0)$, (ii) a mobile liquid state $(|\mathbf{n}_i|=1,s_i=0)$, and (iii) an immobile crystal state $(|\mathbf{n}_i|=0,s_i \neq 0)$. In the atomistic picture, $C[\mathbf{n}_i,s_i]$ represents the idea that crystal clusters, by definition, do not possess liquid-like mobility. 
\begin{figure}[t]
\centering
\includegraphics[width=\linewidth]{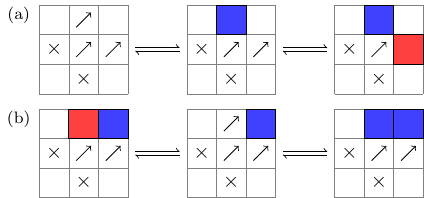}
\caption{(Top) An illustration of possible moves for liquid-crystal transitions elucidating the first kinetic constraint, where mobile liquid state facilitates crystallization. (Bottom) An illustration of the second kinetic constraint, where a crystal must transition into a mobile state before it can change into another crystal state. $\times$ indicate sites where any kinetic transition is forbidden.}
\label{fig:arrowpotts_gb}
\end{figure}

According to MD simulations of hard spheres, crystal nucleation is more likely to occur in regions with high liquid mobility \citep{Puser2009,Sanz2011}. Furthermore, these clusters grow through a percolation-like process where an "avalanche" of particle motions trigger crystalline ordering \citep{valeriani2012compact,Sanz2014}. Inspired by these observations, we model related phenomena with the addition of two kinetic constraints:
\begin{enumerate}
    \item Mobile liquid states facilitate crystallization, which means that the transition between a mobile liquid and crystal state is kinetically constrained by the nearest mobile liquid state of the same direction. Moreover, an immobile liquid state cannot directly transition into or arise from a crystal state; see Fig.~\ref{fig:arrowpotts_gb}a. 
    \item Two crystal clusters with different orientations cannot spontaneously re-organize to obtain the same orientation. Consequently, a crystal state must first transition into a mobile liquid state, subject to the same kinetic constraint as the first one, before it can transition into another crystal state; see Fig.~\ref{fig:arrowpotts_gb}b. 
\end{enumerate}
Note that the second kinetic constraint introduces glassy dynamics into the grain boundaries, which has been observed in colloidal systems \citep{Nagamanasa2011}. However, such coupling is not necessarily true at high annealing temperatures and for low-angle grain boundaries, where grain boundary dynamics can be understood from dislocation dynamics \citep{anderson2017theory}.  

Lastly, we need an equation of state for the chemical potential $h(T)$. A minimal form for $h(T)$ can be deduced from the bulk free energy $f_\mathrm{b}$ obtained by mean-field theory (MFT) analysis of Eq.~\eqref{eq:arrowpotts_fullhamilton}, which is given by 
\begin{eqnarray}
f_\mathrm{b} =&& \left(h_{\mathrm{eff}}(T)-\frac{z \Delta \epsilon}{2} \right)x_0+\frac{z}{2}\left(\Delta\epsilon x_0^2-\epsilon\sum_{k=0}^qx_k^2\right) \nonumber
\\
&&+k_\mathrm{B} T\sum_{k=0}^qx_k \ln x_k
\label{eq:arrowpotts_freenergy}
\end{eqnarray}
where $h_\mathrm{eff}(T) = h(T)-k_\mathrm{B} T \ln\left(1+2^d e^{-\beta J_0}\right)$, $x_k$ denotes the fraction of liquid ($k=0$) or crystal ($k>0$) states present in the system, and $z$ is the coordination number. The logarithmic term in $h_\mathrm{eff}(T)$ is the ensemble-averaged contribution of $\mathcal{H}_\mathrm{liq}$ in Eq.~\eqref{eq:arrowpotts_arrowpart}. 

Note the dependence of $h_\mathrm{eff}$ on $J_0$. This implies that $J_0$, a kinetic parameter, 
can influence the thermodynamics including its melting temperature $T_m$. However, it is desirable to study the influence of glassy dynamics (by varying $J_0$) under the same thermodynamic conditions. In other words, we want the melting temperature $T_m$ to be independent of $J_0$. To achieve this, we introduce a new parameter $\lambda$ defined as
\begin{equation}
\lambda  \equiv \frac{1}{T} \Big(h_\mathrm{eff}(T)-\frac{z \Delta \epsilon}{2}\Big) \,. \label{eq:arrowpotts_lambdadef}
\end{equation}
Substituting Eq.~\eqref{eq:arrowpotts_lambdadef} into Eq.~\eqref{eq:arrowpotts_freenergy}, the mean-field bulk free energy can be rewritten as 
\begin{align}
f_\mathrm{b} =& \lambda T  x_0+\frac{z}{2}\left(\Delta\epsilon x_0^2-\epsilon\sum_{k=0}^qx_k^2\right) \nonumber
\\
& + k_\mathrm{B} T\sum_{k=0}^qx_k \ln x_k \,.
\label{eq:arrowpotts_freenergy1}
\end{align} 
This change of variables in terms of $\lambda$ eliminates $J_0$ from the bulk free energy $f_\mathrm{b}$, allowing the thermodynamics to be independent of the model's kinetic parameter. Using Eq.~\eqref{eq:arrowpotts_lambdadef}, we can finally write the equation of state for $h(T)$ as  
\begin{equation}
h(T)=\frac{z \Delta \epsilon}{2}+T\left[\lambda+k_\mathrm{B} \ln\left(1+2^d e^{-\beta J_0}\right)\right] \,. \label{eq:arrowpotts_finalchempot}
\end{equation}
As can be seen from Eq.~\eqref{eq:arrowpotts_finalchempot}, at low temperatures ($\beta J_0 \gg 1$) the parameter $\lambda$ sets the slope of the chemical potential allowing us to favor the crystalline or liquid phase at different temperatures as long as  $\lambda < 0$. \ed{A more detailed explanation of the MFT and derivation of Eq.~\eqref{eq:arrowpotts_freenergy1} is given in \ed{SM \footnote{See Supplementary Material.}, Sec. 2.1-2.3}}. 

\section{Thermodynamics \& Phase Diagram} \label{sec:arrowpotts_thermodynamics}
\begin{figure}
    \centering
    \includegraphics[width=0.85\linewidth]{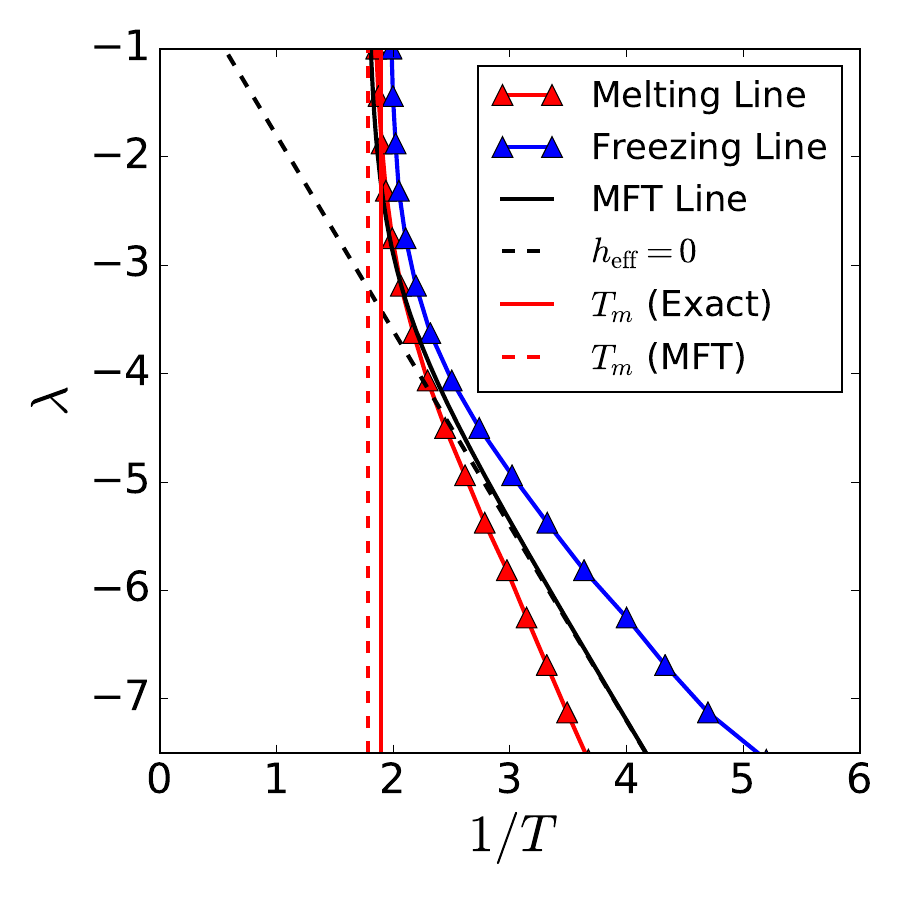}
    \caption{Phase diagram of the Arrow-Potts model confirming both the Ising and Potts regimes. Parameters: $q=32$, $\Delta \epsilon = 1.00$, $\epsilon = 0.90$, $J_0=0.25$, $N=100^2$.}
    \label{fig:arrowpotts_pd}
\end{figure}

To study the crystallization kinetics, we first compute the phase diagrams and corresponding melting temperatures $T_m$ for the Arrow-Potts model. This can be done through both the mean field theory (MFT) and cooling/melting Monte Carlo (MC) simulations as shown in Fig.~\ref{fig:arrowpotts_pd} for a two-dimensional (2D) system. $T_m$ can be obtained from MFT by finding the temperature for which the minimum of free-energy in Eq.~\eqref{eq:arrowpotts_freenergy1} of the liquid phase is equal to one of the $q$ degenerate minima of the solid phase.

The phase diagram of the Arrow-Potts model shown in Fig.~\ref{fig:arrowpotts_pd} is amenable to analytical results, which can be used to corroborate the MFT and MC results. These analytical results rely on an exact mapping of the model Hamiltonian in Eq.~\eqref{eq:arrowpotts_pottspart} to the \textit{Potts lattice gas} Hamiltonian \cite{Berker1978a,Nienhuis1979} given by
\begin{equation}
\mathcal{H}_\mathrm{PG} = -\sum_{\langle ij \rangle} [J_1+J_2\delta_{s_i,s_j}]\phi_i\phi_j  - \mu^\prime \sum_i (1-\phi_i) \label{eq:arrowpotts_pg} 
\end{equation}
where $s_i$ is the scalar Potts spin variable with $q+1$ states, $\phi_i=\{0,1\}$ is the lattice gas variable with $\phi_i=0$ and $\phi_i=1$ corresponding to liquid and crystal states respectively. The Potts lattice gas model in Eq.~\eqref{eq:arrowpotts_pg} is amenable to exact results and yields useful limiting cases \citep{Berker1976b,Berker1978a,Nienhuis1979}. 

To begin with, transforming the model in Eq.~\eqref{eq:arrowpotts_pottspart} to Eq.~\eqref{eq:arrowpotts_pg} yields $J_1 = \epsilon-\Delta \epsilon$, $J_2 = \epsilon$, and $\mu^\prime = -h_{\mathrm{eff}}(T)+z\epsilon-z\Delta \epsilon/2$; \ed{see SM \cite{Note1}, Sec. 2.3}. 
In the limit of $\lambda\to -\infty$, the Potts lattice gas transforms into the Ising model, and $T_m$ can be found when $h_\mathrm{eff}=0$
\begin{equation}
T_m \approx -\frac{z\Delta \epsilon}{2\lambda} \label{eq:arrowpotts_tm} \,,
\end{equation}  
where $z=4$ for a 2D square lattice. We shall call this limit \textit{the Ising regime}. In the opposite limit of $\lambda \to +\infty$, the Potts lattice gas transforms into \textit{the standard Potts model} \citep{Potts1952,Wu1982} and thus the melting temperature is the critical point, which is exactly known for a 2D square lattice and in MFT for any dimensions \citep{Kihara1954,Hintermann1978}
\begin{equation}
T_m \approx  \begin{cases}
\dfrac{\epsilon}{k_\mathrm{B} \ln(1+\sqrt{q})} & \text{(Exact)}  
\\
\dfrac{z\epsilon(q-1)}{2k_\mathrm{B} q\ln(q)} & \text{(MFT)} \end{cases}
\ .
\label{eq:arrowpotts_exactTc}
\end{equation}
The critical point stays as a first-order phase transition for $q>4$ \citep{Baxter1973}, and both exact and MFT results in Eq.~\eqref{eq:arrowpotts_exactTc} coincide with each other for $q \gg 1$ \citep{Mittag1974,Pearce1980}. We shall call this limit \textit{the Potts regime}. An extended discussion of the mapping of Arrow-Potts model to the Potts lattice gas and the analytical results is given \ed{in SM \cite{Note1}, Sec 2.3}. 

\begin{table}
\centering
\begin{ruledtabular}
\begin{tabular}{lcccccccc}
Name   & $d$ & $N$ & $q$ & $J_0$ & $\epsilon$ & $\Delta \epsilon$ & $\lambda$  \\ \hline
Set 1     & 2 & $100^2$ & 8 & 0.25 & 0.4545 & 0.3864 & -4.2849 \\ 
Set 2     & 2 & $100^2$ & 16 & 0.25 & 0.5357 & 0.4553 & -5.0498 \\ 
Set 3     & 2 & $100^2$ & 24 & 0.25 & 0.6155 & 0.5231 & -5.8019  \\ 
\end{tabular}
\end{ruledtabular}
\caption{List of sets of model parameters for all subsequent crystallization studies.}\label{table:arrowpotts_params}
\end{table}

The existence of Ising and Potts regimes can be readily confirmed with the phase diagram plotted on the $\lambda$ vs. $T$ plane in Fig.~\ref{fig:arrowpotts_pd}.
As $\lambda$ tends to zero, the phase boundaries from MFT and MC simulations approach the line corresponding to Eq.~\eqref{eq:arrowpotts_exactTc}, validating the prediction of the Potts regime. As $\lambda \ll -1$, the MFT phase boundary coincides exactly with Eq.~\eqref{eq:arrowpotts_tm}, validating the prediction of the Ising regime. Note that the model displays different melting/cooling lines with hysteresis effects from finite rate protocols. However, the analytical predictions are always in between these lines.

Given the thermodynamics and phase diagram 
of the Arrow-Potts model, in what follows we focus our study on aspects of crystallization vs. vitrification choosing sets of parameters that fall within the Ising regime. These parameters are tabulated in Table~\ref{table:arrowpotts_params}, and will be used to test the unified theory developed later. Crystallization studies corresponding to the Potts regime will be left to future work.

\section{TTT Diagrams and Microstructure} \label{sec:arrowpotts_results}
As mentioned before, TTT diagrams are sensitive to protocols \citep{christian2002theory}. To explore this aspect of TTT diagrams, we study two idealized crystallization protocols within the model:
\begin{itemize}
\item \textit{Supercooled liquid crystallization:} the supercooled state of the model is prepared at any given temperature $T$ such that no crystal states are initially present. This is achieved using the Arrow model from Section~\ref{sec:arrow_describe}, where one can obtain an equilibrium configuration made purely from the liquid states. This configuration is then taken as the initial metastable configuration for the Arrow-Potts model running at the same temperature, allowing for crystallization.
\item \textit{Quenching:} the liquid state is prepared at the melting temperature $T_m$ in the same way as the first protocol, and crystallization is initiated immediately at some operating temperature $T<T_m$.
\end{itemize}

\begin{figure}
    \centering
    \includegraphics[width=\linewidth]{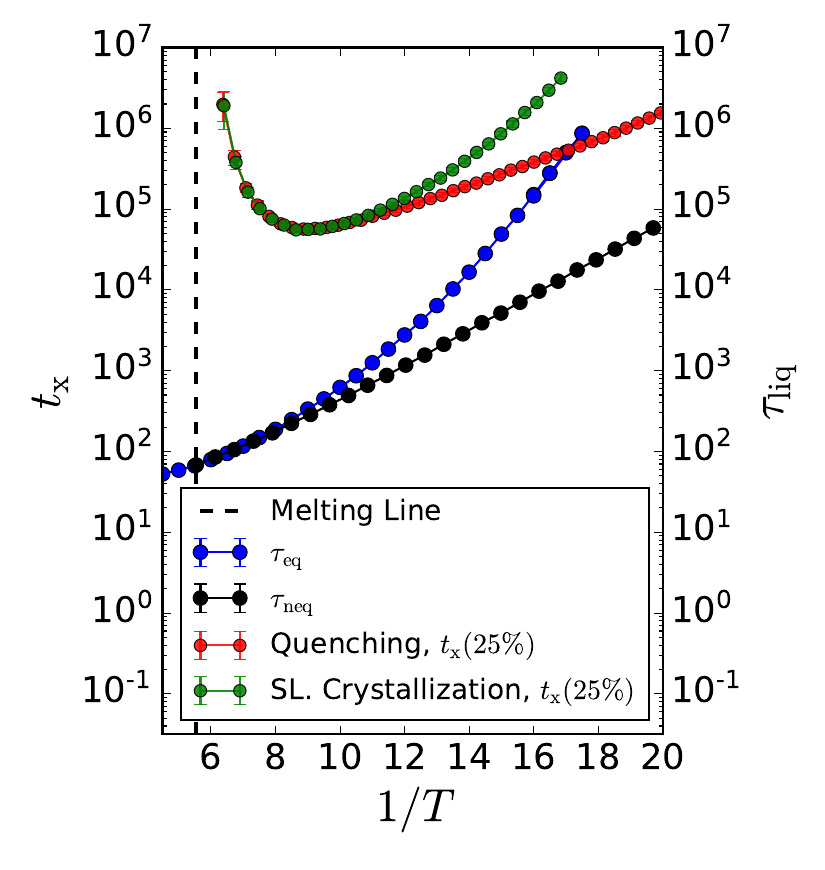}
    \caption{TTT diagram from supercooled liquid crystallization (green) and quenching (red) protocols for Set 2 plotted alongside the Arrow model's liquid relaxation time at equilibrium $\tau_\mathrm{eq}$ (blue) and during quenching $\tau_\mathrm{neq}$ (black). Each temperature point is an average over 50 trajectories and the time scales are reported in Monte Carlo sweeps (MCS).}
    \label{fig:arrowpotts_ttt}
\end{figure}

\begin{figure*}
\centering
\includegraphics[width=\textwidth]{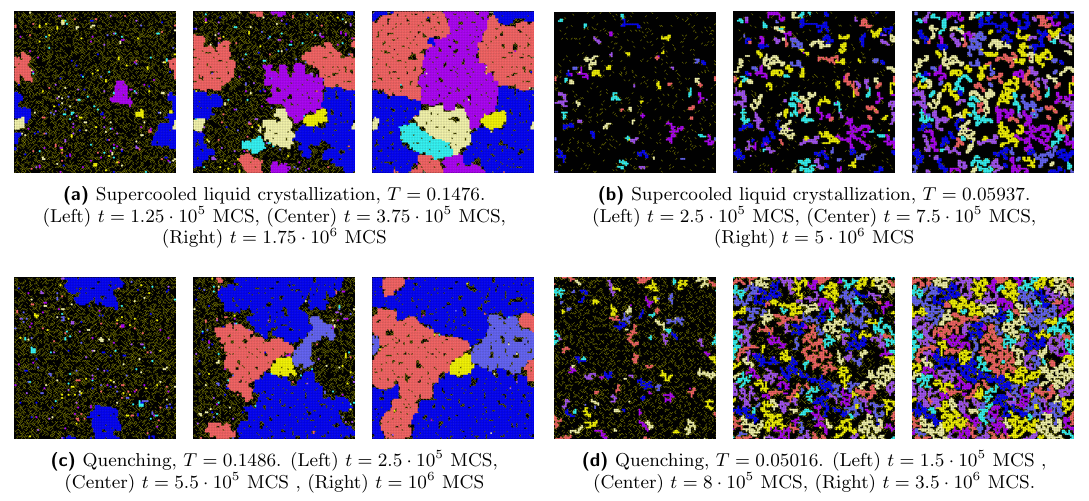}
\caption{Sample trajectory of Set 1 for supercooled liquid crystallization at (a) high-$T$ and (b) low-$T$ and quenching at (c) high-$T$ and (d) low-$T$. Empty black sites represent immobile liquid states while black sites with yellow arrows represent mobile liquid states. The rest of all colored sites are crystal states.}
\label{fig:arrowpotts_microstructure}
\end{figure*}

These two choices represent idealized limits of the protocols, as they correspond to the limiting cases of finite cooling-rates typically employed in experimental studies on crystallization \citep{Masuhr1999,Schroers1999,Hays1999,Legg2007a,Mukherjee2004a,Gallino2007}. Supercooled liquid crystallization represents a protocol where the liquid is prepared carefully and cooled sufficiently slow so that metastability is reached no matter how deep it is being super-cooled. In contrast, the quenching protocol represents a protocol with an infinitely fast cooling rate, which forces the liquid to fall out of metastable equilibrium at any temperature $T<T_m$. Studying the model with these limiting cases constitutes the first step towards understanding protocol effects in crystallization. 
Figure~\ref{fig:arrowpotts_ttt} shows TTT diagrams for the two protocols as computed from the Arrow-Potts model along with the liquid relaxation times. Note that TTT diagrams typically plot the time scales of crystallization and liquid relaxation \citep{Legg2007a,Mukherjee2004a,Gallino2007}. The crystallization time scale $t_\mathrm{x}(x\%)$ is calculated as the mean time to reach a chosen percentage of crystal fraction from the Arrow-Potts model.  
For the supercooled liquid crystallization protocol, the relevant $\tau_\mathrm{liq}$ corresponds to the relaxation timescale of the Arrow model at equilibrium ($\tau_\mathrm{eq}$), which follows the parabolic law in Eq.~\eqref{eq:parablaw}. However, for the quenching protocol, $\tau_\mathrm{liq}$ corresponds to a non-equilibrium timescale ($\tau_\mathrm{neq}$) associated with the irreversible relaxation/equilibration from a high-temperature state to a lower one. The analytical calculation of $\tau_\mathrm{neq}$ will be addressed later in Section~\ref{sec:tau_neq}. 

As can be seen from Fig.~\ref{fig:arrowpotts_ttt}, the time scales for crystallization $t_\mathrm{x}(x\%)$ for both protocols coincide well at high temperatures. However, at low temperatures, $t_\mathrm{x}(x\%)$ from supercooled liquid crystallization exhibits a super-Arrhenius behavior much like $\tau_\mathrm{eq}$, while $t_\mathrm{x}(x\%)$ from the quenching protocol exhibits an Arrhenius behavior, which resembles $\tau_\mathrm{neq}$. The fact that these two protocols produce super-Arrhenius vs. Arrhenius behaviors from the Arrow-Potts model suggests that they may provide a key to understand the cross-over from Trend I and Trend II typically observed in the experimental TTT diagrams (Fig.~\ref{fig:experimental}):
\begin{itemize}
\item \textit{Trend I}: At high enough temperatures and suitable cooling rates, quenches are sufficiently shallow so that the system can equilibrate to its metastable state prior to the onset of crystallization. Thus, Trend I is a result of crystallizing the supercooled liquid at the desired target temperature.  
\item \textit{Trend II}: When the system is being cooled too quickly at some finite rate, the liquid cannot keep up with the large perturbations in temperature over time. Thus, it begins to fall out of metastable equilibrium, and the low-temperature crystallization time exhibits an Arrhenius trend much like the liquid relaxation times corresponding to the quenching protocol of the Arrow model. 
\end{itemize}
In other words, the location of the cross-over is a balance between the cooling rate and the system's intrinsic capability to relax into metastable states.

The final microstructure of the material depends on the temperature at which the material is formed. Figure~\ref{fig:arrowpotts_microstructure} shows representative trajectories of crystallization from the Arrow-Potts model in the two protocols at high and low temperatures. At high temperatures,  as shown in Fig.~\ref{fig:arrowpotts_microstructure}a, the system spends majority of its time nucleating crystals. 
Once a critical nucleus is reached, the crystalline regions grow and finally result in a classic polycrystalline structure (Fig.~\ref{fig:arrowpotts_microstructure}a-Right). 
In contrast, at low temperatures, crystals nucleate immediately 
but grow slowly into fractal and ramified crystal grains  (Fig.~\ref{fig:arrowpotts_microstructure}b,d-Right). This fractal morphology is consistent with the MD simulations of super-compressed hard spheres \citep{valeriani2012compact, Puser2009,Sanz2011,Sanz2014}. Once the mobile liquid states are expended and crystal growth subsided, the final microstructure is a mixture of crystalline and immobile liquid domains. In other words, the system only partially crystallizes/vitrifies at lower temperatures, and produces ramified crystalline domains embedded in a larger disordered matrix (Fig.~\ref{fig:arrowpotts_microstructure}b-Right). Taken together, these observations suggest that at high temperatures, there exist substantial nucleation free energy barriers while growth proceeds quickly. However, at low temperatures, the nucleation barriers are diminished, while the growth is affected because of the enormous slowdown from glassy liquid dynamics.

Furthermore, there are no discernible qualitative differences in the microstructures resulting from both the supercooled liquid crystallization and quenching protocols at both high and low temperatures (Figs.~\ref{fig:arrowpotts_microstructure}a-d). However, as can be seen from the crystallization time scales in Fig.~\ref{fig:arrowpotts_ttt} and visual inspection in Fig.~\ref{fig:arrowpotts_microstructure}b and  Fig.~\ref{fig:arrowpotts_microstructure}d, there exists more crystal fractions at lower temperatures in the quenching protocol than the supercooled liquid crystallization protocol. This can be understood by realizing that crystals in the quenching protocol result from an initially hotter liquid, which has more mobile regions or excitations compared to the equilibrium supercooled liquid at the same temperature. Thus, the ability to crystallize from the mobile regions is greater in the quenching protocol resulting in lower time scales of crystallization compared to the supercooled case (Fig.~\ref{fig:arrowpotts_ttt}). 

\section{A Theory for Crystallization Time} \label{sec:arrowpotts_theory}

In this section, we derive an analytical formula for the crystallization time $t_\mathrm{x}(x\%)$ for the two protocols as plotted in the TTT diagram  in Fig.~\ref{fig:arrowpotts_ttt}. To this end, we begin by realizing from the aforementioned observations that there exist two important timescales controlling crystallization: (1) the nucleation time $\tau_\mathrm{nuc}$ corresponding to a critical crystal cluster, and (2) the liquid relaxation time $\tau_\mathrm{liq}$ related to the growth of the crystal regions. At high temperatures, crystallization is nucleation-dominated despite the rapid kinetics and thus, 
\begin{equation}
\tau_\mathrm{nuc} \gg \tau_\mathrm{liq} \quad \text{for} \ T\to T_m \,.
\end{equation}
At low temperatures, crystallization is growth-dominated and is controlled by the glassy dynamics and thus,  
\begin{equation}
\tau_\mathrm{nuc} \ll \tau_\mathrm{liq} \quad \text{for} \ T \ll T_m \,.
\end{equation}
In the intermediate temperatures, both the nucleation and relaxation timescales are comparable to each other. 

The analytical formula for crystallization times requires unification of theories of nucleation, crystal growth, glassy relaxation dynamics, and phase transformation. 
In what follows, we briefly summarize the four key theories utilized and developed in our work:
\begin{enumerate}
\item \textit{The classical and field theories of nucleation} \citep{Fisher1967,Langer1992,langer1969statistical} provide an accurate formula for the crystal nucleation rate $\bar{I}$, and therefore the nucleation time $\tau_{\mathrm{nuc}} = 1/\bar{I}$. This is achieved by relating the nucleation free-energy barrier to the size of the critical cluster and the capillary fluctuations of its interface.
\item \textit{Dynamical facilitation theory} provides a formula for the relaxation time $\tau_{\mathrm{neq}}$ of quenched liquids and explains the origin of the Arrhenius energy barrier in the quenching protocol. This formula complements the parabolic law (Eq.~\eqref{eq:parablaw}) derived for equilibrium relaxation from Section~\ref{sec:arrow_describe}.
\item \textit{A random walk theory of crystal growth} provides a formula for the effective radius of the growing crystal as a function of time, $R_g(t)$, by accounting for the influence of dynamical heterogeneity on the crystal-growth pathways. 
\item The above three theories are finally combined using the \textit{Kolmogorov-Johnson-Mehl-Avrami} (KJMA) theory \citep{kolmogorov1937statistical,william1939reaction,avrami1940kinetics}, which gives us an explicit formula for the crystallization time $t_\mathrm{x}$ as a function of temperature. With the final theory, we quantitatively capture the non-monotonic TTT diagrams for different protocols.
\end{enumerate}

We note that while the field theories of nucleation and the KJMA theory have been developed in the past, our work presented here introduces new theories related to crystal growth, non-equilibrium liquid relaxation times for the quenching protocols, and subsequent extensions of the KJMA theory to unify the different theories, which culminates into the final formula for the crystallization time $t_\mathrm{x}$. We further note that only ideas and minimal sets of derivations are presented in this section; \ed{more detailed calculations can be found in the SM \cite{Note1}}. 

\subsection{Nucleation Theory} 
We begin by a brief description of the classical nucleation theory (CNT) \citep{Langer1992} to calculate the nucleation free energy barrier, review the Becker-D{\"o}ring theory \citep{Becker1935} to obtain the nucleation rate, and end with the field theory of nucleation \citep{langer1969statistical} to obtain the corrections to the nucleation rate due to fluctuations of the crystal-liquid interface.

\subsubsection{Classical Nucleation Theory (CNT)}  \label{subsec:arrowpotts_slc}
In CNT, the nucleation free-energy barrier $\Delta F_\mathrm{CNT}$ is the result of two competing effects: (1) the chemical potential difference between crystal and liquid states $\Delta \mu$, which thermodynamically drives the system to crystallize, and (2) the interfacial tension $\gamma$, which provides a free-energy cost to have a crystal-liquid interface. 

For lattice-based models like the Arrow-Potts model,  the free-energy of a nucleating cluster is given by
\begin{equation}
\Delta F(N) =  -\Delta \mu (N-1)+ \gamma  (N-1)^{(d-1)/d} \label{eq:nucgrowth_cntfreenergy} \,,
\end{equation}
where $N$ is the number of sites occupied by the crystal states making up the cluster. 
Note that Eq.~\eqref{eq:nucgrowth_cntfreenergy} is similar to the free energy of a liquid droplet written in the context of vapor-liquid equilibrium \citep{Fisher1967,Langer1992}, where $N$ indicates the number of atoms attached to the droplet. The free-energy barrier  $\Delta F_\mathrm{CNT}$ can then be computed from Eq.~\eqref{eq:nucgrowth_cntfreenergy} by finding its maximum. In other words, $\Delta F_\mathrm{CNT}=\Delta F(N_c)$ where $N_c$ is the critical size of the cluster given by
\begin{equation}
N_c = 1+\left(\frac{(d-1)\gamma}{d \Delta \mu}\right)^d \,. \label{eq:nucgrowth_cntcriticalsize}  
\end{equation}

To obtain the nucleation rate, the CNT can be combined with
\textit{the Becker-D{\"o}ring theory} \citep{Becker1935}, which assumes that nucleation proceeds by step-wise addition and removal of the smallest constituent of the nucleating cluster. This theory also assumes that the addition/removal processes occur within large enough clusters ($N \gg 1$) so that the population of clusters of size $N$, denoted as $\nu_N$, evolve in time through a Fokker-Planck equation
\begin{eqnarray}
\frac{\partial \nu_N}{\partial t} =&&-\frac{\partial I}{\partial N}=\frac{\partial}{\partial N}\left[a(N) \langle \nu_N \rangle \frac{\partial}{\partial N}\left(\frac{\nu_N}{\langle \nu_N \rangle}\right)\right] \,, \label{eq:beckerdoring_newfokker}
\\
\langle  \nu_N \rangle =&& \langle  \nu_1 \rangle e^{-\beta \Delta F(N)} \,,
\end{eqnarray} 
where $I$ is the flux of population of nucleating clusters of size $N$, $a(N)$ is the rate of growth of a cluster from size $N$ to $N+1$, and $\langle  \nu_N \rangle$ is the equilibrium population of clusters of size $N$ with $\Delta F(N)$ computed from Eq.~\eqref{eq:nucgrowth_cntfreenergy}. \ed{See SM \cite{Note1}, Sec. 3.1 for a complete derivation of Eq.~\eqref{eq:beckerdoring_newfokker}}.

In principle, nucleation can be a time-independent phenomenon, a scenario supported by Eq.~\eqref{eq:beckerdoring_newfokker}. To demonstrate this, we set up the system so that it stays metastable even at long times. This is achieved by depleting the populations of large clusters ($N \gg N_c$) and equilibrating the populations of small clusters ($N \ll N_c$) 
\begin{equation}
\lim_{N \to \infty}\nu_N = 0  \ \ \text{and} \ \ \lim_{N\to 0}\nu_N = \langle  \nu_N \rangle \label{eq:becker_bc} \,.
\end{equation}
Equations~\eqref{eq:becker_bc} serve as boundary conditions for Eq.~\eqref{eq:beckerdoring_newfokker}. Solving Eq.~\eqref{eq:beckerdoring_newfokker} yields the nucleation rate from the Becker-D{\"o}ring theory $\bar{I}_\mathrm{BD}$ as 
\begin{eqnarray}
\bar{I}_\mathrm{BD} = && \left[\int_{0}^\infty\dfrac{1}{a(N^\prime)\langle \nu_{N^\prime} \rangle}\diff N^\prime\right]^{-1} \label{eq:cnt_integralrate}
\\
= && \kappa_0 A_\mathrm{c}(T) Z_0(T)e^{-\beta \Delta F_\mathrm{CNT}} \label{eq:final_cnt} \,,
\end{eqnarray}
where $Z(T) = \sqrt{-\Delta F_\mathrm{CNT}^{\prime\prime}(N_c)/2\pi k_\mathrm{B} T}$ is the Zel'dovich factor \citep{Zeldovich1943},  $A_\mathrm{c}(T)$ is the surface area of the critical cluster, and $\kappa_0$ is some kinetic factor; \ed{see SM \cite{Note1}, Sec. 3.1 for the derivation of Eq.~\eqref{eq:cnt_integralrate}-\eqref{eq:final_cnt}}. 
For an application of the Becker-D{\"o}ring theory to the Ising model, see Ref.~\citep{Maibaum2008}.

\subsubsection{Field Theory of Nucleation}
While CNT is intuitive, the real nucleation process is a hopping process between one basin of attraction to another in a higher-dimensional free-energy landscape. In the case of crystallization, one basin corresponds to the liquid phase while the other corresponds to the crystalline phase. The critical cluster will then be the configuration of the system at the saddle point separating these two basins. In other words, a theory which relies upon a single reaction coordinate like CNT might miss important degrees of freedom relevant for understanding nucleation correctly. \textit{The field theory of nucleation} \citep{langer1969statistical} resolves this issue by assuming that there exists a description of the system in terms of a continuous order parameter field $\phi(\* x,t)$. This theory incorporates Gaussian fluctuation modes of $\phi(\* x,t)$ at the transition state, and leads to a more accurate formula for the nucleation rate.

The theory can be applied to any first-order phase transition, as long as the system is amenable to a field-theoretic description; for example, an Ising model \citep{Goldenfeld2018,Amit2005}. As a result, we can apply it to study nucleation in the Arrow-Potts model, whose field-theoretic description can be deduced from its mapping to the Potts lattice gas \cite{Berker1978a,Nienhuis1979, Wu1982}, which in turn, recall from Section~\ref{sec:arrowpotts_thermodynamics}, behaves like the Ising model within the current set of model parameters (Table~\ref{table:arrowpotts_params}). This implies that the field-theoretic description of the Ising model is directly applicable to the Arrow-Potts model.

The field-theoretic description of the Ising model is the Landau-Ginzburg (LG) model; see Ref.~\citep{Goldenfeld2018,Amit2005} for details on how to map the Ising model onto the LG model. Furthermore, the field theory of nucleation has been applied to the LG model. In fact, this was the first intended application of the original work \citep{langer1969statistical}, although later works \citep{gunther1980goldstone,Zwerger1985,munster2000analytical,munster2003classical} showed that the mathematical derivations can be further simplified. As a result, we can apply the nucleation rate formula derived for the LG model directly to the Arrow-Potts model. 

We will now sketch the derivation of the nucleation rate formula, as applied to the LG model; \ed{see SM, Sec. 3.2 for a complete derivation} which follows closely Ref.~\citep{gunton1983introduction}, the original work \citep{langer1969statistical}, and other past works \citep{gunther1980goldstone,Zwerger1985,munster2000analytical,munster2003classical}. To begin, we write the Langevin equation for the order parameter field $\phi(\* x,t)$ under non-conserving dynamics as a stochastic partial differential equation
\begin{equation}
\frac{\partial \phi(\* x,t)}{\partial t} =  -\Gamma_0 \frac{\delta F}{\delta \phi} + \eta(\* x,t) \label{eq:modela_langevin} \,,
\end{equation}
where $\Gamma_0$ is a constant mobility factor, $\eta(\*x,t)$ is a space-time Gaussian white-noise, and the free-energy functional $F[\phi]$ for the LG model is given by 
\begin{equation}
F[\phi] = \int \diff^d \*x\left[\frac{1}{2}|\nabla \phi|^2-\frac{1}{2}\tau \phi^2 +\frac{1}{4!}g \phi^4-h\phi \right] \,,
\end{equation}
where $h$ is an external field, and $\tau$ and $g$ are phenomenological parameters. Now, let $P(\{\phi(\* x)\},t)$ be the probability of finding an order parameter field $\phi(\*x)$ at time $t$. The Fokker-Planck equation associated with Eq.~\eqref{eq:modela_langevin} is 
\begin{equation}
\frac{\partial P(\{\phi(\* x)\},t)}{\partial t} = -\int \diff^d \*x^\prime \frac{\delta J}{\delta \phi(\*x^\prime)} \label{eq:modela_fokkerplanck}
\end{equation}
where $J$ is the probability current density given by 
\begin{equation}
J = -\Gamma_0\left(P\frac{\delta F}{\delta \phi(\*x^\prime)}+k_\mathrm{B} T \frac{\delta P}{\delta \phi(\*x^\prime)}\right)\,. \label{eq:langer_probcurrent}
\end{equation}
A derivation of Eq.~\eqref{eq:modela_fokkerplanck} from Eq.~\eqref{eq:modela_langevin} can be found in standard textbooks; see Ref.~\citep{Goldenfeld2018}.

The field theory of nucleation starts by obtaining the mean-field solution for the saddle point given by $\frac{\delta F}{\delta \phi} = 0$ and denoted as $\bar{\phi}(\* x)$. A detailed derivation for $\bar{\phi}(\* x)$ can be found in Ref. \citep{munster2000analytical,munster2003classical}. In the LG model, $\bar{\phi}(\* x)$ physically represents a spherical cluster of the stable phase in a background of the metastable phase. Expanding the free-energy functional $F[\phi]$ about $\bar{\phi}(\* x)$ up to quadratic order yields
\begin{equation}
F  =  F[\bar{\phi}] + \frac{1}{2}\int \diff^d \* x \int \diff^d \*x^\prime \ \hat{\phi}(\*x) \mathcal{M}(\* x,\* x^\prime) \hat{\phi}(\*x^\prime) + \cdots \,,   \label{eq:langer_freenergyexpand}
\end{equation}
where $\hat{\phi}(\* x) = \phi(\* x)-\bar{\phi}(\* x)$ and $\mathcal{M}(\* x,\* x^\prime)$ is defined as
\begin{equation}
\mathcal{M}(\* x,\* x^\prime)=\left. \frac{\delta^2 F}{\delta \phi(\* x)\delta \phi(\* x^\prime)}\right|_{\phi=\bar{\phi}} \,.
\end{equation}
If we view $\mathcal{M}$ as the "Hessian" of the LG model evaluated at the saddle point, we can use its eigenfunction expansion to reduce Eq.~\eqref{eq:langer_freenergyexpand} into
\begin{equation}
F =  F[\bar{\phi}]  + \frac{1}{2}\sum_n \lambda_n\xi_n^2 + \cdots \,, \label{eq:langer_freenergyexpand1}
\end{equation}
where $\lambda_n$ is the $n$-th eigenvalue and $\xi_n$ is the coefficient of the $n$-th eigenfunction. The details of this eigenfunction expansion can be found \ed{in SM \cite{Note1}, Sec 3.2.1}. Note that in the original work \citep{langer1969statistical}, the author developed the expansion on a finite-dimensional system so that the eigenfunction expansion transforms into an eigendecomposition of a Hessian matrix.

The eigenvalues $\lambda_n$ characterize the fluctuation modes at the saddle point and they can be broadly classified into three types:
\begin{enumerate}
\item \textit{Negative eigenvalue}, $\lambda_0 < 0$, represents a fluctuation mode which lowers the free-energy and is responsible for bringing the system towards the basin of the stable state. 
\item \textit{Zero eigenvalues}, $\lambda_1 = 0$, are fluctuation modes with no energetic penalty. In the LG model, these fluctuations move the droplet's center of mass, which costs no energy due to translation-invariance of $F[\phi]$. There are as many zero modes as the real-space dimension to reflect the total translational degrees of freedom.  
\item \textit{Positive eigenvalues}, $\lambda_{n>1}>0$, are fluctuation modes which increase the free-energy. For the LG model, these are capillary wave excitations. The physical connection between capillary wave excitations to the fluctuation modes at the saddle point was first shown in Ref. \citep{gunther1980goldstone}, which was instrumental in generalizing the nucleation rate formula to any dimension.
\end{enumerate}

At this stage, the Fokker-Planck equation can be solved by setting up boundary conditions to maintain the metastable state at long times and combine them with the eigenfunction expansion used in Eq.~\eqref{eq:langer_freenergyexpand1}.  Subsequently, one can use the solution of the Fokker-Planck equation to obtain the nucleation rate through the probability current density $J$ in Eq.~\eqref{eq:langer_probcurrent}; \ed{see SM \cite{Note1}, Sec. 3.2.1}. The final result for the nucleation rate derived from the field theory, $\bar{I}_\mathrm{FT}$, can be written in terms of the CNT free-energy barrier, $\Delta F_\mathrm{CNT}$, as 
\begin{equation}
\bar{I}_\mathrm{FT} = \frac{|\kappa|}{2\pi} \Omega_0 e^{-\beta \Delta F_\mathrm{CNT}} \,, \label{eq:langer_nucleationformula}
\end{equation}
where $\kappa$ is a kinetic prefactor, and $\Omega_0$ is an effective statistical prefactor  that lumps the fluctuation modes together given by 
\begin{equation}
\Omega_0 = \mathcal{V} \left( \frac{\lambda_1^0}{2 \pi k_\mathrm{B} T} \right)^{d/2}\prod_{n \neq 1} \left(\frac{\lambda_n^0}{|\lambda_n|} \right)^{\nu_n/2}\,, \label{eq:langer_statisticalprefactor} 
\end{equation}
with $\lambda^{(0)}_n$ being the $n$-th eigenvalue of the fluctuation mode at the metastable basin, $\nu_n$ the degeneracy of each $n$-th eigenvalue, and $\mathcal{V}$ is the contribution from zero-eigenvalue modes.

To apply Eqs.~\eqref{eq:langer_nucleationformula}-\eqref{eq:langer_statisticalprefactor} to the Arrow-Potts model, we need to express Eqs.~\eqref{eq:langer_nucleationformula}-\eqref{eq:langer_statisticalprefactor} in terms of the chemical potential $\Delta \mu$ and interfacial tension $\gamma$. Specializing to 2D, for which we later test the Arrow-Potts model, the nucleation rate is given by
\begin{eqnarray}
\bar{I}_\mathrm{FT} = \kappa_0 \sqrt{\frac{\Delta \mu}{k_\mathrm{B} T}}\exp\left(-\beta \Delta F\right) \,, \label{eq:arrowpotts_nucleationformula}
\\
\Delta F = \underbrace{\frac{\pi \gamma^2}{\Delta \mu}}_{\Delta F_\mathrm{CNT}}+\underbrace{\frac{5}{2}k_\mathrm{B} T \ln \left(\frac{\sqrt{\pi} \gamma}{\Delta \mu}  \right)}_{\Delta F_\mathrm{corr}}, \label{eq:arrowpotts_finaldeltfformula}
\end{eqnarray}
where $\Delta F_\mathrm{corr}$ can be physically thought of as a correction to CNT due to capillary wave fluctuations. Note that the derivation of Eqs.~\eqref{eq:arrowpotts_nucleationformula}-\eqref{eq:arrowpotts_finaldeltfformula} from Eqs.~\eqref{eq:langer_nucleationformula}-\eqref{eq:langer_statisticalprefactor} is the most difficult part of the theory, due the product of eigenvalues present in Eq.~\eqref{eq:langer_statisticalprefactor}; \ed{see SM, Sec. 3.2.2 for a treatment of this product using cutoff regularization \citep{ gunther1980goldstone,Zwerger1985,gunton1983introduction}}. Other regularization techniques can be used to simplify Eq.~\eqref{eq:langer_statisticalprefactor}, some of which include analytic continuation \citep{Langer1967}, zeta-function regularization \citep{munster2000analytical}, and dimensional regularization \citep{munster2003classical}.

We end this section by making two important remarks. First, we have neglected the influence of glassy dynamics since Eqs.~\eqref{eq:arrowpotts_nucleationformula}-\eqref{eq:arrowpotts_finaldeltfformula} will only be applied to the nucleation-dominated regime, which only exists at high temperatures. Second, the fluctuation correction is necessary for the Ising model in 2D  \citep{Jacucci1983,Ryu2010}. Thus, it will be a crucial correction to the nucleation barrier in the Arrow-Potts model as will be shown later. 

\subsection{Theory for Liquid Relaxation Timescales}\label{sec:tau_neq}
Recall from Fig.~\ref{fig:arrowpotts_ttt} that at low temperatures, the TTT diagram of the two protocols deviate from one another. As mentioned before, the crystallization time $t_\mathrm{x}$ from the quenching protocol exhibits Arrhenius behavior while the supercooled liquid crystallization protocol exhibits  super-Arrhenius behavior. Moreover, these behaviors are consistent with the corresponding trends in liquid relaxation times. The super-Arrhenius behaviors in $t_\mathrm{x}$ from the supercooled crystallization protocol can be expected given that the glassy dynamics of the metastable equilibrium state, already discussed in Section~\ref{sec:arrow_describe}, limits the crystal growth rate. However, understanding the Arrhenius behaviors in $t_\mathrm{x}$ is contingent upon decoding the origin of the Arrhenius behaviors in the non-equilibrium relaxation times $\tau_\mathrm{neq}$. In this section, we derive analytical formulas for the Arrhenius behaviors of $\tau_\mathrm{neq}$ for the Arrow model. 

Note that the Arrhenius behavior of $\tau_\mathrm{neq}$ is not uncommon, and has been observed in both MD simulations and kinetically constrained models of glass formers \cite{Hudson2016,Hudson2018,limmer2014length}. In fact, these simulations have revealed that glassy systems, when subjected to finite-rate cooling protocols, exhibit a cross-over from super-Arrhenius at high temperatures to Arrhenius behaviors at low temperatures around the cooling rate dependent glass transition temperature $T_\mathrm{g}$. At high temperatures the system still exhibits hierarchical relaxation until it falls out of equilibrium around $T_\mathrm{g}$, which sets a finite length scale for the separation between excitations $\ell_\mathrm{neq} \approx \ell_\mathrm{eq}(T_\mathrm{g})$ \cite{Hudson2018}. This then sets a temperature independent barrier $E_\mathrm{neq}$, which results in the Arrhenius behaviors at low temperatures. Inspired by these observations, we now identify the non-equilibrium energy barrier $E_\mathrm{neq}$ for the quenching protocol relevant for this work.  

\begin{figure}
    \centering
    \includegraphics[height=0.85\linewidth]{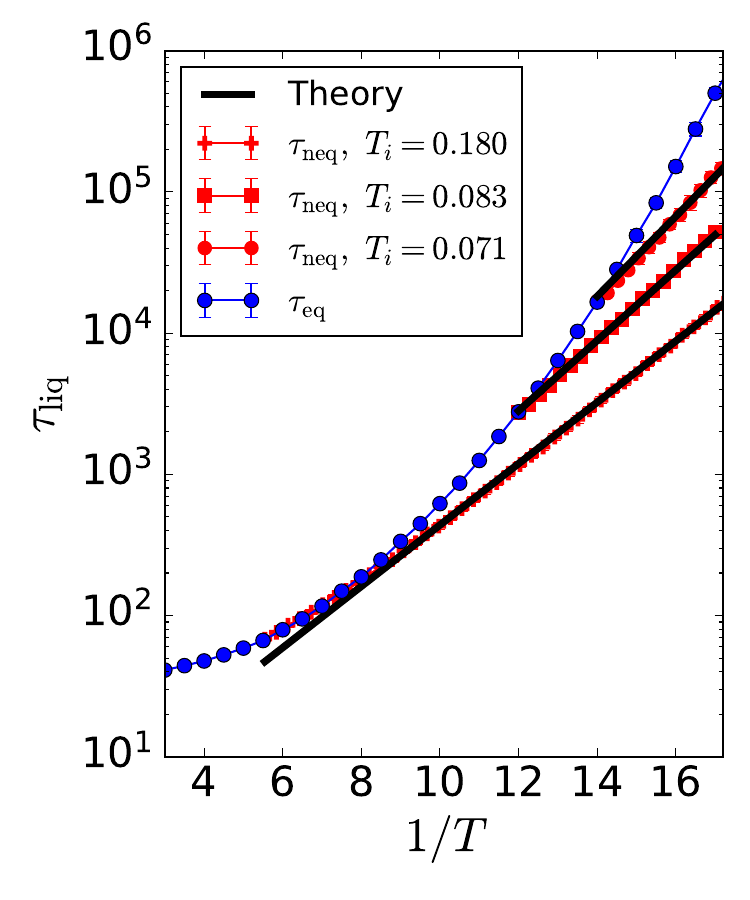}
    \caption{Arrhenius plot of the Arrow model's relaxation time obtained through quenching from different initial temperatures $T_i$. Black lines are the theoretical predictions from Eq.~\eqref{eq:noneqtime}.} 
    \label{fig:arrow_eneqwithfits}
\end{figure}

\begin{figure}
    \centering
    \includegraphics[height=0.85\linewidth]{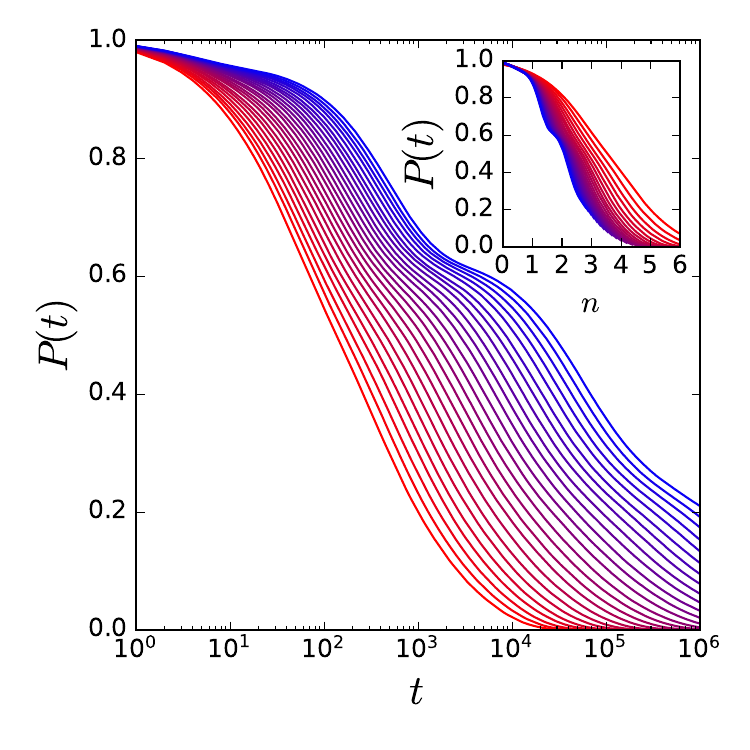}
    \caption{The decay of persistence function $P(t)$ as a function of time, measured in Monte Carlo sweeps (MCS) for a quenching protocol from an initial temperature $T_i = 0.180$. Inset shows the same data using a new rescaled time, defined by $n=(k_\mathrm{B} T/J_0) \ln t$. Note the collapse of low temperature data to a staircase master curve.} 
    \label{fig:arrow_persistnoneq}
\end{figure}

Figure~\ref{fig:arrow_eneqwithfits} shows the relaxation times $\tau_\mathrm{neq}$ as a function of temperature when quenched from different initial temperatures $T_i$, exhibiting Arrhenius behaviors with different energy barriers. To understand these relaxation behaviors, we plot the time-evolution of the persistence function $P(t)$ in Fig.~\ref{fig:arrow_persistnoneq} for instantaneous quenches from one particular high temperature ($T_i=0.180$) to lower temperatures. Note that as we quench to deeper temperatures, $P(t)$ develops a sequential staircase decay with different stages of relaxation. The staircase decay arises from the inherent distribution of relaxation timescales \citep{Sollich1999,Sollich2003}, which in turn arises due to hierarchical relaxations of pairs of excitations with their own energy barriers. During quenching, this distribution is non-stationary due to a decrease in the concentration of excitations as a function of time. Given the hierarchical nature of relaxation, the relaxation time for a pair of excitations separated by $\ell$ can be built from a single spin relaxation $\tau_1 \sim e^{\beta J_0}$ using the following relation 
\begin{equation}
\tau_n = (\tau_1)^n \quad \text{for} \ 2^{n-1} < \ell \leq 2^n \,, \label{eq:liqrelax_timescale}
\end{equation}
where $n$ is an integer; \ed{see SM \cite{Note1}, Sec. 1.3}. 
\ed{A derivation of Eq.~\eqref{eq:liqrelax_timescale} for the simpler one-dimensional East model can be found in Refs.~\citep{Sollich1999,Sollich2003}.}
Using Eq.~\eqref{eq:liqrelax_timescale}, one may rescale time $t$ in  terms of the relaxation stages of the staircase as $\ln t = n\beta J_0$, which then allows the persistence function $P(t)$ to follow a master curve (see Fig.~\ref{fig:arrow_persistnoneq} inset).  
Since the relaxation time $\tau_\mathrm{neq}$ is given by the relation $P(\tau_\mathrm{neq})=1/e$, the energy barrier $E_\mathrm{neq}$ can be calculated by identifying the corresponding relaxation stage in the staircase decay. For quenching relaxations from high $T_i$, the inset in Fig.~\ref{fig:arrow_persistnoneq} shows a relaxation stage corresponding to $n\approx 2$, which then yields 
\begin{equation}
E_\mathrm{neq} \approx E_{n=2} =  2 J_0 \,, \label{eq:barrier_neq}
\end{equation}
where $E_n=n J_0$ consistent with Eq.~\eqref{eq:liqrelax_timescale} \citep{Faggionato2012,Sollich1999,Sollich2003}; see also \ed{SM \cite{Note1}, Sec. 1.3}. 

\begin{figure}
    \centering
    \includegraphics[height=0.85\linewidth]{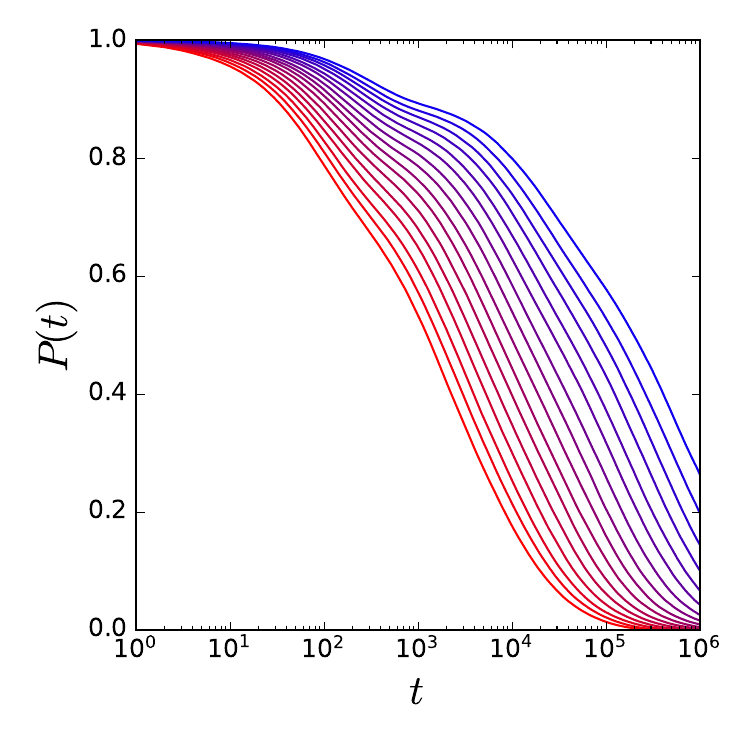}
    \caption{The decay of persistence function $P(t)$ as a function of time, measured in Monte Carlo sweeps (MCS) for a quenching protocol from an initial temperature $T_i = 0.083$. Note the lack of relaxation stages in comparison to quenches from high temperature in Fig.~\ref{fig:arrow_persistnoneq}.}
    \label{fig:arrow_persistnoneqdeeper}
\end{figure}

For quenching relaxation times from lower initial temperatures (for example $T_i=(0.083, 0.071)$), Eq.~\eqref{eq:barrier_neq} is no longer valid as the persistence times simply broaden with no apparent relaxation stages; see Fig.~\ref{fig:arrow_persistnoneqdeeper}. Instead, $E_\mathrm{neq}$ corresponds to the quenched energy barrier arising from relaxing excitations that are separated by the equilibrium length scale at $T_i$ similar to finite-rate cooling protocols in \cite{Hudson2018}. This then yields, 
\begin{equation}
E_\mathrm{neq} \approx E_\mathrm{eq}(T_i) \approx J_0 \log_2 \ell_\mathrm{eq}(T_i) \,, \label{eq:arrowpotts_quenchedbarrier}
\end{equation}
where $\ell_\mathrm{eq}(T_i)$ is the initial equilibrium length-scale. Using Eq.~\eqref{eq:barrier_neq} and Eq.~\eqref{eq:arrowpotts_quenchedbarrier}, the final formula for $\tau_\mathrm{neq}$ can be written as
\begin{equation}
\tau_\mathrm{neq} \sim \begin{cases}
\exp(2 \beta J_0) & \text{for} \ 2 < \ell_\mathrm{eq}(T_i) \leq  2^2
\\
\exp\left(\beta J_0 \log_2 \ell_\mathrm{eq}(T_i)\right) & \text{for} \ \ell_\mathrm{eq}(T_i) >  2^2
\end{cases} \, .
\label{eq:noneqtime}
\end{equation}   
Equation~\eqref{eq:noneqtime} shows excellent agreement with the relaxation times plotted in Fig.~\ref{fig:arrow_eneqwithfits}.

\subsection{Theory of Crystal Growth} \label{sec:arrowpotts_growth}

\begin{figure}
\includegraphics[width=\linewidth]{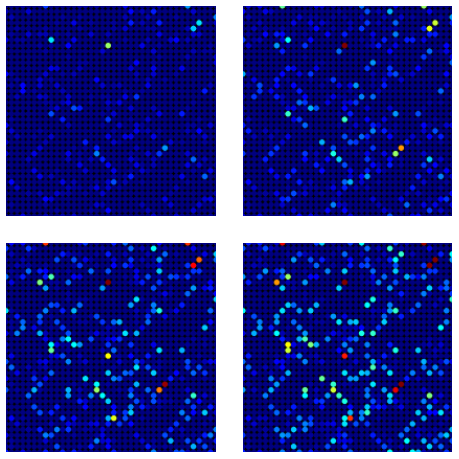}
\caption{Visualization of  dynamical  heterogeneity in the Arrow model using the frequency of spin flips per lattice site at $T=  0.33$ where red and blue sites indicate sites with higher and lower activity, respectively. Snapshots are shown for a system at equilibrium at 4 different times: (Upper left) $t=6.25\tau_\mathrm{eq}$, (upper right) $t= 12.5\tau_\mathrm{eq}$, (lower left) $t=18.75\tau_\mathrm{eq}$, and (lower right) $t=25\tau_\mathrm{eq}$. Chains of excitations emerge due to facilitation, and they act as pathways for crystal growth in the Arrow-Potts model.}
\label{fig:arrowpotts_dynhetero}
\end{figure}

\begin{figure}
\includegraphics[width=\linewidth]{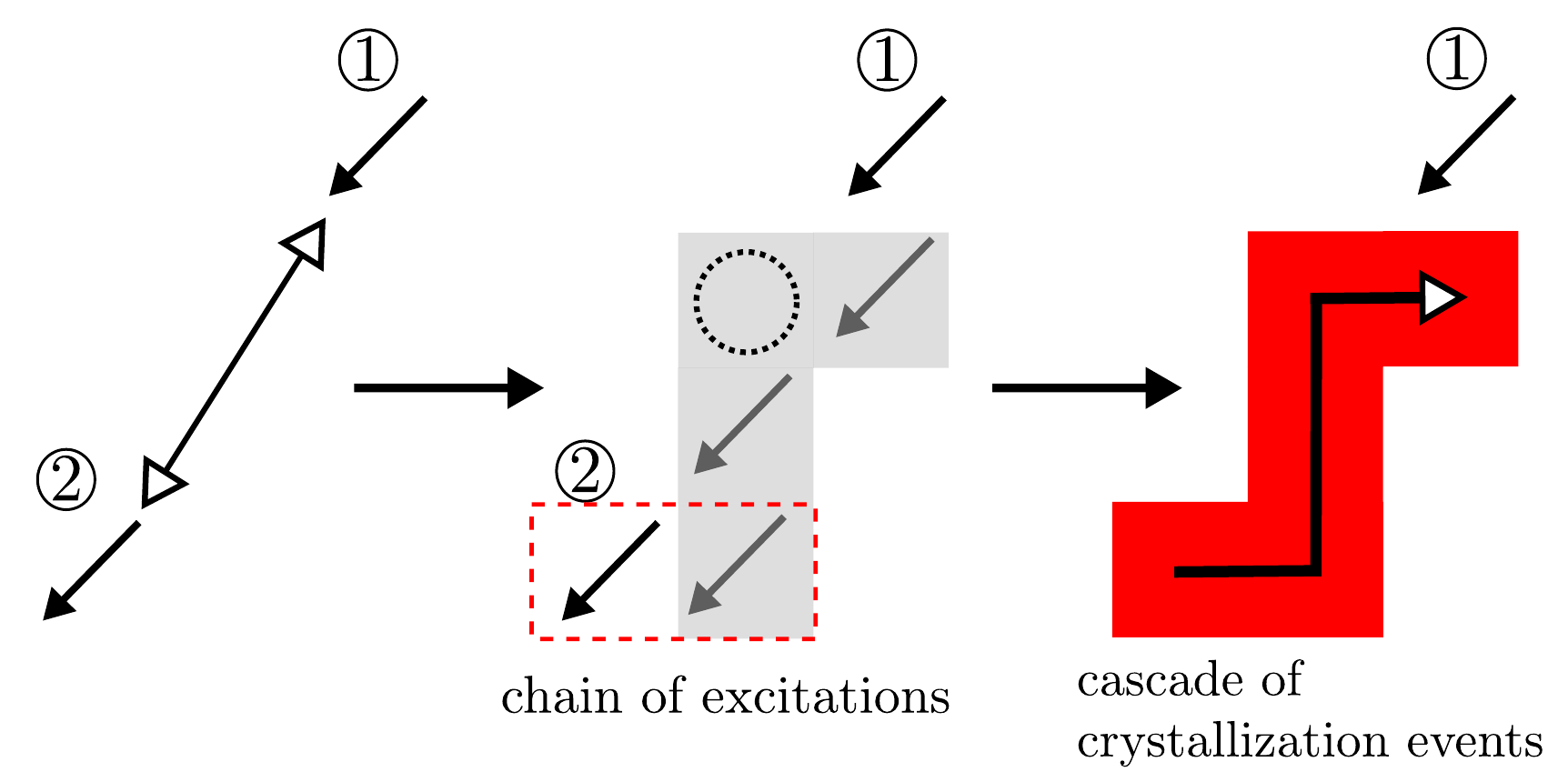}
\caption{An illustration for the crystal-branching mechanism in the Arrow-Potts model. Similar to Fig.~\ref{fig:arrowpotts_liqchain}, the liquid creates a chain of excitations but instead of relaxing to immobile liquid states, the system performs a cascade of crystallization events. This produces crystals with the same orientation, colored as red, along the original chain of excitations.} %
\label{fig:arrowpotts_cryschain}
\end{figure}

\begin{figure*}
    \centering
    \includegraphics[width=0.45\linewidth]{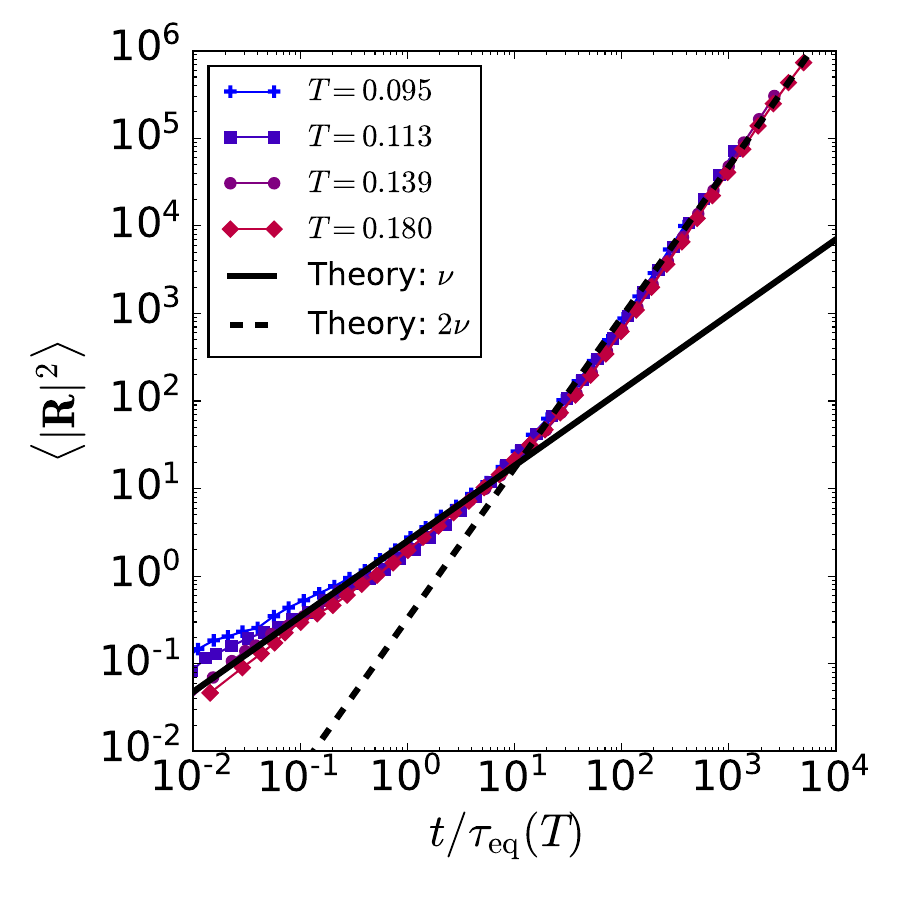}
    \quad
    \includegraphics[width=0.45\linewidth]{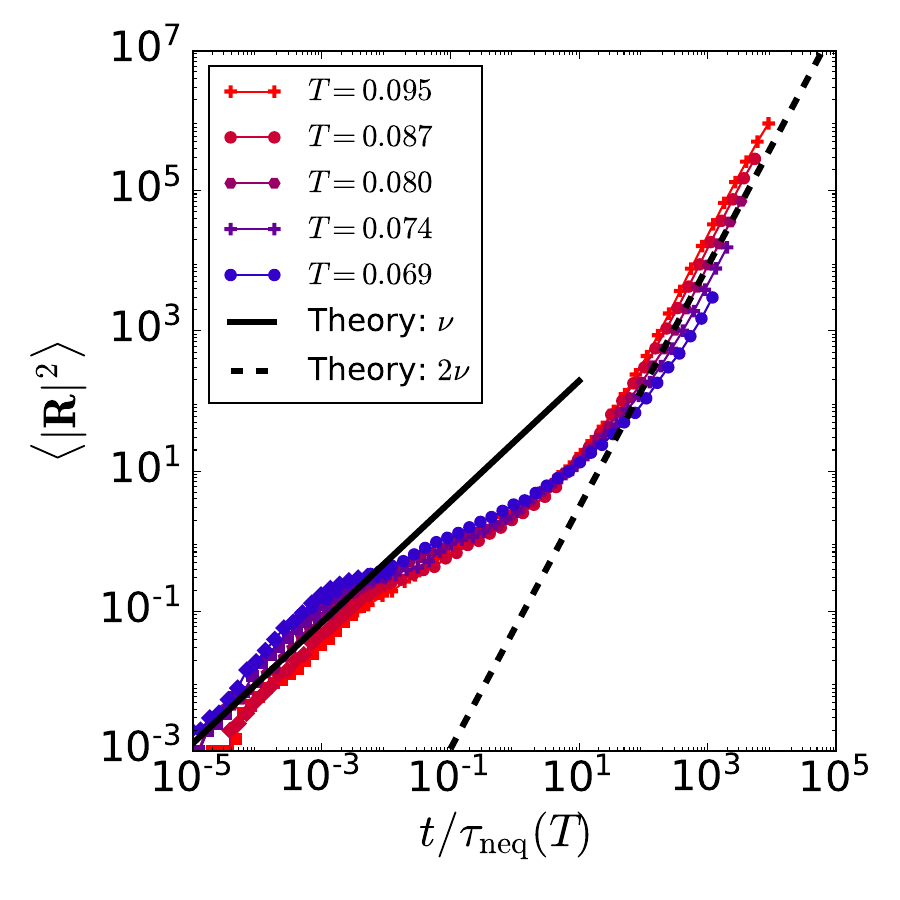}
    \caption{The collapsed mean-squared displacement (MSD) of the probe obtained from the Arrow model for a supercooled liquid at equilibrium (left) and during quenching (right). Data from the quenching simulations are obtained from an initial temperature $T_i \approx 0.180$. Bold and dashed lines are the scaling predictions in Eq.~\eqref{eq:arrow_finalscaling}.}
    \label{fig:arrow_probe_theory}
\end{figure*}

Why do crystals grow by branching? Recall from Section~\ref{sec:arrow_describe} that facilitation causes a cascade of relaxation events followed by the creation of a chain of excitations (Fig.~\ref{fig:arrowpotts_liqchain}). In fact, by tracking spin flips in each lattice site in the Arrow model, one may observe that most active sites form chain-like structures as shown in Fig.~\ref{fig:arrowpotts_dynhetero}. Now, in the Arrow-Potts model, the system can use these chains of mobile excitations as crystal growth pathways where relaxation to immobile liquid states are now replaced by crystallization. As a result, nucleated crystals prefer to grow by branching, as illustrated in Fig.~\ref{fig:arrowpotts_cryschain}, thus providing an intuitive explanation for the observations from MD simulations of super-compressed hard spheres \citep{Puser2009,Sanz2011,Sanz2014}.

We now proceed to estimate the characteristic radius of the growing crystal $R_g(t)$, which will be useful in obtaining the crystallization time scale from the KJMA theory discussed in the next section. This can be achieved by constructing an artificial probe, which traverses through the cascade process by moving only along the sites that facilitate the relaxation events. To this end, the dynamics of the probe is subjected to the following dynamical rules:
\begin{enumerate}
\item Place the probe at some initial excitation. We label this position as $\* R(0)$.
\item Update the probe position at time $t$ if and only if the initial excitation is relaxed to an immobile liquid state. The new position $\* R(t)$ is equal to the position of the nearest-neighboring excitation that facilitates the relaxation.
\item Repeat Step 2.
\end{enumerate}
With these rules, the probe provides a measure of the boundaries of the growing crystal. Alternatively, in an equilibrium supercooled liquid, the probe dynamics can also be used to track the spreading of clustered mobile regions across the liquid, which is distinct from the self-diffusion of individual atoms/particles. The resulting mean-squared displacement (MSD) of the probe also provides a way to map the cascade of relaxation events to a random walk model, which we proceed to develop below. 

Figure~\ref{fig:arrow_probe_theory} shows the MSD of the probe embedded in the Arrow model, both at equilibrium and during quenching. For the equilibrium dynamics, nondimensionalizing time with the equilibrium relaxation time $\tau_\mathrm{eq}$ yields a data collapse, where a short-time ($t < \tau_\mathrm{eq}$) and a long-time ($t > \tau_\mathrm{eq}$) regime can be observed. A similar nondimensionalization for quenching simulations, however, with the non-equilibrium relaxation time $\tau_\mathrm{neq}$ again yields data collapse with the two asymptotic regimes but with a new intermediate regime. Furthermore, the MSDs in both protocols follow non-diffusive behavior. These phenomena can be explained by a random walk theory, which couples the dynamics of the probe to the underlying dynamical heterogeneity. In this section, we will only sketch the derivation of the analytical formulas for the MSD, but a complete discussion of the theory will be presented in \ed{SM \cite{Note1}, Sec. 4.2}. 

To begin, there are two noteworthy observations about the random walk process. First, facilitation is directional, which causes the probe to behave like a biased random walker at long times. Second, excitations surrounding the probe form a random environment, which may act as obstacles for the probe. This is because excitations may possess a different directionality than the probe's walking direction, and we know from Section~\ref{sec:arrow_describe} that excitations possessing one directionality do not participate in the dynamics of excitations possessing a different directionality. Coupled with the fractal nature of the environment, this causes anomalous diffusion for the probe at all times. 

Random walks in fractal environments can be modeled using a scaling theory \citep{Rammal1983, Strichartz2006}. This is done by proposing a  scaling hypothesis given by
\begin{equation}
t(L) = b^{z} t(L/b)\,, \label{eq:scalinghypo}
\end{equation}
where $t(L)$ is the characteristic timescale for biased diffusion over some length-scale $L$, $b$ is some scaling factor, and $z$ is the dynamic exponent. By choosing $L=b$ and $L \sim \sqrt{\langle |\* R(t)|^2 \rangle}$, 
where $\sqrt{\langle |\* R(t)|^2 \rangle}$ is the root-mean-squared (RMS) displacement, Eq.~\eqref{eq:scalinghypo} can be re-written as another scaling relation
\begin{equation}
\langle |\* R(t)|^2 \rangle \sim \left(t/t(1)\right)^{2/z} \label{eq:generalmsdscaling} \,,
\end{equation}
where $t(1)$ is the timescale to jump a single step. Since the probe's first step occurs after the relaxation of a chain of excitations, $t(1)$ is given by $\tau_\mathrm{eq}$ at equilibrium and $\tau_\mathrm{neq}$ during quenching. 

Dynamical heterogeneity in the Arrow model approximately follows a Sierpinski gasket \citep{Garrahan2002,Berthier2005}. This simple fractal allows us to compute the exponent $z$ analytically, which we summarize as follows but provide the derivations in \ed{SM \cite{Note1}, Sec. 4.2}. Let $\tau_\mathrm{liq}$ be the liquid relaxation timescale, either at equilibrium ($\tau_\mathrm{liq}=\tau_\mathrm{eq}$) or during quenching ($\tau_\mathrm{liq}=\tau_\mathrm{neq}$). We can write two scaling relations governing the short-time and long-time regime respectively as
\begin{equation}
t(L) = \begin{cases}
5^{m} t(L/2^{m}) & t \ll \tau_\mathrm{liq}
\\
(5/2^z)^{m} t(L/2^{m}) & t \gg \tau_\mathrm{liq} 
\end{cases} \,, \label{eq:scalinginsg}
\end{equation}
where $m$ is the $m$-th iteration of the Sierpinski gasket construction \citep{Strichartz2006}.
Using Eqs.~\eqref{eq:scalinghypo}-\eqref{eq:scalinginsg}, we then obtain the following equation for the MSD
\begin{equation}
\left\langle |\*  R(t)|^2 \right\rangle = B (t/\tau_\mathrm{liq})^{\alpha} \sim \begin{cases}
(t/\tau_\mathrm{liq})^\nu \ & t \ll \tau_\mathrm{liq}
\\
(t/\tau_\mathrm{liq})^{2\nu} \ & t \gg \tau_\mathrm{liq} 
\end{cases} \,,
\label{eq:arrow_finalscaling}
\end{equation}
where $\nu= 2/z = 2\ln 2/\ln 5$, $B$ is some kinetic pre-factor, and $\alpha$ is a general scaling exponent, which can either be $\nu$ or $2\nu$. 
The scaling relations in Eq.~\eqref{eq:arrow_finalscaling} are  verified in Fig.~\ref{fig:arrow_probe_theory}, but note that they do not cover the intermediate regime for the quenching dynamics. This, however, is not an issue as Eq.~\eqref{eq:arrow_finalscaling} is sufficient for predicting crystal growth rates in all regimes of interest, as we will show later.

To apply Eq.~\eqref{eq:arrow_finalscaling} for crystal growth, we assume that the growing crystal radius is equal to the RMS displacement of the probe, i.e., $R_g(t)=\sqrt{\left\langle |\*  R(t)|^2 \right\rangle}$. In the nucleation-dominated regime, only a fractional amount of time is spent on growing crystals, so that the short-time exponent $\nu$ governs the time-evolution of $R_g(t)$. In the growth-dominated regime, the long-time exponent $2\nu$ is more appropriate as the system spends more of its time to grow crystals. Altogether, we can write the final formula for $R_g(t)$ by applying the exponent $\nu$ and $2\nu$ to high-temperature and low-temperature crystal growth, respectively, as 
\begin{equation}
R_g(t) = \sqrt{B}(t/\tau_\mathrm{liq})^{\alpha/2} \sim \begin{cases}
(t/\tau_\mathrm{liq})^{\nu/2} \ & T \to T_m
\\
(t/\tau_\mathrm{liq})^{\nu} \ & T \ll T_m
\end{cases} \,, \label{eq:arrowpotts_rgformula}
\end{equation}
where $T_m$ is the melting temperature. Equation~\eqref{eq:arrowpotts_rgformula} along with the nucleation rate in Eq.~\eqref{eq:arrowpotts_nucleationformula} provide us enough information to predict the overall crystallization timescale.  

\subsection{The Kolmogorov-Johnson-Mehl-Avrami Theory} \label{sec:arrowpotts_kjma}
With the formulas for nucleation rate in Eq.~\eqref{eq:arrowpotts_nucleationformula} and growth phenomena in Eq.~\eqref{eq:arrowpotts_rgformula}, we now proceed to derive an expression for crystallization timescale $t_\mathrm{x}$ using the Kolmogorov-Johnson-Mehl-Avrami (KJMA) theory \citep{kolmogorov1937statistical,william1939reaction,avrami1940kinetics}. 
The KJMA theory describes the time evolution of the volume fraction of the crystalline phase $x(t)$ using a probabilistic approach, where the arrival of nucleation events is modeled as a spatial point process \cite{van2007stochastic}. Using the mathematical framework of spatial point processes, we can derive an analytical formula for $x(t)$ that yields $t_\mathrm{x}$. In what follows, we review the formulation of the KJMA theory, and sketch its modification to analyze the case of crystallization under the influence of glassy dynamics as well as the derivation for $t_\mathrm{x}$.

\begin{figure}
    \centering
    \includegraphics[width=\linewidth]{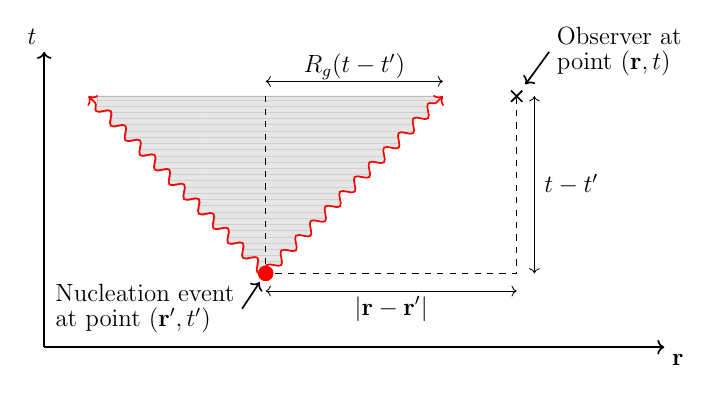}
    \includegraphics[width=\linewidth]{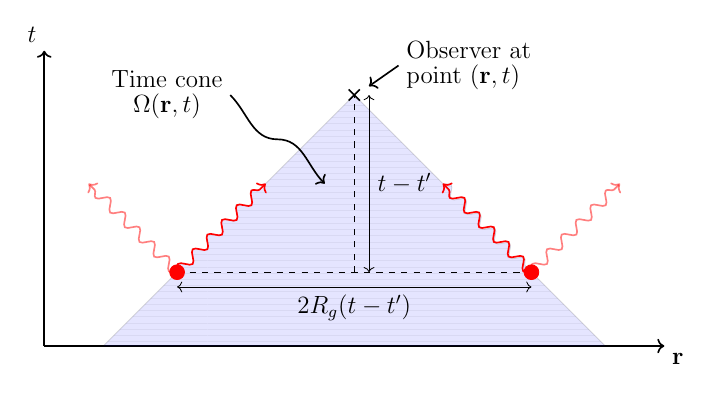}
    \caption{(Top) A schematic of the space-time picture with an illustration of the inequality (Eq~\eqref{eq:kjma_upperbound}), where a nucleation event (red dot) is too far away from the observer (cross). Red wavy lines represent the growth front of the crystal. (Bottom) An illustration of the time-cone $\Omega(\* r,t)$ with two nucleation events placed at its boundary, representing the furthest any nucleation event can exist before they become undetectable. Note that random nucleation events can still arrive outside and inside the time-cone.}
    \label{fig:arrowpotts_timecone}
\end{figure}

The theory begins by constructing an indicator function $\chi$ to detect random nucleation events in the system. The indicator function is associated with a spatio-temporal observation range $\Omega$ in space-time, where any nucleation event that occurs outside of $\Omega$ is ignored. This region $\Omega$ corresponds to a time-cone in space-time \citep{cahn1995time,ohta1987domain,sekimoto1986evolution}, whose radius is set by the characteristic radius of a growing crystal $R_g(t)$. To see this, consider the indicator function as an observer located at a position in space-time $(\* r,t)$, whose task is to count crystals that have arrived at its location. If a past nucleation event is too far from the observer, then the crystal originating from that event will not be detected; see Fig.~\ref{fig:arrowpotts_timecone} (top). In fact, a crystal will not be detected if the location of its past nucleation event at $(\* r^\prime,  t')$ obeys the following inequality 
\begin{equation}
|\* r-\* r^\prime| > \int_0^{t-t^\prime} v_\mathrm{xtl}(\tau) \diff \tau = R_g(t-t^\prime) \, , \label{eq:kjma_upperbound}
\end{equation}
where $v_\mathrm{xtl}(t)$ is a time-dependent growth rate given by $v_\mathrm{xtl}(t) = dR_g(t)/dt$. Using Eq.~\eqref{eq:kjma_upperbound}, we can define $\Omega$ as the following set of points in space-time
\begin{equation}
\Omega(\*r,t) \equiv \{\, (\* r^\prime,t^\prime) \in \mathbb{R}^{d}\times[0,t] \ \mid \ |\* r-\* r^\prime| \leq R_g(t-t^\prime) \,\} \,, \label{eq:kjma_omegaadef}
\end{equation}
which, indeed, forms a time-cone of radius $R_g(t)$ and height $t$; see Fig.~\ref{fig:arrowpotts_timecone} (bottom). We can further use Eq.~\eqref{eq:kjma_omegaadef} to express the indicator function as
\begin{equation}\label{eq:indicator-function}
\chi(\omega; \Omega(\*r,t)) \equiv 
\begin{cases}
1 & \text{if a crystal arrives at a point}
\\
& \text{$\omega \in \Omega(\*r,t)$ in space-time}
\\
0 & \text{otherwise}
\end{cases} \, ,
\end{equation}
where $\omega$ is a point in space-time.

Given the indicator function definition Eq.~\eqref{eq:indicator-function}, we now evaluate the volume fraction of crystalline phase $x(t)$. Instead of the crystal fraction, it is easier to evaluate the volume fraction of the remaining liquid phase $x_0(\* r,t)$ as the probability for no crystals to arrive at point $(\* r,t)$ that can then be used to calculate $x(t) = 1-x_0(t)$. In the theory of spatial point processes \citep{van2007stochastic}, $x_0(\* r,t)$ can be computed as a series expansion in $\chi(\omega; \Omega(\*r,t))$. Abbreviating the indicator function as $\chi(\omega)$, we can write $x_0(t)$ in terms of this expansion as
\begin{eqnarray}
x_0(\*r, t) =  1 + &&  \sum_{n=1}^{\infty} \frac{(-1)^n}{n !} \left( \int  \chi\left(\omega_{1}\right) \chi\left(\omega_{2}\right) \cdots\chi\left(\omega_{n}\right) \right.  \nonumber
\\
&& \left. f_{n}\left(\omega_{1}, \ldots, \omega_{n}\right) \diff \omega_{1} \diff \omega_{2} \cdots \diff \omega_{n} \vphantom{\int} \right) \,,
\label{eq:kjma_seriesexpand}
\end{eqnarray}
where $f_n(\omega_1,\ldots,\omega_n)$ is the $n$-point probability distribution functions governing the arrival of $n$-many random nucleation events; \ed{see SM \cite{Note1}, Sec. 5.1}. Note that $f_1(\omega_1) \diff \omega_1 $ is the probability for a crystal to emerge inside a small space-time volume $\diff \omega_1$ \citep{van2007stochastic}, so that $f_1(\omega_1)$ is simply equal to the nucleation rate $\bar{I}$
\begin{equation}
f_1(\omega_1) \equiv \bar{I} \label{eq:kjma_nucisdist} \,.
\end{equation}
At this stage, the theory is made simpler by ignoring space-time correlations between nucleation events so that every $n$-th order distribution is expressed as a product of $f_1(\omega_1)$, \emph{i.e.,}
\begin{eqnarray}
f_n(\omega_1,\omega_2,\cdots,\omega_n) =& f_1(\omega_1)f_1(\omega_2)\cdots f_1(\omega_n) \,. \label{eq:kjma_uncorrel}
\end{eqnarray}
Together with Eq.~\eqref{eq:kjma_nucisdist} and Eq.~\eqref{eq:kjma_uncorrel}, one may exactly evaluate the series expansion in Eq.~\eqref{eq:kjma_seriesexpand} to obtain
\begin{eqnarray}
\nonumber x_0(\*r, t) &&= \exp \left[ -\int \chi(\omega; \Omega(\*r,t)) \bar{I} \diff\omega \right]
\\
&& = \exp \left[ - \int_{\Omega(\* r,t)} \! \! \bar{I} \diff\omega \right] \,,
\label{eq:kjma_almostfinal}
\end{eqnarray}
where the second equality is obtained because the indicator function is non-zero only in $\Omega(\*r,t)$; \ed{see SM \cite{Note1}, Sec. 5.1 for a derivation of Eq.~\eqref{eq:kjma_almostfinal} from Eq.~\eqref{eq:kjma_seriesexpand}}. 

To obtain the crystal volume fraction, we must compute the integral in Eq.~\eqref{eq:kjma_almostfinal}, which is equal to the space-time volume of the time cone spanned by $\Omega(\*r,t)$. This implies that the integral should not depend on the position $\* r$, since it only acts as the spatial origin for the time-cone. Coupled with the formulas for nucleation (Eq.~\eqref{eq:arrowpotts_nucleationformula}) and growth (Eq.~\eqref{eq:arrowpotts_rgformula}) phenomena, we can write the time-evolution of the crystal volume fraction $x(t)$ from Eq.~\eqref{eq:kjma_almostfinal} as 
\begin{eqnarray}
x(t) =1- \exp \left[ - (t/\tau_\mathrm{xtl})^{\frac{\alpha d}{2}+1} \right] \label{eq:arrowpotts_kjma} \,,
\end{eqnarray}
where $\tau_\mathrm{xtl}$ is the characteristic crystallization timescale
\begin{eqnarray}
\tau_\mathrm{xtl} && = \left[\frac{\left(\frac{\alpha d}{2}+1\right) \tau_\mathrm{nuc} (\tau_\mathrm{liq})^{\frac{\alpha d}{2}} }{A(d) B^{d/2} } \right]^{\frac{1}{\frac{\alpha d}{2}+1}}
 \label{eq:arrowpotts_tauxtl} \,,
\end{eqnarray}
with $B$ a pre-factor arising from $R_g(t)$ (see Eq.~\eqref{eq:arrowpotts_rgformula}), and $\tau_\mathrm{nuc} = 1/\bar{I}$, and $A(d)$ a geometrical factor associated with a $d$-dimensional sphere given by 
\begin{equation}
A(d) = \frac{\pi ^{\frac {d}{2}}}{\Gamma \left(\frac{d}{2}+1\right)}\, .
\end{equation}
As can be seen from Eq.~\eqref{eq:arrowpotts_tauxtl}, $\tau_\mathrm{xtl}$ scales with the nucleation and liquid relaxation timescales with two distinct exponents as 
\begin{eqnarray}
\tau_\mathrm{xtl} && \sim \left[\tau_\mathrm{nuc}(\tau_\mathrm{liq})^{\frac{\alpha d}{2}}  \right]^{\frac{1}{\frac{\alpha d}{2}+1}}
 \label{eq:arrowpotts_proportional} \,.
\end{eqnarray}
This scaling relation constitutes an important result of our work that will be tested in Section~\ref{sec:arrowpotts_theoryvssim}.

\begin{figure*}[t]
    \centering
    \includegraphics[width=\linewidth]{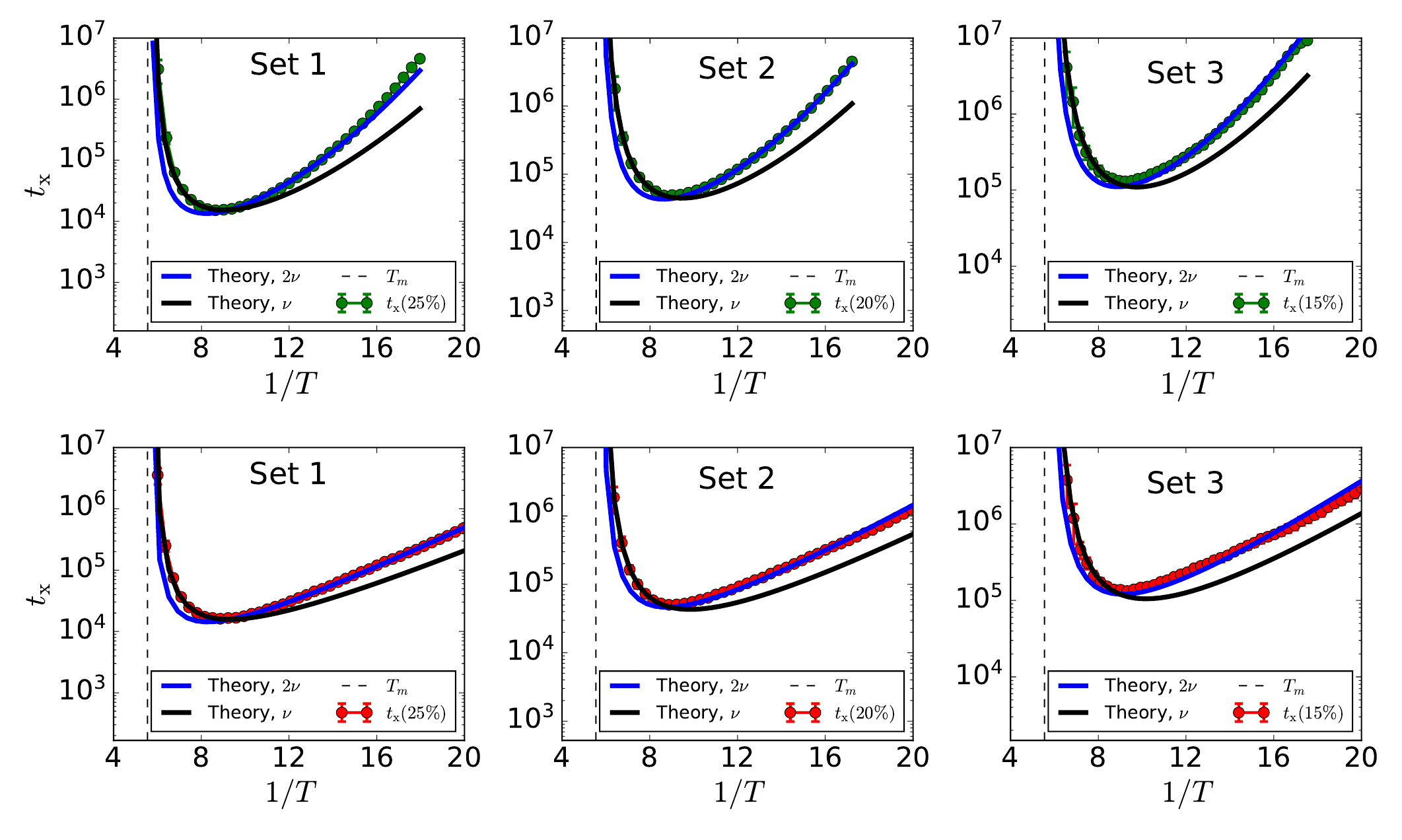}
    \caption{The TTT diagrams for all sets of model parameters listed in Table~\ref{table:arrowpotts_params} and both protocols: (top row) supercooled liquid crystallization and (bottom row) quenching. 
    Also plotted are the theoretical curves from Eq.~\eqref{eq:arrowpotts_fittingeq} shown in black and blue lines for the nucleation-dominated and growth-dominated regimes, respectively. Note that the non-monotonic shape of the theoretical curves requires no adjustable parameters when applied to the Arrow-Potts model.}
    \label{fig:arrowpotts_fitted1}
\end{figure*}

Finally, the formula for the crystallization time $t_\mathrm{x}(x \%)$ for a chosen crystal fraction of $x\%$ can be obtained from Eq.~\eqref{eq:arrowpotts_kjma} as 
\begin{eqnarray}
t_\mathrm{x}(x\%) = B_\mathrm{fit} C(d,x\%)\left[\tilde{\tau}_\mathrm{nuc}(\tau_\mathrm{liq})^{\frac{d\alpha}{2}}  \right]^{\frac{1}{\frac{d\alpha}{2}+1}} \,,
\label{eq:arrowpotts_fittingeq}
\end{eqnarray}
where $\tilde{\tau}_\mathrm{nuc}\equiv \kappa_0 \tau_\mathrm{nuc}$ is the dimensionless nucleation time with $\kappa_0$ being the kinetic pre-factor of the nucleation rate, $C(d,x)$ is a pre-factor given by 
\begin{equation}
C(d,x\%) = \left[-\frac{\left(\frac{d\alpha}{2}+1 \right)\ln (1-x)}{A(d)}\right]^{\frac{1}{\frac{d\alpha}{2}+1}} \, ,
\end{equation}
and $B_\mathrm{fit}$ is a lumped fitting parameter
\begin{equation}
B_\mathrm{fit} =\left(\kappa_0B^{d/2}\right)^{-\frac{1}{\frac{d\alpha}{2}+1}} \, .
\end{equation}
With Eq.~\eqref{eq:arrowpotts_fittingeq} and the scaling prediction \eqref{eq:arrowpotts_proportional}, we are now ready to test the theory with the simulation data from the Arrow-Potts model.

\section{Theory vs. Simulations } \label{sec:arrowpotts_theoryvssim}
Figure~\ref{fig:arrowpotts_fitted1} shows the results of the TTT diagrams from both supercooled liquid crystallization (top row) and quenching protocols (bottom row) for sets of model parameters listed in Table~\ref{table:arrowpotts_params}. Both protocols show non-monotonous behaviors.  In this section, we test the formula for the crystallization time $t_\mathrm{x}(x\%)$  in Eq.~\eqref{eq:arrowpotts_fittingeq} and the scaling prediction from Eq.~\eqref{eq:arrowpotts_proportional} with the results presented in Fig.~\ref{fig:arrowpotts_fitted1}. By verifying Eqs.~\eqref{eq:arrowpotts_proportional} and \eqref{eq:arrowpotts_fittingeq}, we demonstrate how the theory accurately captures the crystallization energy barriers in the Arrow-Potts model, which also allow us to account for the non-monotonic shape of the TTT diagrams. 
Such a test also demonstrates how the crystallization time $t_\mathrm{x}$ encodes the universality of the glassy dynamics. 

To apply Eqs.~\eqref{eq:arrowpotts_proportional}-\eqref{eq:arrowpotts_fittingeq} to the Arrow-Potts model, we first need analytical expressions for the surface tension $\gamma$ and the chemical potential $\Delta \mu$, since $\tau_\mathrm{nuc}$ depends on these parameters. To this end, recall from Section~\ref{sec:arrowpotts_thermodynamics} that the Arrow-Potts model behaves like the Ising model within the current set of model parameters (Table~\ref{table:arrowpotts_params}). This implies that the surface tension $\gamma$ is approximately equal to that of the Ising model in a 2D square lattice \citep{Onsager1944}
\begin{equation}
\gamma \approx 2J +k_\mathrm{B} T \ln \left( \tanh(\beta J) \right) \,,
\end{equation} 
where $J = \left(2\epsilon -\Delta \epsilon \right)/4$. The chemical potential $\Delta \mu$ is equal to the effective chemical potential $h_\mathrm{eff}(T)$ \ed{(SM \cite{Note1}, Sec. 5.1)}, obtained by solving Eq.~\eqref{eq:arrowpotts_lambdadef} and setting the coordination number $z$ to that of a square lattice ($z=4$) yielding
\begin{equation}
\Delta \mu = h_\mathrm{eff}(T) = \lambda T+2\Delta \epsilon \,.   
\end{equation}
These exact expressions for $\gamma$ and $\Delta \mu$ are precisely the reason why test our model in 2D as they reduce the number of fitting parameters.
For the liquid relaxation time $\tau_\mathrm{liq}$, we have the formulas for the supercooled liquid crystallization protocol (Eq.~\eqref{eq:parablaw}) and the quenching protocol (Eq.~\eqref{eq:noneqtime}), but they are only applicable at low temperatures ($T \ll T_m$). To maintain accuracy across all temperatures, we will instead interpolate from the relaxation time data of the Arrow model. Altogether, the combined analytical expressions specify the crystallization energy barriers with no adjustable parameters, and thus fix the non-monotonic shape of the TTT diagram. While the theory is still left with a single fitting parameter $B_\mathrm{fit}$, its value only shifts the theoretical curve up and down on a TTT diagram, and therefore bears no significant consequence in predicting the non-monotonic behaviors.

\begin{figure}
\centering
\includegraphics[width=0.9\linewidth]{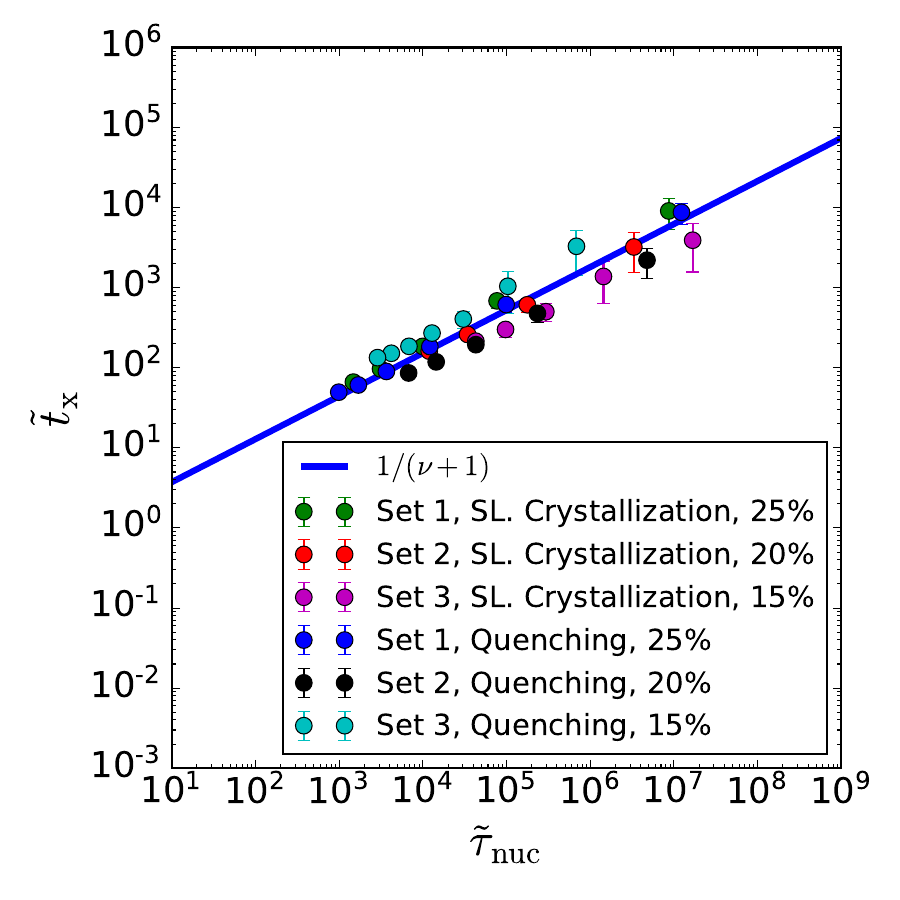}
\includegraphics[width=0.9\linewidth]{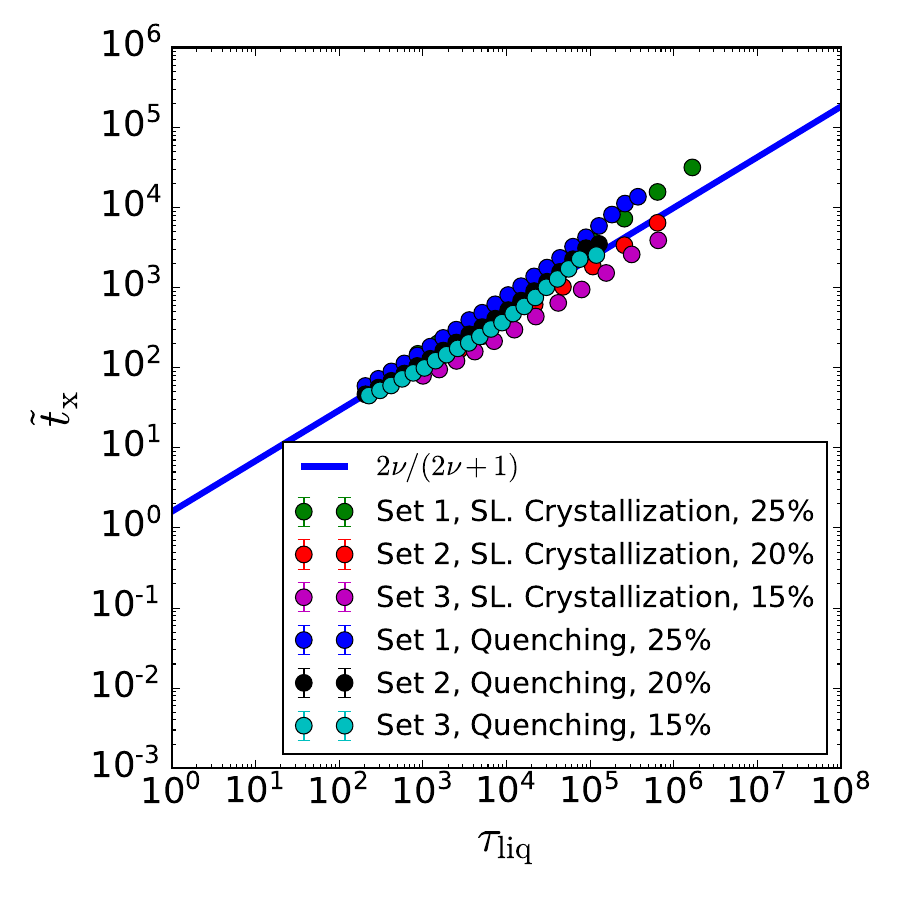}
\vspace{-12pt}
\caption{Verifying the scaling relations and corresponding exponents between overall crystallization time and the nucleation time (top) in the nucleation-dominated regime, and the liquid relaxation time (bottom) in the growth-dominated regime.}
\label{fig:arrowpotts_scaling}
\end{figure}

Recall that $\alpha$ in Eqs.~\eqref{eq:arrowpotts_proportional} and \eqref{eq:arrowpotts_fittingeq} is related to the growth dynamics in Eq.~\eqref{eq:arrowpotts_rgformula}, and can either be $\nu$
or $2\nu$ depending on the temperature at which the crystal grows. 
Therefore,  we test Eq.~\eqref{eq:arrowpotts_fittingeq} in the nucleation-dominated regime by setting the scaling exponent $\alpha$ to $\nu=2 \ln 2/\ln 5$, and adjust $B_\mathrm{fit}$ to match the crystallization time data at a temperature range of $T^* < T <T_m$ where $T^*$ is the temperature for the minimum crystallization time. In the growth-dominated regime ($T < T^*$), we do a similar procedure but with the scaling exponent $\alpha$ set to $2\nu$. The results for both the supercooled liquid crystallization and quenching protocol are shown in Fig.~\ref{fig:arrowpotts_fitted1} where excellent agreement can be found between the theory's prediction for the non-monotonic behavior and that of the simulation data.

To unveil the universality encoded in the crystallization time $t_\mathrm{x}$, we investigate the scaling prediction from Eq.~\eqref{eq:arrowpotts_proportional} in different regimes for crystallization. In the nucleation-dominated regime, we expect that $\tau_\mathrm{nuc} \gg \tau_\mathrm{liq}$ so that the crystallization time $t_\mathrm{x}$ scales as
\begin{equation}
t_\mathrm{x} \sim \left(\tau_\mathrm{nuc}\right)^{\frac{1}{\nu+1}} \quad T \to T_m \,.
\label{eq:arrowpotts_scalinghight}
\end{equation}
In the growth-dominated regime, $\tau_\mathrm{nuc} \ll \tau_\mathrm{liq}$ so that a new scaling relation is obtained
\begin{equation}
t_\mathrm{x} \sim \left(\tau_\mathrm{liq}\right)^{\frac{2\nu}{2\nu+1}}, \quad T \ll T_m  \,.
\label{eq:arrowpotts_scalinglowt}
\end{equation}
Equations~\eqref{eq:arrowpotts_scalinghight} and Eq.~\eqref{eq:arrowpotts_scalinglowt} provide a way to test the exponents with respect to $\tau_\mathrm{nuc}$ and $\tau_\mathrm{liq}$, respectively. We now test them by collapsing different crystallization time data onto a master curve. However, it is convenient to test the scaling predictions by collapsing the data for different regimes of crystallization. To that end, for the nucleation-dominated regime ($T^* < T < T_m$), the crystallization time data is first plotted on a log-log scale with respect to the dimensionless nucleation time $\tilde{\tau}_\mathrm{nuc}\equiv \kappa_0 \tau_\mathrm{nuc}$. Next, we fit the $\log t_\mathrm{x}$ vs. $\log \tilde{\tau}_\mathrm{nuc}$ data with a straight-line and use its $y$-intercept to shift the origin to zero, so that $t_\mathrm{x}$ is now dimensionless. The result of this data-collapse procedure in the nucleation regime is shown in Fig.~\ref{fig:arrowpotts_scaling}, showing agreement with the scaling exponent $1/(\nu+1)$. Using a similar procedure, we perform another data-collapse procedure for the growth-dominated regime ($T < T^*$), by plotting the dimensionless $t_\mathrm{x}$ with respect to $\tau_\mathrm{liq}$ in Fig.~\ref{fig:arrowpotts_scaling}, again showing agreement with the derived scaling exponent $2\nu/(2\nu+1)$.

\begin{figure}
\centering
\includegraphics[width=\linewidth]{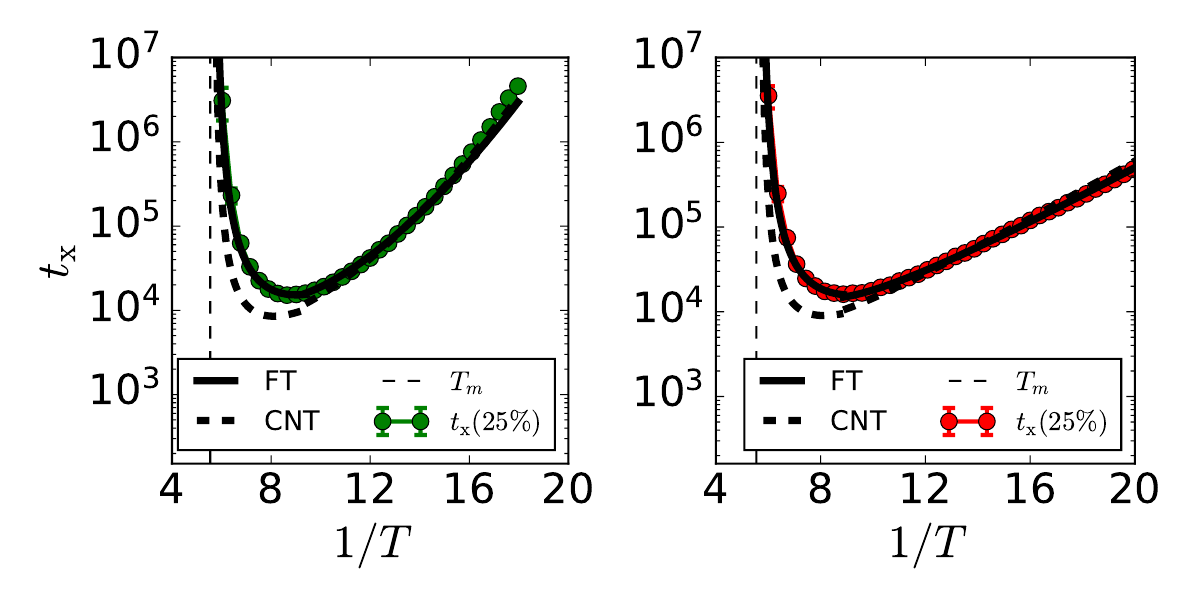}
\caption{A comparison of the original theory (bold line) with the one whose nucleation rate is replaced by CNT. Left and right figures correspond to supercooled liquid crystallization and quenching protocol data of Set 1 model parameters.}
\label{fig:arrowpotts_cntcompare}
\end{figure}

We end this section by asking what happens if we ignore the fluctuation correction in the nucleation rate? To answer this question, we first construct a piece-wise curve that predicts the entire range of $t_\mathrm{x}$ presented in Fig.~\ref{fig:arrowpotts_cntcompare}. To this end, if $T_m < T < T^*$, the piece-wise curve corresponds to the curve for the nucleation-dominated regime. Otherwise ($T \leq T^*$), it is the curve for the growth-dominated regime. Such a piece-wise curve is shown as a bold line in  Fig.~\ref{fig:arrowpotts_cntcompare}. Next, we construct another curve from the first piece-wise curve, where the fluctuation correction term $\Delta F_\mathrm{corr}$ is removed from the nucleation rate formula  in Eq.~\eqref{eq:arrowpotts_nucleationformula}. This second curve becomes a new prediction for the TTT diagram, with the nucleation described by classical nucleation theory (CNT). The CNT prediction is also shown in Fig.~\ref{fig:arrowpotts_cntcompare} (dashed line), which under predicts the minimum crystallization time by roughly half of an order of magnitude, thus showing the importance of fluctuations in the overall crystallization time.

\section{Conclusions} \label{sec:arrowpotts_conclusion}
In summary, the Arrow-Potts model demonstrates several key features of crystallization in glass-formers that can be found in both experiments and MD simulations. The theory we developed uses the KJMA theory to combine the field theory of nucleation, a random walk theory for crystal growth, and dynamical facilitation theory for glassy dynamics. When crystallizing the supercooled state of the liquid, which can be achieved in practice with finite-rate cooling protocols and at high enough temperatures, the crystallization time follows a super-Arrhenius behavior of Trend I. At deeper quenches, the liquid begins to fall out of equilibrium and produces an Arrhenius behavior instead (Trend II), which emerges as a signature of out-of-equilibrium relaxation in the liquid. Fractal crystal morphology emerges as a result of dynamical facilitation providing a cascade of facilitated structural reordering. The pathways that are followed are string-like that causes the crystals to grow through a branching or percolation-like process. This cascade spreads at a speed which obeys scaling properties that are universal to the self-similar properties of mobility field fluctuations, i.e., its dynamical heterogeneity, in space and time.

\ed{
We also note that the crystal-growth mechanism and microstructure found in the Arrow-Potts model are consistent with the observations found in MD simulations. First, the crystal morphology, which consists of fractal ramified clusters, can be found in super-compressed hard spheres \citep{valeriani2012compact}, and the growth mechanism also proceeds through a cascade of structural reordering \citep{Sanz2014}. Second, the growth-dominated regime of the Arrow-Potts model is also found in the Kob-Andersen model \citep{Kob1994}, a binary atomistic model of glass former, where a recent study \citep{Ingebrigtsen2018} has shown that small compositional fluctuations can trigger immediate and irreversible crystal growth in deeply supercooled conditions. Lastly, a recent study has demonstrated a transition point between a poly-crystal and a glass in a binary glass former through quasi-static compression \citep{Zhang2018}, where the microstructure near the transition point is composed of ramified crystalline domains surrounded by a larger disordered matrix similar to the microstructure found in the Arrow-Potts model. Altogether, these observations showcase the ability of the Arrow-Potts model in producing all relevant aspects of nucleation and growth in real glass formers.
}

Future work should focus on experimental tests. In particular, it would be interesting to see how well scaling tests for Eq.~\eqref{eq:arrowpotts_scalinglowt} and Eq.~\eqref{eq:arrowpotts_scalinghight} hold. To perform the scaling tests, one needs (1) $\tau_\mathrm{nuc}$, (2) $\tau_\mathrm{liq}$ (such as viscosity $\eta$), and (3) some measure of crystallization time $\tau_\mathrm{xtl}$ (such as $t_\mathrm{x}(x\%)$ or the one used for Fig.~\ref{fig:experimental}). These quantities, though less direct than the actual parameters entering the theory's final formula (Eq.~\eqref{eq:arrowpotts_proportional}), are more directly measurable \citep{Debenedetti1996} and hence more amenable to scaling tests. With that being said, there appear to be glassy systems where experimental data is available. Some of them include bulk metallic glass alloys where crystallization kinetics are studied through TTT diagrams and viscosity measurements of the supercooled liquid \citep{Masuhr1999,Schroers1999,Hays1999}.  A recent study has also collected data for nucleation and growth rates as well as viscosity simultaneously for various organic compounds \citep{Huang2018}, which makes it amenable for scaling tests. Colloidal suspensions may also be a great choice because both length- and times-scales of the system are amenable to real-time imaging analysis \citep{Weeks2000,Tan2014}. Using these imaging techniques, one may quantify directly how the size of a growing crystal evolves in time and verify if Eq.~\eqref{eq:arrowpotts_rgformula} is obeyed. Alternatively, one may probe the relaxation cascade mechanism in the supercooled or super-compressed suspensions. This could be achieved by imaging the mobile regions, similar to how dynamical heterogeneity is imaged in past studies \citep{Gao2007,Kegel2000,Fris2011}, and observe how such regions spread from one spot to the next. Once the radius of gyration of this growing mobile region is computed, one may verify whether the analogous scaling from Eq.~\eqref{eq:arrow_finalscaling} is obeyed. We leave such an analysis for future study.
$\\$

\begin{acknowledgements}
MRH and KKM are supported by the Director, Office of Science, Office of Basic Energy Sciences, of the U.S. Department of Energy under contract No. DEAC02-05CH11231. 
KKM deeply acknowledges late Prof. David Chandler for insightful and useful discussions on the model, and the work was initiated during the late stages of his postdoctoral stay with him. KKM is also indebted to Dr. Kelsey Schuster for collaboration in the initial stages of this work, and with whom early calculations on the phase diagram of the Arrow-Potts model and crystallization time were performed. The authors thank Cory Hargus for a careful reading of the manuscript.
\end{acknowledgements}

\bibliography{apssamp}

\begin{thebibliography}{90}%
\makeatletter
\providecommand \@ifxundefined [1]{%
 \@ifx{#1\undefined}
}%
\providecommand \@ifnum [1]{%
 \ifnum #1\expandafter \@firstoftwo
 \else \expandafter \@secondoftwo
 \fi
}%
\providecommand \@ifx [1]{%
 \ifx #1\expandafter \@firstoftwo
 \else \expandafter \@secondoftwo
 \fi
}%
\providecommand \natexlab [1]{#1}%
\providecommand \enquote  [1]{``#1''}%
\providecommand \bibnamefont  [1]{#1}%
\providecommand \bibfnamefont [1]{#1}%
\providecommand \citenamefont [1]{#1}%
\providecommand \href@noop [0]{\@secondoftwo}%
\providecommand \href [0]{\begingroup \@sanitize@url \@href}%
\providecommand \@href[1]{\@@startlink{#1}\@@href}%
\providecommand \@@href[1]{\endgroup#1\@@endlink}%
\providecommand \@sanitize@url [0]{\catcode `\\12\catcode `\$12\catcode
  `\&12\catcode `\#12\catcode `\^12\catcode `\_12\catcode `\%12\relax}%
\providecommand \@@startlink[1]{}%
\providecommand \@@endlink[0]{}%
\providecommand \url  [0]{\begingroup\@sanitize@url \@url }%
\providecommand \@url [1]{\endgroup\@href {#1}{\urlprefix }}%
\providecommand \urlprefix  [0]{URL }%
\providecommand \Eprint [0]{\href }%
\providecommand \doibase [0]{https://doi.org/}%
\providecommand \selectlanguage [0]{\@gobble}%
\providecommand \bibinfo  [0]{\@secondoftwo}%
\providecommand \bibfield  [0]{\@secondoftwo}%
\providecommand \translation [1]{[#1]}%
\providecommand \BibitemOpen [0]{}%
\providecommand \bibitemStop [0]{}%
\providecommand \bibitemNoStop [0]{.\EOS\space}%
\providecommand \EOS [0]{\spacefactor3000\relax}%
\providecommand \BibitemShut  [1]{\csname bibitem#1\endcsname}%
\let\auto@bib@innerbib\@empty
\bibitem [{\citenamefont {Masuhr}\ \emph {et~al.}(1999)\citenamefont {Masuhr},
  \citenamefont {Waniuk}, \citenamefont {Busch},\ and\ \citenamefont
  {Johnson}}]{Masuhr1999}%
  \BibitemOpen
  \bibfield  {author} {\bibinfo {author} {\bibfnamefont {A.}~\bibnamefont
  {Masuhr}}, \bibinfo {author} {\bibfnamefont {T.~A.}\ \bibnamefont {Waniuk}},
  \bibinfo {author} {\bibfnamefont {R.}~\bibnamefont {Busch}},\ and\ \bibinfo
  {author} {\bibfnamefont {W.~L.}\ \bibnamefont {Johnson}},\ }\bibfield
  {title} {\emph {\bibinfo {title} {Time Scales for Viscous Flow, Atomic
  Transport, and Crystallization in the Liquid and Supercooled Liquid States of
  $\mathrm{Zr}_{41.2}\mathrm{Ti}_{13.8}\mathrm{Cu}_{12.5}\mathrm{Ni}_{10.0}\mathrm{Be}_{22.5}$}},\
  }\href {https://doi.org/10.1103/PhysRevLett.82.2290} {\bibfield  {journal}
  {\bibinfo  {journal} {Phys. Rev. Lett.}\ }\textbf {\bibinfo {volume} {82}},\
  \bibinfo {pages} {2290} (\bibinfo {year} {1999})}\BibitemShut {NoStop}%
\bibitem [{\citenamefont {Schroers}\ \emph {et~al.}(1999)\citenamefont
  {Schroers}, \citenamefont {Masuhr}, \citenamefont {Johnson},\ and\
  \citenamefont {Busch}}]{Schroers1999}%
  \BibitemOpen
  \bibfield  {author} {\bibinfo {author} {\bibfnamefont {J.}~\bibnamefont
  {Schroers}}, \bibinfo {author} {\bibfnamefont {A.}~\bibnamefont {Masuhr}},
  \bibinfo {author} {\bibfnamefont {W.~L.}\ \bibnamefont {Johnson}},\ and\
  \bibinfo {author} {\bibfnamefont {R.}~\bibnamefont {Busch}},\ }\bibfield
  {title} {\emph {\bibinfo {title} {Pronounced Asymmetry in the Crystallization
  Behavior during Constant Heating and Cooling of a Bulk Metallic Glass-Forming
  Liquid}},\ }\href {https://doi.org/10.1103/PhysRevB.60.11855} {\bibfield
  {journal} {\bibinfo  {journal} {Phys. Rev. B}\ }\textbf {\bibinfo {volume}
  {60}},\ \bibinfo {pages} {11855} (\bibinfo {year} {1999})}\BibitemShut
  {NoStop}%
\bibitem [{\citenamefont {Hays}\ \emph {et~al.}(1999)\citenamefont {Hays},
  \citenamefont {Kim},\ and\ \citenamefont {Johnson}}]{Hays1999}%
  \BibitemOpen
  \bibfield  {author} {\bibinfo {author} {\bibfnamefont {C.~C.}\ \bibnamefont
  {Hays}}, \bibinfo {author} {\bibfnamefont {C.~P.}\ \bibnamefont {Kim}},\ and\
  \bibinfo {author} {\bibfnamefont {W.~L.}\ \bibnamefont {Johnson}},\
  }\bibfield  {title} {\emph {\bibinfo {title} {Large Supercooled Liquid Region
  and Phase Separation in the
  $\mathrm{Zr}$-$\mathrm{Ti}$-$\mathrm{Ni}$-$\mathrm{Cu}$-$\mathrm{Be}$ Bulk
  Metallic Glasses}},\ }\href {https://doi.org/10.1063/1.124606} {\bibfield
  {journal} {\bibinfo  {journal} {Appl. Phys. Lett.}\ }\textbf {\bibinfo
  {volume} {75}},\ \bibinfo {pages} {1089} (\bibinfo {year}
  {1999})}\BibitemShut {NoStop}%
\bibitem [{\citenamefont {Kim}\ \emph {et~al.}(2002)\citenamefont {Kim},
  \citenamefont {Choi}, \citenamefont {Suresh},\ and\ \citenamefont
  {Argon}}]{Kim2002}%
  \BibitemOpen
  \bibfield  {author} {\bibinfo {author} {\bibfnamefont {J.~J.}\ \bibnamefont
  {Kim}}, \bibinfo {author} {\bibfnamefont {Y.}~\bibnamefont {Choi}}, \bibinfo
  {author} {\bibfnamefont {S.}~\bibnamefont {Suresh}},\ and\ \bibinfo {author}
  {\bibfnamefont {A.~S.}\ \bibnamefont {Argon}},\ }\bibfield  {title} {\emph
  {\bibinfo {title} {Nanocrystallization During Nanoindentation of a Bulk
  Amorphous Metal Alloy at Room Temperature}},\ }\href
  {https://doi.org/10.1126/science.1067453} {\bibfield  {journal} {\bibinfo
  {journal} {Science}\ }\textbf {\bibinfo {volume} {295}},\ \bibinfo {pages}
  {654} (\bibinfo {year} {2002})}\BibitemShut {NoStop}%
\bibitem [{\citenamefont {Huang}\ \emph {et~al.}(2018)\citenamefont {Huang},
  \citenamefont {Chen}, \citenamefont {Gui}, \citenamefont {Shi}, \citenamefont
  {Zhang},\ and\ \citenamefont {Yu}}]{Huang2018}%
  \BibitemOpen
  \bibfield  {author} {\bibinfo {author} {\bibfnamefont {C.}~\bibnamefont
  {Huang}}, \bibinfo {author} {\bibfnamefont {Z.}~\bibnamefont {Chen}},
  \bibinfo {author} {\bibfnamefont {Y.}~\bibnamefont {Gui}}, \bibinfo {author}
  {\bibfnamefont {C.}~\bibnamefont {Shi}}, \bibinfo {author} {\bibfnamefont
  {G.~G.~Z.}\ \bibnamefont {Zhang}},\ and\ \bibinfo {author} {\bibfnamefont
  {L.}~\bibnamefont {Yu}},\ }\bibfield  {title} {\emph {\bibinfo {title}
  {Crystal Nucleation Rates in Glass-Forming Molecular Liquids: D-sorbitol,
  D-arabitol, D-xylitol, and Glycerol}},\ }\href
  {https://doi.org/10.1063/1.5042112} {\bibfield  {journal} {\bibinfo
  {journal} {J. Chem. Phys.}\ }\textbf {\bibinfo {volume} {149}},\ \bibinfo
  {pages} {054503} (\bibinfo {year} {2018})}\BibitemShut {NoStop}%
\bibitem [{\citenamefont {Karmwar}\ \emph {et~al.}(2011)\citenamefont
  {Karmwar}, \citenamefont {Boetker}, \citenamefont {Graeser}, \citenamefont
  {Strachan}, \citenamefont {Rantanen},\ and\ \citenamefont
  {Rades}}]{Karmwar2011}%
  \BibitemOpen
  \bibfield  {author} {\bibinfo {author} {\bibfnamefont {P.}~\bibnamefont
  {Karmwar}}, \bibinfo {author} {\bibfnamefont {J.}~\bibnamefont {Boetker}},
  \bibinfo {author} {\bibfnamefont {K.}~\bibnamefont {Graeser}}, \bibinfo
  {author} {\bibfnamefont {C.}~\bibnamefont {Strachan}}, \bibinfo {author}
  {\bibfnamefont {J.}~\bibnamefont {Rantanen}},\ and\ \bibinfo {author}
  {\bibfnamefont {T.}~\bibnamefont {Rades}},\ }\bibfield  {title} {\emph
  {\bibinfo {title} {Investigations on the Effect of Different Cooling Rates on
  the Stability of Amorphous Indomethacin}},\ }\href
  {https://doi.org/10.1016/J.EJPS.2011.08.010} {\bibfield  {journal} {\bibinfo
  {journal} {Eur. J. Pharm. Sci.}\ }\textbf {\bibinfo {volume} {44}},\ \bibinfo
  {pages} {341} (\bibinfo {year} {2011})}\BibitemShut {NoStop}%
\bibitem [{\citenamefont {Blaabjerg}\ \emph {et~al.}(2016)\citenamefont
  {Blaabjerg}, \citenamefont {Lindenberg}, \citenamefont {L{\"{o}}bmann},
  \citenamefont {Grohganz},\ and\ \citenamefont {Rades}}]{Blaabjerg2016}%
  \BibitemOpen
  \bibfield  {author} {\bibinfo {author} {\bibfnamefont {L.~I.}\ \bibnamefont
  {Blaabjerg}}, \bibinfo {author} {\bibfnamefont {E.}~\bibnamefont
  {Lindenberg}}, \bibinfo {author} {\bibfnamefont {K.}~\bibnamefont
  {L{\"{o}}bmann}}, \bibinfo {author} {\bibfnamefont {H.}~\bibnamefont
  {Grohganz}},\ and\ \bibinfo {author} {\bibfnamefont {T.}~\bibnamefont
  {Rades}},\ }\bibfield  {title} {\emph {\bibinfo {title} {Glass Forming
  Ability of Amorphous Drugs Investigated by Continuous Cooling and Isothermal
  Transformation}},\ }\href {https://doi.org/10.1021/acs.molpharmaceut.6b00650}
  {\bibfield  {journal} {\bibinfo  {journal} {Mol. Pharm.}\ }\textbf {\bibinfo
  {volume} {13}},\ \bibinfo {pages} {3318} (\bibinfo {year}
  {2016})}\BibitemShut {NoStop}%
\bibitem [{\citenamefont {Iezzi}\ \emph {et~al.}(2011)\citenamefont {Iezzi},
  \citenamefont {Mollo}, \citenamefont {Torresi}, \citenamefont {Ventura},
  \citenamefont {Cavallo},\ and\ \citenamefont {Scarlato}}]{Iezzi2011}%
  \BibitemOpen
  \bibfield  {author} {\bibinfo {author} {\bibfnamefont {G.}~\bibnamefont
  {Iezzi}}, \bibinfo {author} {\bibfnamefont {S.}~\bibnamefont {Mollo}},
  \bibinfo {author} {\bibfnamefont {G.}~\bibnamefont {Torresi}}, \bibinfo
  {author} {\bibfnamefont {G.}~\bibnamefont {Ventura}}, \bibinfo {author}
  {\bibfnamefont {A.}~\bibnamefont {Cavallo}},\ and\ \bibinfo {author}
  {\bibfnamefont {P.}~\bibnamefont {Scarlato}},\ }\bibfield  {title} {\emph
  {\bibinfo {title} {Experimental Solidification of an Andesitic Melt by
  Cooling}},\ }\href {https://doi.org/10.1016/J.CHEMGEO.2011.01.024} {\bibfield
   {journal} {\bibinfo  {journal} {Chem. Geol.}\ }\textbf {\bibinfo {volume}
  {283}},\ \bibinfo {pages} {261} (\bibinfo {year} {2011})}\BibitemShut
  {NoStop}%
\bibitem [{\citenamefont {Limmer}\ and\ \citenamefont
  {Chandler}(2011)}]{Limmer2011}%
  \BibitemOpen
  \bibfield  {author} {\bibinfo {author} {\bibfnamefont {D.~T.}\ \bibnamefont
  {Limmer}}\ and\ \bibinfo {author} {\bibfnamefont {D.}~\bibnamefont
  {Chandler}},\ }\bibfield  {title} {\emph {\bibinfo {title} {The Putative
  Liquid-Liquid Transition Is a Liquid-Solid Transition in Atomistic Models of
  Water}},\ }\href {https://doi.org/10.1063/1.3643333} {\bibfield  {journal}
  {\bibinfo  {journal} {J. Chem. Phys.}\ }\textbf {\bibinfo {volume} {135}},\
  \bibinfo {pages} {134503} (\bibinfo {year} {2011})}\BibitemShut {NoStop}%
\bibitem [{\citenamefont {Limmer}\ and\ \citenamefont
  {Chandler}(2013)}]{limmer2013corresponding}%
  \BibitemOpen
  \bibfield  {author} {\bibinfo {author} {\bibfnamefont {D.~T.}\ \bibnamefont
  {Limmer}}\ and\ \bibinfo {author} {\bibfnamefont {D.}~\bibnamefont
  {Chandler}},\ }\bibfield  {title} {\emph {\bibinfo {title} {Corresponding
  States for Mesostructure and Dynamics of Supercooled Water}},\ }\href@noop {}
  {\bibfield  {journal} {\bibinfo  {journal} {Faraday Discuss.}\ }\textbf
  {\bibinfo {volume} {167}},\ \bibinfo {pages} {485} (\bibinfo {year}
  {2013})}\BibitemShut {NoStop}%
\bibitem [{\citenamefont {Donati}\ \emph {et~al.}(1998)\citenamefont {Donati},
  \citenamefont {Douglas}, \citenamefont {Kob}, \citenamefont {Plimpton},
  \citenamefont {Poole},\ and\ \citenamefont {Glotzer}}]{Donati1998}%
  \BibitemOpen
  \bibfield  {author} {\bibinfo {author} {\bibfnamefont {C.}~\bibnamefont
  {Donati}}, \bibinfo {author} {\bibfnamefont {J.~F.}\ \bibnamefont {Douglas}},
  \bibinfo {author} {\bibfnamefont {W.}~\bibnamefont {Kob}}, \bibinfo {author}
  {\bibfnamefont {S.~J.}\ \bibnamefont {Plimpton}}, \bibinfo {author}
  {\bibfnamefont {P.~H.}\ \bibnamefont {Poole}},\ and\ \bibinfo {author}
  {\bibfnamefont {S.~C.}\ \bibnamefont {Glotzer}},\ }\bibfield  {title} {\emph
  {\bibinfo {title} {Stringlike Cooperative Motion in a Supercooled Liquid}},\
  }\href {https://doi.org/10.1103/PhysRevLett.80.2338} {\bibfield  {journal}
  {\bibinfo  {journal} {Phys. Rev. Lett.}\ }\textbf {\bibinfo {volume} {80}},\
  \bibinfo {pages} {2338} (\bibinfo {year} {1998})}\BibitemShut {NoStop}%
\bibitem [{\citenamefont {Gebremichael}\ \emph {et~al.}(2004)\citenamefont
  {Gebremichael}, \citenamefont {Vogel},\ and\ \citenamefont
  {Glotzer}}]{Gebremichael2004}%
  \BibitemOpen
  \bibfield  {author} {\bibinfo {author} {\bibfnamefont {Y.}~\bibnamefont
  {Gebremichael}}, \bibinfo {author} {\bibfnamefont {M.}~\bibnamefont
  {Vogel}},\ and\ \bibinfo {author} {\bibfnamefont {S.~C.}\ \bibnamefont
  {Glotzer}},\ }\bibfield  {title} {\emph {\bibinfo {title} {Particle Dynamics
  and the Development of String-Like Motion in a Simulated Monoatomic
  Supercooled Liquid}},\ }\href {https://doi.org/10.1063/1.1644539} {\bibfield
  {journal} {\bibinfo  {journal} {J. Chem. Phys.}\ }\textbf {\bibinfo {volume}
  {120}},\ \bibinfo {pages} {4415} (\bibinfo {year} {2004})}\BibitemShut
  {NoStop}%
\bibitem [{\citenamefont {Hurley}\ and\ \citenamefont
  {Harrowell}(1995)}]{hurley1995kinetic}%
  \BibitemOpen
  \bibfield  {author} {\bibinfo {author} {\bibfnamefont {M.}~\bibnamefont
  {Hurley}}\ and\ \bibinfo {author} {\bibfnamefont {P.}~\bibnamefont
  {Harrowell}},\ }\bibfield  {title} {\emph {\bibinfo {title} {Kinetic
  Structure of a Two-Dimensional Liquid}},\ }\href@noop {} {\bibfield
  {journal} {\bibinfo  {journal} {Phys. Rev. E}\ }\textbf {\bibinfo {volume}
  {52}},\ \bibinfo {pages} {1694} (\bibinfo {year} {1995})}\BibitemShut
  {NoStop}%
\bibitem [{\citenamefont {Valeriani}\ \emph {et~al.}(2012)\citenamefont
  {Valeriani}, \citenamefont {Sanz}, \citenamefont {Pusey}, \citenamefont
  {Poon}, \citenamefont {Cates},\ and\ \citenamefont
  {Zaccarelli}}]{valeriani2012compact}%
  \BibitemOpen
  \bibfield  {author} {\bibinfo {author} {\bibfnamefont {C.}~\bibnamefont
  {Valeriani}}, \bibinfo {author} {\bibfnamefont {E.}~\bibnamefont {Sanz}},
  \bibinfo {author} {\bibfnamefont {P.~N.}\ \bibnamefont {Pusey}}, \bibinfo
  {author} {\bibfnamefont {W.~C.~K.}\ \bibnamefont {Poon}}, \bibinfo {author}
  {\bibfnamefont {M.~E.}\ \bibnamefont {Cates}},\ and\ \bibinfo {author}
  {\bibfnamefont {E.}~\bibnamefont {Zaccarelli}},\ }\bibfield  {title} {\emph
  {\bibinfo {title} {From Compact to Fractal Crystalline Clusters in
  Concentrated Systems of Monodisperse Hard Spheres}},\ }\href
  {https://doi.org/10.1039/C2SM25121C} {\bibfield  {journal} {\bibinfo
  {journal} {Soft Matter}\ }\textbf {\bibinfo {volume} {8}},\ \bibinfo {pages}
  {4960} (\bibinfo {year} {2012})}\BibitemShut {NoStop}%
\bibitem [{\citenamefont {Pusey}\ \emph {et~al.}(2009)\citenamefont {Pusey},
  \citenamefont {Zaccarelli}, \citenamefont {Valeriani}, \citenamefont {Sanz},
  \citenamefont {Poon},\ and\ \citenamefont {Cates}}]{Puser2009}%
  \BibitemOpen
  \bibfield  {author} {\bibinfo {author} {\bibfnamefont {P.~N.}\ \bibnamefont
  {Pusey}}, \bibinfo {author} {\bibfnamefont {E.}~\bibnamefont {Zaccarelli}},
  \bibinfo {author} {\bibfnamefont {C.}~\bibnamefont {Valeriani}}, \bibinfo
  {author} {\bibfnamefont {E.}~\bibnamefont {Sanz}}, \bibinfo {author}
  {\bibfnamefont {W.~C.~K.}\ \bibnamefont {Poon}},\ and\ \bibinfo {author}
  {\bibfnamefont {M.~E.}\ \bibnamefont {Cates}},\ }\bibfield  {title} {\emph
  {\bibinfo {title} {Hard Spheres: Crystallization and Glass Formation}},\
  }\href {https://doi.org/10.1098/rsta.2009.0181} {\bibfield  {journal}
  {\bibinfo  {journal} {Philos. Trans. R. Soc. A}\ }\textbf {\bibinfo {volume}
  {367}},\ \bibinfo {pages} {4993} (\bibinfo {year} {2009})}\BibitemShut
  {NoStop}%
\bibitem [{\citenamefont {Sanz}\ \emph {et~al.}(2011)\citenamefont {Sanz},
  \citenamefont {Valeriani}, \citenamefont {Zaccarelli}, \citenamefont {Poon},
  \citenamefont {Pusey},\ and\ \citenamefont {Cates}}]{Sanz2011}%
  \BibitemOpen
  \bibfield  {author} {\bibinfo {author} {\bibfnamefont {E.}~\bibnamefont
  {Sanz}}, \bibinfo {author} {\bibfnamefont {C.}~\bibnamefont {Valeriani}},
  \bibinfo {author} {\bibfnamefont {E.}~\bibnamefont {Zaccarelli}}, \bibinfo
  {author} {\bibfnamefont {W.~C.~K.}\ \bibnamefont {Poon}}, \bibinfo {author}
  {\bibfnamefont {P.~N.}\ \bibnamefont {Pusey}},\ and\ \bibinfo {author}
  {\bibfnamefont {M.~E.}\ \bibnamefont {Cates}},\ }\bibfield  {title} {\emph
  {\bibinfo {title} {Crystallization Mechanism of Hard Sphere Glasses}},\
  }\href {https://doi.org/10.1103/PhysRevLett.106.215701} {\bibfield  {journal}
  {\bibinfo  {journal} {Phys. Rev. Lett.}\ }\textbf {\bibinfo {volume} {106}},\
  \bibinfo {pages} {215701} (\bibinfo {year} {2011})}\BibitemShut {NoStop}%
\bibitem [{\citenamefont {Sanz}\ \emph {et~al.}(2014)\citenamefont {Sanz},
  \citenamefont {Valeriani}, \citenamefont {Zaccarelli}, \citenamefont {Poon},
  \citenamefont {Cates},\ and\ \citenamefont {Pusey}}]{Sanz2014}%
  \BibitemOpen
  \bibfield  {author} {\bibinfo {author} {\bibfnamefont {E.}~\bibnamefont
  {Sanz}}, \bibinfo {author} {\bibfnamefont {C.}~\bibnamefont {Valeriani}},
  \bibinfo {author} {\bibfnamefont {E.}~\bibnamefont {Zaccarelli}}, \bibinfo
  {author} {\bibfnamefont {W.~C.~K.}\ \bibnamefont {Poon}}, \bibinfo {author}
  {\bibfnamefont {M.~E.}\ \bibnamefont {Cates}},\ and\ \bibinfo {author}
  {\bibfnamefont {P.~N.}\ \bibnamefont {Pusey}},\ }\bibfield  {title} {\emph
  {\bibinfo {title} {Avalanches Mediate Crystallization in a Hard-Sphere
  Glass}},\ }\href {https://doi.org/10.1073/PNAS.1308338110} {\bibfield
  {journal} {\bibinfo  {journal} {Proc. Natl. Acad. Sci. U.S.A.}\ }\textbf
  {\bibinfo {volume} {111}},\ \bibinfo {pages} {75} (\bibinfo {year}
  {2014})}\BibitemShut {NoStop}%
\bibitem [{\citenamefont {Angell}\ \emph {et~al.}(2000)\citenamefont {Angell},
  \citenamefont {Ngai}, \citenamefont {McKenna}, \citenamefont {McMillan},\
  and\ \citenamefont {Martin}}]{Angell2000}%
  \BibitemOpen
  \bibfield  {author} {\bibinfo {author} {\bibfnamefont {C.~A.}\ \bibnamefont
  {Angell}}, \bibinfo {author} {\bibfnamefont {K.~L.}\ \bibnamefont {Ngai}},
  \bibinfo {author} {\bibfnamefont {G.~B.}\ \bibnamefont {McKenna}}, \bibinfo
  {author} {\bibfnamefont {P.~F.}\ \bibnamefont {McMillan}},\ and\ \bibinfo
  {author} {\bibfnamefont {S.~W.}\ \bibnamefont {Martin}},\ }\bibfield  {title}
  {\emph {\bibinfo {title} {Relaxation in Glassforming Liquids and Amorphous
  Solids}},\ }\href {https://doi.org/10.1063/1.1286035} {\bibfield  {journal}
  {\bibinfo  {journal} {J. Appl. Phys}\ }\textbf {\bibinfo {volume} {88}},\
  \bibinfo {pages} {3113} (\bibinfo {year} {2000})}\BibitemShut {NoStop}%
\bibitem [{\citenamefont {Kegel}\ and\ \citenamefont {van
  Blaaderen}(2000)}]{Kegel2000}%
  \BibitemOpen
  \bibfield  {author} {\bibinfo {author} {\bibfnamefont {W.~K.}\ \bibnamefont
  {Kegel}}\ and\ \bibinfo {author} {\bibfnamefont {A.}~\bibnamefont {van
  Blaaderen}},\ }\bibfield  {title} {\emph {\bibinfo {title} {Direct
  Observation of Dynamical Heterogeneities in Colloidal Hard-Sphere
  Suspensions}},\ }\href {https://doi.org/10.1126/science.287.5451.290}
  {\bibfield  {journal} {\bibinfo  {journal} {Science}\ }\textbf {\bibinfo
  {volume} {287}},\ \bibinfo {pages} {290} (\bibinfo {year}
  {2000})}\BibitemShut {NoStop}%
\bibitem [{\citenamefont {Weeks}\ \emph {et~al.}(2000)\citenamefont {Weeks},
  \citenamefont {Crocker}, \citenamefont {Levitt}, \citenamefont {Schofield},\
  and\ \citenamefont {Weitz}}]{Weeks2000}%
  \BibitemOpen
  \bibfield  {author} {\bibinfo {author} {\bibfnamefont {E.~R.}\ \bibnamefont
  {Weeks}}, \bibinfo {author} {\bibfnamefont {J.~C.}\ \bibnamefont {Crocker}},
  \bibinfo {author} {\bibfnamefont {A.~C.}\ \bibnamefont {Levitt}}, \bibinfo
  {author} {\bibfnamefont {A.}~\bibnamefont {Schofield}},\ and\ \bibinfo
  {author} {\bibfnamefont {D.~A.}\ \bibnamefont {Weitz}},\ }\bibfield  {title}
  {\emph {\bibinfo {title} {Three-Dimensional Direct Imaging of Structural
  Relaxation Near the Colloidal Glass Transition}},\ }\href
  {https://doi.org/10.1126/science.287.5453.627} {\bibfield  {journal}
  {\bibinfo  {journal} {Science}\ }\textbf {\bibinfo {volume} {287}},\ \bibinfo
  {pages} {627} (\bibinfo {year} {2000})}\BibitemShut {NoStop}%
\bibitem [{\citenamefont {Gao}\ and\ \citenamefont {Kilfoil}(2007)}]{Gao2007}%
  \BibitemOpen
  \bibfield  {author} {\bibinfo {author} {\bibfnamefont {Y.}~\bibnamefont
  {Gao}}\ and\ \bibinfo {author} {\bibfnamefont {M.~L.}\ \bibnamefont
  {Kilfoil}},\ }\bibfield  {title} {\emph {\bibinfo {title} {Direct Imaging of
  Dynamical Heterogeneities Near the Colloid-Gel Transition}},\ }\href
  {https://doi.org/10.1103/PhysRevLett.99.078301} {\bibfield  {journal}
  {\bibinfo  {journal} {Phys. Rev. Lett.}\ }\textbf {\bibinfo {volume} {99}},\
  \bibinfo {pages} {078301} (\bibinfo {year} {2007})}\BibitemShut {NoStop}%
\bibitem [{\citenamefont {Fris}\ \emph {et~al.}(2011)\citenamefont {Fris},
  \citenamefont {Appignanesi},\ and\ \citenamefont {Weeks}}]{Fris2011}%
  \BibitemOpen
  \bibfield  {author} {\bibinfo {author} {\bibfnamefont {J.~A.~R.}\
  \bibnamefont {Fris}}, \bibinfo {author} {\bibfnamefont {G.~A.}\ \bibnamefont
  {Appignanesi}},\ and\ \bibinfo {author} {\bibfnamefont {E.~R.}\ \bibnamefont
  {Weeks}},\ }\bibfield  {title} {\emph {\bibinfo {title} {Experimental
  Verification of Rapid, Sporadic Particle Motions by Direct Imaging of Glassy
  Colloidal Systems}},\ }\href {https://doi.org/10.1103/PhysRevLett.107.065704}
  {\bibfield  {journal} {\bibinfo  {journal} {Phys. Rev. Lett.}\ }\textbf
  {\bibinfo {volume} {107}},\ \bibinfo {pages} {065704} (\bibinfo {year}
  {2011})}\BibitemShut {NoStop}%
\bibitem [{\citenamefont {Garrahan}\ and\ \citenamefont
  {Chandler}(2002)}]{Garrahan2002}%
  \BibitemOpen
  \bibfield  {author} {\bibinfo {author} {\bibfnamefont {J.~P.}\ \bibnamefont
  {Garrahan}}\ and\ \bibinfo {author} {\bibfnamefont {D.}~\bibnamefont
  {Chandler}},\ }\bibfield  {title} {\emph {\bibinfo {title} {Geometrical
  Explanation and Scaling of Dynamical Heterogeneities in Glass Forming
  Systems}},\ }\href {https://doi.org/10.1103/PhysRevLett.89.035704} {\bibfield
   {journal} {\bibinfo  {journal} {Phys. Rev. Lett.}\ }\textbf {\bibinfo
  {volume} {89}},\ \bibinfo {pages} {035704} (\bibinfo {year}
  {2002})}\BibitemShut {NoStop}%
\bibitem [{\citenamefont {Chandler}\ and\ \citenamefont
  {Garrahan}(2010)}]{Chandler2010}%
  \BibitemOpen
  \bibfield  {author} {\bibinfo {author} {\bibfnamefont {D.}~\bibnamefont
  {Chandler}}\ and\ \bibinfo {author} {\bibfnamefont {J.~P.}\ \bibnamefont
  {Garrahan}},\ }\bibfield  {title} {\emph {\bibinfo {title} {Dynamics on the
  Way to Forming Glass: Bubbles in Space-Time}},\ }\href
  {https://doi.org/10.1146/annurev.physchem.040808.090405} {\bibfield
  {journal} {\bibinfo  {journal} {Annu. Rev. Phys. Chem}\ }\textbf {\bibinfo
  {volume} {61}},\ \bibinfo {pages} {191} (\bibinfo {year} {2010})}\BibitemShut
  {NoStop}%
\bibitem [{\citenamefont {Biroli}\ and\ \citenamefont
  {Garrahan}(2013)}]{Biroli2013}%
  \BibitemOpen
  \bibfield  {author} {\bibinfo {author} {\bibfnamefont {G.}~\bibnamefont
  {Biroli}}\ and\ \bibinfo {author} {\bibfnamefont {J.~P.}\ \bibnamefont
  {Garrahan}},\ }\bibfield  {title} {\emph {\bibinfo {title} {Perspective: The
  Glass Transition}},\ }\href {https://doi.org/10.1063/1.4795539} {\bibfield
  {journal} {\bibinfo  {journal} {J. Chem. Phys.}\ }\textbf {\bibinfo {volume}
  {138}},\ \bibinfo {pages} {12A301} (\bibinfo {year} {2013})}\BibitemShut
  {NoStop}%
\bibitem [{\citenamefont {Keys}\ \emph {et~al.}(2011)\citenamefont {Keys},
  \citenamefont {Hedges}, \citenamefont {Garrahan}, \citenamefont {Glotzer},\
  and\ \citenamefont {Chandler}}]{Keys2011}%
  \BibitemOpen
  \bibfield  {author} {\bibinfo {author} {\bibfnamefont {A.~S.}\ \bibnamefont
  {Keys}}, \bibinfo {author} {\bibfnamefont {L.~O.}\ \bibnamefont {Hedges}},
  \bibinfo {author} {\bibfnamefont {J.~P.}\ \bibnamefont {Garrahan}}, \bibinfo
  {author} {\bibfnamefont {S.~C.}\ \bibnamefont {Glotzer}},\ and\ \bibinfo
  {author} {\bibfnamefont {D.}~\bibnamefont {Chandler}},\ }\bibfield  {title}
  {\emph {\bibinfo {title} {Excitations Are Localized and Relaxation Is
  Hierarchical in Glass-Forming Liquids}},\ }\href
  {https://doi.org/10.1103/PhysRevX.1.021013} {\bibfield  {journal} {\bibinfo
  {journal} {Phys. Rev. X}\ }\textbf {\bibinfo {volume} {1}},\ \bibinfo {pages}
  {021013} (\bibinfo {year} {2011})}\BibitemShut {NoStop}%
\bibitem [{\citenamefont {Ritort}\ and\ \citenamefont
  {Sollich}(2003)}]{Ritort2003}%
  \BibitemOpen
  \bibfield  {author} {\bibinfo {author} {\bibfnamefont {F.}~\bibnamefont
  {Ritort}}\ and\ \bibinfo {author} {\bibfnamefont {P.}~\bibnamefont
  {Sollich}},\ }\bibfield  {title} {\emph {\bibinfo {title} {Glassy Dynamics of
  Kinetically Constrained Models}},\ }\href
  {https://doi.org/10.1080/0001873031000093582} {\bibfield  {journal} {\bibinfo
   {journal} {Adv. Phys.}\ }\textbf {\bibinfo {volume} {52}},\ \bibinfo {pages}
  {219} (\bibinfo {year} {2003})}\BibitemShut {NoStop}%
\bibitem [{\citenamefont {Palmer}\ \emph {et~al.}(1984)\citenamefont {Palmer},
  \citenamefont {Stein}, \citenamefont {Abrahams},\ and\ \citenamefont
  {Anderson}}]{palmer1984models}%
  \BibitemOpen
  \bibfield  {author} {\bibinfo {author} {\bibfnamefont {R.~G.}\ \bibnamefont
  {Palmer}}, \bibinfo {author} {\bibfnamefont {D.~L.}\ \bibnamefont {Stein}},
  \bibinfo {author} {\bibfnamefont {E.}~\bibnamefont {Abrahams}},\ and\
  \bibinfo {author} {\bibfnamefont {P.~W.}\ \bibnamefont {Anderson}},\
  }\bibfield  {title} {\emph {\bibinfo {title} {Models of Hierarchically
  Constrained Dynamics for Glassy Relaxation}},\ }\href@noop {} {\bibfield
  {journal} {\bibinfo  {journal} {Phys. Rev. Lett.}\ }\textbf {\bibinfo
  {volume} {53}},\ \bibinfo {pages} {958} (\bibinfo {year} {1984})}\BibitemShut
  {NoStop}%
\bibitem [{\citenamefont {Garrahan}\ and\ \citenamefont
  {Chandler}(2003)}]{Garrahan2003}%
  \BibitemOpen
  \bibfield  {author} {\bibinfo {author} {\bibfnamefont {J.~P.}\ \bibnamefont
  {Garrahan}}\ and\ \bibinfo {author} {\bibfnamefont {D.}~\bibnamefont
  {Chandler}},\ }\bibfield  {title} {\emph {\bibinfo {title} {Coarse-Grained
  Microscopic Model of Glass Formers}},\ }\href
  {https://doi.org/10.1073/pnas.1233719100} {\bibfield  {journal} {\bibinfo
  {journal} {Proc. Natl. Acad. Sci. U.S.A.}\ }\textbf {\bibinfo {volume}
  {100}},\ \bibinfo {pages} {9710} (\bibinfo {year} {2003})}\BibitemShut
  {NoStop}%
\bibitem [{\citenamefont {Potts}(1952)}]{Potts1952}%
  \BibitemOpen
  \bibfield  {author} {\bibinfo {author} {\bibfnamefont {R.~B.}\ \bibnamefont
  {Potts}},\ }\bibfield  {title} {\emph {\bibinfo {title} {Some Generalized
  Order-Disorder Transformations}},\ }\href
  {https://doi.org/10.1017/S0305004100027419} {\bibfield  {journal} {\bibinfo
  {journal} {Math. Proc. Camb. Philos. Soc}\ }\textbf {\bibinfo {volume}
  {48}},\ \bibinfo {pages} {106} (\bibinfo {year} {1952})}\BibitemShut
  {NoStop}%
\bibitem [{\citenamefont {Wu}(1982)}]{Wu1982}%
  \BibitemOpen
  \bibfield  {author} {\bibinfo {author} {\bibfnamefont {F.~Y.}\ \bibnamefont
  {Wu}},\ }\bibfield  {title} {\emph {\bibinfo {title} {The Potts Model}},\
  }\href {https://doi.org/10.1103/RevModPhys.54.235} {\bibfield  {journal}
  {\bibinfo  {journal} {Rev. Mod. Phys.}\ }\textbf {\bibinfo {volume} {54}},\
  \bibinfo {pages} {235} (\bibinfo {year} {1982})}\BibitemShut {NoStop}%
\bibitem [{\citenamefont {Aldous}\ and\ \citenamefont
  {Diaconis}(2002)}]{Aldous2002}%
  \BibitemOpen
  \bibfield  {author} {\bibinfo {author} {\bibfnamefont {D.}~\bibnamefont
  {Aldous}}\ and\ \bibinfo {author} {\bibfnamefont {P.}~\bibnamefont
  {Diaconis}},\ }\bibfield  {title} {\emph {\bibinfo {title} {The Asymmetric
  One-Dimensional Constrained Ising Model: Rigorous Results}},\ }\href
  {https://doi.org/10.1023/A:1015170205728} {\bibfield  {journal} {\bibinfo
  {journal} {J. Stat. Phys.}\ }\textbf {\bibinfo {volume} {107}},\ \bibinfo
  {pages} {945} (\bibinfo {year} {2002})}\BibitemShut {NoStop}%
\bibitem [{\citenamefont {Chleboun}\ \emph {et~al.}(2014)\citenamefont
  {Chleboun}, \citenamefont {Faggionato},\ and\ \citenamefont
  {Martinelli}}]{chleboun2014influence}%
  \BibitemOpen
  \bibfield  {author} {\bibinfo {author} {\bibfnamefont {P.}~\bibnamefont
  {Chleboun}}, \bibinfo {author} {\bibfnamefont {A.}~\bibnamefont
  {Faggionato}},\ and\ \bibinfo {author} {\bibfnamefont {F.}~\bibnamefont
  {Martinelli}},\ }\bibfield  {title} {\emph {\bibinfo {title} {The Influence
  of Dimension on the Relaxation Process of East-Like Models: Rigorous
  Results}},\ }\href@noop {} {\bibfield  {journal} {\bibinfo  {journal} {EPL}\
  }\textbf {\bibinfo {volume} {107}},\ \bibinfo {pages} {36002} (\bibinfo
  {year} {2014})}\BibitemShut {NoStop}%
\bibitem [{\citenamefont {Sollich}\ and\ \citenamefont
  {Evans}(1999)}]{Sollich1999}%
  \BibitemOpen
  \bibfield  {author} {\bibinfo {author} {\bibfnamefont {P.}~\bibnamefont
  {Sollich}}\ and\ \bibinfo {author} {\bibfnamefont {M.~R.}\ \bibnamefont
  {Evans}},\ }\bibfield  {title} {\emph {\bibinfo {title} {Glassy Time-Scale
  Divergence and Anomalous Coarsening in a Kinetically Constrained Spin
  Chain}},\ }\href {https://doi.org/10.1103/PhysRevLett.83.3238} {\bibfield
  {journal} {\bibinfo  {journal} {Phys. Rev. Lett.}\ }\textbf {\bibinfo
  {volume} {83}},\ \bibinfo {pages} {3238} (\bibinfo {year}
  {1999})}\BibitemShut {NoStop}%
\bibitem [{\citenamefont {Sollich}\ and\ \citenamefont
  {Evans}(2003)}]{Sollich2003}%
  \BibitemOpen
  \bibfield  {author} {\bibinfo {author} {\bibfnamefont {P.}~\bibnamefont
  {Sollich}}\ and\ \bibinfo {author} {\bibfnamefont {M.~R.}\ \bibnamefont
  {Evans}},\ }\bibfield  {title} {\emph {\bibinfo {title} {Glassy dynamics in
  the Asymmetrically Constrained Kinetic Ising Chain}},\ }\href
  {https://doi.org/10.1103/PhysRevE.68.031504} {\bibfield  {journal} {\bibinfo
  {journal} {Phys. Rev. E}\ }\textbf {\bibinfo {volume} {68}},\ \bibinfo
  {pages} {031504} (\bibinfo {year} {2003})}\BibitemShut {NoStop}%
\bibitem [{\citenamefont {Buhot}\ and\ \citenamefont
  {Garrahan}(2001)}]{Buhot2001}%
  \BibitemOpen
  \bibfield  {author} {\bibinfo {author} {\bibfnamefont {A.}~\bibnamefont
  {Buhot}}\ and\ \bibinfo {author} {\bibfnamefont {J.~P.}\ \bibnamefont
  {Garrahan}},\ }\bibfield  {title} {\emph {\bibinfo {title} {Crossover from
  Fragile to Strong Glassy Behavior in Kinetically Constrained Systems}},\
  }\href {https://doi.org/10.1103/PhysRevE.64.021505} {\bibfield  {journal}
  {\bibinfo  {journal} {Phys. Rev. E}\ }\textbf {\bibinfo {volume} {64}},\
  \bibinfo {pages} {021505} (\bibinfo {year} {2001})}\BibitemShut {NoStop}%
\bibitem [{\citenamefont {J{\"{a}}ckle}\ and\ \citenamefont
  {Eisinger}(1991)}]{Jackle1991}%
  \BibitemOpen
  \bibfield  {author} {\bibinfo {author} {\bibfnamefont {J.}~\bibnamefont
  {J{\"{a}}ckle}}\ and\ \bibinfo {author} {\bibfnamefont {S.}~\bibnamefont
  {Eisinger}},\ }\bibfield  {title} {\emph {\bibinfo {title} {A Hierarchically
  Constrained Kinetic Ising Model}},\ }\href
  {https://doi.org/10.1007/BF01453764} {\bibfield  {journal} {\bibinfo
  {journal} {Z. Phys. B}\ }\textbf {\bibinfo {volume} {84}},\ \bibinfo {pages}
  {115} (\bibinfo {year} {1991})}\BibitemShut {NoStop}%
\bibitem [{\citenamefont {Faggionato}\ \emph {et~al.}(2012)\citenamefont
  {Faggionato}, \citenamefont {Martinelli}, \citenamefont {Roberto},\ and\
  \citenamefont {Toninelli}}]{Faggionato2012}%
  \BibitemOpen
  \bibfield  {author} {\bibinfo {author} {\bibfnamefont {A.}~\bibnamefont
  {Faggionato}}, \bibinfo {author} {\bibfnamefont {F.}~\bibnamefont
  {Martinelli}}, \bibinfo {author} {\bibfnamefont {C.}~\bibnamefont
  {Roberto}},\ and\ \bibinfo {author} {\bibfnamefont {C.}~\bibnamefont
  {Toninelli}},\ }\bibfield  {title} {\emph {\bibinfo {title} {Universality in
  One-Dimensional Hierarchical Coalescence Processes}},\ }\href
  {https://doi.org/10.1214/11-AOP654} {\bibfield  {journal} {\bibinfo
  {journal} {Ann. Probab.}\ }\textbf {\bibinfo {volume} {40}},\ \bibinfo
  {pages} {1377} (\bibinfo {year} {2012})}\BibitemShut {NoStop}%
\bibitem [{\citenamefont {Elmatad}\ \emph {et~al.}(2009)\citenamefont
  {Elmatad}, \citenamefont {Chandler},\ and\ \citenamefont
  {Garrahan}}]{Elmatad2009}%
  \BibitemOpen
  \bibfield  {author} {\bibinfo {author} {\bibfnamefont {Y.~S.}\ \bibnamefont
  {Elmatad}}, \bibinfo {author} {\bibfnamefont {D.}~\bibnamefont {Chandler}},\
  and\ \bibinfo {author} {\bibfnamefont {J.~P.}\ \bibnamefont {Garrahan}},\
  }\bibfield  {title} {\emph {\bibinfo {title} {Corresponding States of
  Structural Glass Formers}},\ }\href {https://doi.org/10.1021/jp810362g}
  {\bibfield  {journal} {\bibinfo  {journal} {J. Phys. Chem. B}\ }\textbf
  {\bibinfo {volume} {113}},\ \bibinfo {pages} {5563} (\bibinfo {year}
  {2009})}\BibitemShut {NoStop}%
\bibitem [{\citenamefont {Katira}\ \emph {et~al.}(2019)\citenamefont {Katira},
  \citenamefont {Garrahan},\ and\ \citenamefont {Mandadapu}}]{Katira2019}%
  \BibitemOpen
  \bibfield  {author} {\bibinfo {author} {\bibfnamefont {S.}~\bibnamefont
  {Katira}}, \bibinfo {author} {\bibfnamefont {J.~P.}\ \bibnamefont
  {Garrahan}},\ and\ \bibinfo {author} {\bibfnamefont {K.~K.}\ \bibnamefont
  {Mandadapu}},\ }\bibfield  {title} {\emph {\bibinfo {title} {Theory for
  Glassy Behavior of Supercooled Liquid Mixtures}},\ }\href@noop {} {\bibfield
  {journal} {\bibinfo  {journal} {Phys. Rev. Lett.}\ }\textbf {\bibinfo
  {volume} {123}},\ \bibinfo {pages} {100602} (\bibinfo {year}
  {2019})}\BibitemShut {NoStop}%
\bibitem [{\citenamefont {Nagamanasa}\ \emph {et~al.}(2011)\citenamefont
  {Nagamanasa}, \citenamefont {Gokhale}, \citenamefont {Ganapathy},\ and\
  \citenamefont {Sood}}]{Nagamanasa2011}%
  \BibitemOpen
  \bibfield  {author} {\bibinfo {author} {\bibfnamefont {K.~H.}\ \bibnamefont
  {Nagamanasa}}, \bibinfo {author} {\bibfnamefont {S.}~\bibnamefont {Gokhale}},
  \bibinfo {author} {\bibfnamefont {R.}~\bibnamefont {Ganapathy}},\ and\
  \bibinfo {author} {\bibfnamefont {A.~K.}\ \bibnamefont {Sood}},\ }\bibfield
  {title} {\emph {\bibinfo {title} {Confined Glassy Dynamics at Grain
  Boundaries in Colloidal Crystals}},\ }\href
  {https://doi.org/10.1073/pnas.1101858108} {\bibfield  {journal} {\bibinfo
  {journal} {Proc. Natl. Acad. Sci. U.S.A.}\ }\textbf {\bibinfo {volume}
  {108}},\ \bibinfo {pages} {11323} (\bibinfo {year} {2011})}\BibitemShut
  {NoStop}%
\bibitem [{\citenamefont {Anderson}\ \emph {et~al.}(2017)\citenamefont
  {Anderson}, \citenamefont {Hirth},\ and\ \citenamefont
  {Lothe}}]{anderson2017theory}%
  \BibitemOpen
  \bibfield  {author} {\bibinfo {author} {\bibfnamefont {P.~M.}\ \bibnamefont
  {Anderson}}, \bibinfo {author} {\bibfnamefont {J.~P.}\ \bibnamefont
  {Hirth}},\ and\ \bibinfo {author} {\bibfnamefont {J.}~\bibnamefont {Lothe}},\
  }\href@noop {} {\emph {\bibinfo {title} {Theory of Dislocations}}}\ (\bibinfo
   {publisher} {Cambridge University Press},\ \bibinfo {year}
  {2017})\BibitemShut {NoStop}%
\bibitem [{Note1()}]{Note1}%
  \BibitemOpen
  \bibinfo {note} {See Supplementary Material.}\BibitemShut {Stop}%
\bibitem [{\citenamefont {Berker}\ \emph {et~al.}(1978)\citenamefont {Berker},
  \citenamefont {Ostlund},\ and\ \citenamefont {Putnam}}]{Berker1978a}%
  \BibitemOpen
  \bibfield  {author} {\bibinfo {author} {\bibfnamefont {A.~N.}\ \bibnamefont
  {Berker}}, \bibinfo {author} {\bibfnamefont {S.}~\bibnamefont {Ostlund}},\
  and\ \bibinfo {author} {\bibfnamefont {F.~A.}\ \bibnamefont {Putnam}},\
  }\bibfield  {title} {\emph {\bibinfo {title} {Renormalization-Group Treatment
  of a Potts Lattice Gas for Krypton Adsorbed onto Graphite}},\ }\href
  {https://doi.org/10.1103/PhysRevB.17.3650} {\bibfield  {journal} {\bibinfo
  {journal} {Phys. Rev. B}\ }\textbf {\bibinfo {volume} {17}},\ \bibinfo
  {pages} {3650} (\bibinfo {year} {1978})}\BibitemShut {NoStop}%
\bibitem [{\citenamefont {Nienhuis}\ \emph {et~al.}(1979)\citenamefont
  {Nienhuis}, \citenamefont {Berker}, \citenamefont {Riedel},\ and\
  \citenamefont {Schick}}]{Nienhuis1979}%
  \BibitemOpen
  \bibfield  {author} {\bibinfo {author} {\bibfnamefont {B.}~\bibnamefont
  {Nienhuis}}, \bibinfo {author} {\bibfnamefont {A.~N.}\ \bibnamefont
  {Berker}}, \bibinfo {author} {\bibfnamefont {E.~K.}\ \bibnamefont {Riedel}},\
  and\ \bibinfo {author} {\bibfnamefont {M.}~\bibnamefont {Schick}},\
  }\bibfield  {title} {\emph {\bibinfo {title} {First- and Second-Order Phase
  Transitions in Potts Models: Renormalization-Group Solution}},\ }\href
  {https://doi.org/10.1103/PhysRevLett.43.737} {\bibfield  {journal} {\bibinfo
  {journal} {Phys. Rev. Lett.}\ }\textbf {\bibinfo {volume} {43}},\ \bibinfo
  {pages} {737} (\bibinfo {year} {1979})}\BibitemShut {NoStop}%
\bibitem [{\citenamefont {Berker}\ and\ \citenamefont
  {Wortis}(1976)}]{Berker1976b}%
  \BibitemOpen
  \bibfield  {author} {\bibinfo {author} {\bibfnamefont {A.~N.}\ \bibnamefont
  {Berker}}\ and\ \bibinfo {author} {\bibfnamefont {M.}~\bibnamefont
  {Wortis}},\ }\bibfield  {title} {\emph {\bibinfo {title}
  {Blume-Emery-Griffiths-Potts Model in Two Dimensions: Phase Diagram and
  Critical Properties from a Position-Space Renormalization Group}},\ }\href
  {https://doi.org/10.1103/PhysRevB.14.4946} {\bibfield  {journal} {\bibinfo
  {journal} {Phys. Rev. B}\ }\textbf {\bibinfo {volume} {14}},\ \bibinfo
  {pages} {4946} (\bibinfo {year} {1976})}\BibitemShut {NoStop}%
\bibitem [{\citenamefont {Kihara}\ \emph {et~al.}(1954)\citenamefont {Kihara},
  \citenamefont {Midzuno},\ and\ \citenamefont {Shizume}}]{Kihara1954}%
  \BibitemOpen
  \bibfield  {author} {\bibinfo {author} {\bibfnamefont {T.}~\bibnamefont
  {Kihara}}, \bibinfo {author} {\bibfnamefont {Y.}~\bibnamefont {Midzuno}},\
  and\ \bibinfo {author} {\bibfnamefont {T.}~\bibnamefont {Shizume}},\
  }\bibfield  {title} {\emph {\bibinfo {title} {Statistics of Two-Dimensional
  Lattices with Many Components}},\ }\href {https://doi.org/10.1143/JPSJ.9.681}
  {\bibfield  {journal} {\bibinfo  {journal} {J. Phys. Soc. Jpn.}\ }\textbf
  {\bibinfo {volume} {9}},\ \bibinfo {pages} {681} (\bibinfo {year}
  {1954})}\BibitemShut {NoStop}%
\bibitem [{\citenamefont {Hintermann}\ \emph {et~al.}(1978)\citenamefont
  {Hintermann}, \citenamefont {Kunz},\ and\ \citenamefont
  {Wu}}]{Hintermann1978}%
  \BibitemOpen
  \bibfield  {author} {\bibinfo {author} {\bibfnamefont {A.}~\bibnamefont
  {Hintermann}}, \bibinfo {author} {\bibfnamefont {H.}~\bibnamefont {Kunz}},\
  and\ \bibinfo {author} {\bibfnamefont {F.~Y.}\ \bibnamefont {Wu}},\
  }\bibfield  {title} {\emph {\bibinfo {title} {Exact Results for the Potts
  Model in Two Dimensions}},\ }\href {https://doi.org/10.1007/BF01011773}
  {\bibfield  {journal} {\bibinfo  {journal} {J. Stat. Phys.}\ }\textbf
  {\bibinfo {volume} {19}},\ \bibinfo {pages} {623} (\bibinfo {year}
  {1978})}\BibitemShut {NoStop}%
\bibitem [{\citenamefont {Baxter}(1973)}]{Baxter1973}%
  \BibitemOpen
  \bibfield  {author} {\bibinfo {author} {\bibfnamefont {R.~J.}\ \bibnamefont
  {Baxter}},\ }\bibfield  {title} {\emph {\bibinfo {title} {Potts Model at the
  Critical Temperature}},\ }\href {https://doi.org/10.1088/0022-3719/6/23/005}
  {\bibfield  {journal} {\bibinfo  {journal} {J. Phys. C}\ }\textbf {\bibinfo
  {volume} {6}},\ \bibinfo {pages} {L445} (\bibinfo {year} {1973})}\BibitemShut
  {NoStop}%
\bibitem [{\citenamefont {Mittag}\ and\ \citenamefont
  {Stephen}(1974)}]{Mittag1974}%
  \BibitemOpen
  \bibfield  {author} {\bibinfo {author} {\bibfnamefont {L.}~\bibnamefont
  {Mittag}}\ and\ \bibinfo {author} {\bibfnamefont {M.~J.}\ \bibnamefont
  {Stephen}},\ }\bibfield  {title} {\emph {\bibinfo {title} {Mean-Field Theory
  of the Many Component Potts Model}},\ }\href
  {https://doi.org/10.1088/0305-4470/7/9/003} {\bibfield  {journal} {\bibinfo
  {journal} {J. Phys. A: Math. Nuc. Gen.}\ }\textbf {\bibinfo {volume} {7}},\
  \bibinfo {pages} {L109} (\bibinfo {year} {1974})}\BibitemShut {NoStop}%
\bibitem [{\citenamefont {Pearce}\ and\ \citenamefont
  {Griffiths}(1980)}]{Pearce1980}%
  \BibitemOpen
  \bibfield  {author} {\bibinfo {author} {\bibfnamefont {P.~A.}\ \bibnamefont
  {Pearce}}\ and\ \bibinfo {author} {\bibfnamefont {R.~B.}\ \bibnamefont
  {Griffiths}},\ }\bibfield  {title} {\emph {\bibinfo {title} {Potts Model in
  the Many-Component Limit}},\ }\href
  {https://doi.org/10.1088/0305-4470/13/6/035} {\bibfield  {journal} {\bibinfo
  {journal} {J. Phys. A: Math. Gen.}\ }\textbf {\bibinfo {volume} {13}},\
  \bibinfo {pages} {2143} (\bibinfo {year} {1980})}\BibitemShut {NoStop}%
\bibitem [{\citenamefont {Christian}(2002)}]{christian2002theory}%
  \BibitemOpen
  \bibfield  {author} {\bibinfo {author} {\bibfnamefont {J.~W.}\ \bibnamefont
  {Christian}},\ }\href@noop {} {\emph {\bibinfo {title} {The Theory of
  Transformations in Metals and Alloys}}}\ (\bibinfo  {publisher} {Pergamon
  Press},\ \bibinfo {year} {2002})\BibitemShut {NoStop}%
\bibitem [{\citenamefont {Legg}\ \emph {et~al.}(2007)\citenamefont {Legg},
  \citenamefont {Schroers},\ and\ \citenamefont {Busch}}]{Legg2007a}%
  \BibitemOpen
  \bibfield  {author} {\bibinfo {author} {\bibfnamefont {B.~A.}\ \bibnamefont
  {Legg}}, \bibinfo {author} {\bibfnamefont {J.}~\bibnamefont {Schroers}},\
  and\ \bibinfo {author} {\bibfnamefont {R.}~\bibnamefont {Busch}},\ }\bibfield
   {title} {\emph {\bibinfo {title} {Thermodynamics, Kinetics, and
  Crystallization of
  $\mathrm{Pt}_{57.3}\mathrm{Cu}_{14.6}\mathrm{Ni}_{5.3}\mathrm{P}_{22.8}$ Bulk
  Metallic Glass}},\ }\href
  {https://doi.org/https://doi.org/10.1016/j.actamat.2006.09.024} {\bibfield
  {journal} {\bibinfo  {journal} {Acta Mater.}\ }\textbf {\bibinfo {volume}
  {55}},\ \bibinfo {pages} {1109 } (\bibinfo {year} {2007})}\BibitemShut
  {NoStop}%
\bibitem [{\citenamefont {Mukherjee}\ \emph {et~al.}(2004)\citenamefont
  {Mukherjee}, \citenamefont {Kang}, \citenamefont {Johnson},\ and\
  \citenamefont {Rhim}}]{Mukherjee2004a}%
  \BibitemOpen
  \bibfield  {author} {\bibinfo {author} {\bibfnamefont {S.}~\bibnamefont
  {Mukherjee}}, \bibinfo {author} {\bibfnamefont {H.~G.}\ \bibnamefont {Kang}},
  \bibinfo {author} {\bibfnamefont {W.~L.}\ \bibnamefont {Johnson}},\ and\
  \bibinfo {author} {\bibfnamefont {W.~K.}\ \bibnamefont {Rhim}},\ }\bibfield
  {title} {\emph {\bibinfo {title} {Noncontact Measurement of Crystallization
  Behavior, Specific Volume, and Viscosity of Bulk Glass-Forming
  $\mathrm{Zr}$-$\mathrm{Al}$-$\mathrm{Co}$-$\mathrm{(Cu)}$ Alloys}},\ }\href
  {https://doi.org/10.1103/PhysRevB.70.174205} {\bibfield  {journal} {\bibinfo
  {journal} {Phys. Rev. B}\ }\textbf {\bibinfo {volume} {70}},\ \bibinfo
  {pages} {174205} (\bibinfo {year} {2004})}\BibitemShut {NoStop}%
\bibitem [{\citenamefont {Gallino}\ \emph {et~al.}(2007)\citenamefont
  {Gallino}, \citenamefont {Shah},\ and\ \citenamefont {Busch}}]{Gallino2007}%
  \BibitemOpen
  \bibfield  {author} {\bibinfo {author} {\bibfnamefont {I.}~\bibnamefont
  {Gallino}}, \bibinfo {author} {\bibfnamefont {M.~B.}\ \bibnamefont {Shah}},\
  and\ \bibinfo {author} {\bibfnamefont {R.}~\bibnamefont {Busch}},\ }\bibfield
   {title} {\emph {\bibinfo {title} {Enthalpy Relaxation and Its Relation to
  the Thermodynamics and Crystallization of the
  $\mathrm{Zr}_{58.5}\mathrm{Cu}_{15.6}\mathrm{Ni}_{12.8}\mathrm{Al}_{10.3}\mathrm{Nb}_{2.8}$
  Bulk Metallic Glass-Forming Alloy}},\ }\href
  {https://doi.org/https://doi.org/10.1016/j.actamat.2006.09.040} {\bibfield
  {journal} {\bibinfo  {journal} {Acta Mater.}\ }\textbf {\bibinfo {volume}
  {55}},\ \bibinfo {pages} {1367 } (\bibinfo {year} {2007})}\BibitemShut
  {NoStop}%
\bibitem [{\citenamefont {Fisher}(1967)}]{Fisher1967}%
  \BibitemOpen
  \bibfield  {author} {\bibinfo {author} {\bibfnamefont {M.~E.}\ \bibnamefont
  {Fisher}},\ }\bibfield  {title} {\emph {\bibinfo {title} {The Theory of
  Condensation and the Critical Point}},\ }\href
  {https://doi.org/10.1103/PhysicsPhysiqueFizika.3.255} {\bibfield  {journal}
  {\bibinfo  {journal} {Phys. Phys. Fiz.}\ }\textbf {\bibinfo {volume} {3}},\
  \bibinfo {pages} {255} (\bibinfo {year} {1967})}\BibitemShut {NoStop}%
\bibitem [{\citenamefont {Langer}(1992)}]{Langer1992}%
  \BibitemOpen
  \bibfield  {author} {\bibinfo {author} {\bibfnamefont {J.}~\bibnamefont
  {Langer}},\ }in\ \href@noop {} {\emph {\bibinfo {booktitle} {Solids Far from
  Equilibrium}}},\ \bibinfo {editor} {edited by\ \bibinfo {editor}
  {\bibfnamefont {C.}~\bibnamefont {Godr{\'{e}}che}}}\ (\bibinfo  {publisher}
  {Cambridge University Press},\ \bibinfo {year} {1992})\ \bibinfo {edition}
  {1st}\ ed.,\ Chap.~\bibinfo {chapter} {3}, pp.\ \bibinfo {pages}
  {297--362}\BibitemShut {NoStop}%
\bibitem [{\citenamefont {Langer}(1969)}]{langer1969statistical}%
  \BibitemOpen
  \bibfield  {author} {\bibinfo {author} {\bibfnamefont {J.}~\bibnamefont
  {Langer}},\ }\bibfield  {title} {\emph {\bibinfo {title} {Statistical Theory
  of the Decay of Metastable States}},\ }\href
  {https://doi.org/https://doi.org/10.1016/0003-4916(69)90153-5} {\bibfield
  {journal} {\bibinfo  {journal} {Ann. Phys. (N. Y.)}\ }\textbf {\bibinfo
  {volume} {54}},\ \bibinfo {pages} {258 } (\bibinfo {year}
  {1969})}\BibitemShut {NoStop}%
\bibitem [{\citenamefont {Kolmogorov}(1937)}]{kolmogorov1937statistical}%
  \BibitemOpen
  \bibfield  {author} {\bibinfo {author} {\bibfnamefont {A.~N.}\ \bibnamefont
  {Kolmogorov}},\ }\bibfield  {title} {\emph {\bibinfo {title} {On the
  Statistical Theory of the Crystallization of Metals}},\ }\href@noop {}
  {\bibfield  {journal} {\bibinfo  {journal} {Bull. Acad. Sci. USSR, Math.
  Ser}\ }\textbf {\bibinfo {volume} {1}},\ \bibinfo {pages} {355} (\bibinfo
  {year} {1937})}\BibitemShut {NoStop}%
\bibitem [{\citenamefont {William}\ and\ \citenamefont
  {Mehl}(1939)}]{william1939reaction}%
  \BibitemOpen
  \bibfield  {author} {\bibinfo {author} {\bibfnamefont {J.}~\bibnamefont
  {William}}\ and\ \bibinfo {author} {\bibfnamefont {R.}~\bibnamefont {Mehl}},\
  }\bibfield  {title} {\emph {\bibinfo {title} {Reaction Kinetics in Processes
  of Nucleation and Growth}},\ }\href@noop {} {\bibfield  {journal} {\bibinfo
  {journal} {Trans. Metall. AIME}\ }\textbf {\bibinfo {volume} {135}},\
  \bibinfo {pages} {416} (\bibinfo {year} {1939})}\BibitemShut {NoStop}%
\bibitem [{\citenamefont {Avrami}(1940)}]{avrami1940kinetics}%
  \BibitemOpen
  \bibfield  {author} {\bibinfo {author} {\bibfnamefont {M.}~\bibnamefont
  {Avrami}},\ }\bibfield  {title} {\emph {\bibinfo {title} {Kinetics of Phase
  Change. II Transformation‐Time Relations for Random Distribution of
  Nuclei}},\ }\href {https://doi.org/10.1063/1.1750631} {\bibfield  {journal}
  {\bibinfo  {journal} {J. Chem. Phys.}\ }\textbf {\bibinfo {volume} {8}},\
  \bibinfo {pages} {212} (\bibinfo {year} {1940})}\BibitemShut {NoStop}%
\bibitem [{\citenamefont {Becker}\ and\ \citenamefont
  {D{\"{o}}ring}(1935)}]{Becker1935}%
  \BibitemOpen
  \bibfield  {author} {\bibinfo {author} {\bibfnamefont {R.}~\bibnamefont
  {Becker}}\ and\ \bibinfo {author} {\bibfnamefont {W.}~\bibnamefont
  {D{\"{o}}ring}},\ }\bibfield  {title} {\emph {\bibinfo {title} {Kinetische
  Behandlung der Keimbildung in {\"{u}}bers{\"{a}}ttigten D{\"{a}}mpfen}},\
  }\href {https://doi.org/10.1002/andp.19354160806} {\bibfield  {journal}
  {\bibinfo  {journal} {Ann. Phys. (Berl.)}\ }\textbf {\bibinfo {volume}
  {416}},\ \bibinfo {pages} {719} (\bibinfo {year} {1935})}\BibitemShut
  {NoStop}%
\bibitem [{\citenamefont {Zeldovich}(1943)}]{Zeldovich1943}%
  \BibitemOpen
  \bibfield  {author} {\bibinfo {author} {\bibfnamefont {Y.~B.}\ \bibnamefont
  {Zeldovich}},\ }\bibfield  {title} {\emph {\bibinfo {title} {On the Theory of
  New Phase Formation: Cavitation}},\ }\href@noop {} {\bibfield  {journal}
  {\bibinfo  {journal} {Acta Physicochimica USSR}\ }\textbf {\bibinfo {volume}
  {18}},\ \bibinfo {pages} {17} (\bibinfo {year} {1943})}\BibitemShut {NoStop}%
\bibitem [{\citenamefont {Maibaum}(2008)}]{Maibaum2008}%
  \BibitemOpen
  \bibfield  {author} {\bibinfo {author} {\bibfnamefont {L.}~\bibnamefont
  {Maibaum}},\ }\bibfield  {title} {\emph {\bibinfo {title} {Phase
  Transformation Near the Classical Limit of Stability}},\ }\href
  {https://doi.org/10.1103/PhysRevLett.101.256102} {\bibfield  {journal}
  {\bibinfo  {journal} {Phys. Rev. Lett.}\ }\textbf {\bibinfo {volume} {101}},\
  \bibinfo {pages} {256102} (\bibinfo {year} {2008})}\BibitemShut {NoStop}%
\bibitem [{\citenamefont {Goldenfeld}(2018)}]{Goldenfeld2018}%
  \BibitemOpen
  \bibfield  {author} {\bibinfo {author} {\bibfnamefont {N.}~\bibnamefont
  {Goldenfeld}},\ }\href {https://doi.org/10.1201/9780429493492} {\emph
  {\bibinfo {title} {Lectures on Phase Transitions and the Renormalization
  Group}}}\ (\bibinfo  {publisher} {CRC Press},\ \bibinfo {year}
  {2018})\BibitemShut {NoStop}%
\bibitem [{\citenamefont {Amit}\ and\ \citenamefont
  {Martin-Mayor}(2005)}]{Amit2005}%
  \BibitemOpen
  \bibfield  {author} {\bibinfo {author} {\bibfnamefont {D.~J.}\ \bibnamefont
  {Amit}}\ and\ \bibinfo {author} {\bibfnamefont {V.}~\bibnamefont
  {Martin-Mayor}},\ }\href {https://doi.org/10.1142/5715} {\emph {\bibinfo
  {title} {Field Theory, the Renormalization Group, and Critical Phenomena}}},\
  \bibinfo {edition} {3rd}\ ed.\ (\bibinfo  {publisher} {World Scientific},\
  \bibinfo {year} {2005})\BibitemShut {NoStop}%
\bibitem [{\citenamefont {Gunther}\ \emph {et~al.}(1980)\citenamefont
  {Gunther}, \citenamefont {Wallace},\ and\ \citenamefont
  {Nicole}}]{gunther1980goldstone}%
  \BibitemOpen
  \bibfield  {author} {\bibinfo {author} {\bibfnamefont {N.~J.}\ \bibnamefont
  {Gunther}}, \bibinfo {author} {\bibfnamefont {D.~J.}\ \bibnamefont
  {Wallace}},\ and\ \bibinfo {author} {\bibfnamefont {D.~A.}\ \bibnamefont
  {Nicole}},\ }\bibfield  {title} {\emph {\bibinfo {title} {Goldstone Modes in
  Vacuum Decay and First-Order Phase Transitions}},\ }\href
  {https://doi.org/10.1088/0305-4470/13/5/034} {\bibfield  {journal} {\bibinfo
  {journal} {J. Phys. A: Math. Gen.}\ }\textbf {\bibinfo {volume} {13}},\
  \bibinfo {pages} {1755} (\bibinfo {year} {1980})}\BibitemShut {NoStop}%
\bibitem [{\citenamefont {Zwerger}(1985)}]{Zwerger1985}%
  \BibitemOpen
  \bibfield  {author} {\bibinfo {author} {\bibfnamefont {W.}~\bibnamefont
  {Zwerger}},\ }\bibfield  {title} {\emph {\bibinfo {title} {Dynamical
  Interpretation of a Classical Complex Free Energy}},\ }\href
  {https://doi.org/10.1088/0305-4470/18/11/028} {\bibfield  {journal} {\bibinfo
   {journal} {J. Phys. A: Math. Gen.}\ }\textbf {\bibinfo {volume} {18}},\
  \bibinfo {pages} {2079} (\bibinfo {year} {1985})}\BibitemShut {NoStop}%
\bibitem [{\citenamefont {M{\"{u}}nster}\ and\ \citenamefont
  {Rotsch}(2000)}]{munster2000analytical}%
  \BibitemOpen
  \bibfield  {author} {\bibinfo {author} {\bibfnamefont {G.}~\bibnamefont
  {M{\"{u}}nster}}\ and\ \bibinfo {author} {\bibfnamefont {S.}~\bibnamefont
  {Rotsch}},\ }\bibfield  {title} {\emph {\bibinfo {title} {Analytical
  Calculation of the Nucleation Rate for First Order Phase Transitions Beyond
  the Thin Wall Approximation}},\ }\href
  {https://doi.org/10.1007/s100529900242} {\bibfield  {journal} {\bibinfo
  {journal} {Eur. Phys. J. C}\ }\textbf {\bibinfo {volume} {12}},\ \bibinfo
  {pages} {161} (\bibinfo {year} {2000})}\BibitemShut {NoStop}%
\bibitem [{\citenamefont {M{\"{u}}nster}\ and\ \citenamefont
  {Rutkevich}(2003)}]{munster2003classical}%
  \BibitemOpen
  \bibfield  {author} {\bibinfo {author} {\bibfnamefont {G.}~\bibnamefont
  {M{\"{u}}nster}}\ and\ \bibinfo {author} {\bibfnamefont {S.~B.}\ \bibnamefont
  {Rutkevich}},\ }\bibfield  {title} {\emph {\bibinfo {title} {The Classical
  Nucleation Rate in Two Dimensions}},\ }\href
  {https://doi.org/10.1140/epjc/s2002-01091-4} {\bibfield  {journal} {\bibinfo
  {journal} {Eur. Phys. J. C}\ }\textbf {\bibinfo {volume} {27}},\ \bibinfo
  {pages} {297} (\bibinfo {year} {2003})}\BibitemShut {NoStop}%
\bibitem [{\citenamefont {Gunton}\ and\ \citenamefont
  {Droz}(1983)}]{gunton1983introduction}%
  \BibitemOpen
  \bibfield  {author} {\bibinfo {author} {\bibfnamefont {J.~D.}\ \bibnamefont
  {Gunton}}\ and\ \bibinfo {author} {\bibfnamefont {M.}~\bibnamefont {Droz}},\
  }\href@noop {} {\emph {\bibinfo {title} {Introduction to the Theory of
  Metastable and Unstable States}}},\ \bibinfo {series} {Lecture Notes in
  Physics}, Vol.\ \bibinfo {volume} {183}\ (\bibinfo  {publisher} {Springer},\
  \bibinfo {year} {1983})\BibitemShut {NoStop}%
\bibitem [{\citenamefont {Langer}(1967)}]{Langer1967}%
  \BibitemOpen
  \bibfield  {author} {\bibinfo {author} {\bibfnamefont {J.}~\bibnamefont
  {Langer}},\ }\bibfield  {title} {\emph {\bibinfo {title} {Theory of the
  Condensation Point}},\ }\href
  {https://doi.org/https://doi.org/10.1016/0003-4916(67)90200-X} {\bibfield
  {journal} {\bibinfo  {journal} {Ann. Phys. (N. Y.)}\ }\textbf {\bibinfo
  {volume} {41}},\ \bibinfo {pages} {108 } (\bibinfo {year}
  {1967})}\BibitemShut {NoStop}%
\bibitem [{\citenamefont {Jacucci}\ \emph {et~al.}(1983)\citenamefont
  {Jacucci}, \citenamefont {Perini},\ and\ \citenamefont
  {Martin}}]{Jacucci1983}%
  \BibitemOpen
  \bibfield  {author} {\bibinfo {author} {\bibfnamefont {G.}~\bibnamefont
  {Jacucci}}, \bibinfo {author} {\bibfnamefont {A.}~\bibnamefont {Perini}},\
  and\ \bibinfo {author} {\bibfnamefont {G.}~\bibnamefont {Martin}},\
  }\bibfield  {title} {\emph {\bibinfo {title} {Monte Carlo Computation of
  Cluster Free Energies in the Ising model: A Test for the Validity of the
  Capillarity Approximation}},\ }\href
  {https://doi.org/10.1088/0305-4470/16/2/019} {\bibfield  {journal} {\bibinfo
  {journal} {J. Phys. A: Math. Gen.}\ }\textbf {\bibinfo {volume} {16}},\
  \bibinfo {pages} {369} (\bibinfo {year} {1983})}\BibitemShut {NoStop}%
\bibitem [{\citenamefont {Ryu}\ and\ \citenamefont {Cai}(2010)}]{Ryu2010}%
  \BibitemOpen
  \bibfield  {author} {\bibinfo {author} {\bibfnamefont {S.}~\bibnamefont
  {Ryu}}\ and\ \bibinfo {author} {\bibfnamefont {W.}~\bibnamefont {Cai}},\
  }\bibfield  {title} {\emph {\bibinfo {title} {Numerical Tests of Nucleation
  Theories for the Ising Models}},\ }\href
  {https://doi.org/10.1103/PhysRevE.82.011603} {\bibfield  {journal} {\bibinfo
  {journal} {Phys. Rev. E}\ }\textbf {\bibinfo {volume} {82}},\ \bibinfo
  {pages} {011603} (\bibinfo {year} {2010})}\BibitemShut {NoStop}%
\bibitem [{\citenamefont {Hudson}(2016)}]{Hudson2016}%
  \BibitemOpen
  \bibfield  {author} {\bibinfo {author} {\bibfnamefont {A.}~\bibnamefont
  {Hudson}},\ }\emph {\bibinfo {title} {Statistical Mechanics and Dynamics of
  Liquids in and out of Equilibrium}},\ \href@noop {} {Ph.D. thesis},\ \bibinfo
   {school} {University of California - Berkeley} (\bibinfo {year}
  {2016})\BibitemShut {NoStop}%
\bibitem [{\citenamefont {Hudson}\ and\ \citenamefont
  {Mandadapu}(2018)}]{Hudson2018}%
  \BibitemOpen
  \bibfield  {author} {\bibinfo {author} {\bibfnamefont {A.}~\bibnamefont
  {Hudson}}\ and\ \bibinfo {author} {\bibfnamefont {K.~K.}\ \bibnamefont
  {Mandadapu}},\ }\href@noop {} {\bibinfo {title} {On the Nature of the Glass
  Transition in Atomistic Models of Glass Formers}} (\bibinfo {year} {2018}),\
  \Eprint {https://arxiv.org/abs/1804.03769} {arXiv:1804.03769
  [cond-mat.stat-mech]} \BibitemShut {NoStop}%
\bibitem [{\citenamefont {Limmer}(2014)}]{limmer2014length}%
  \BibitemOpen
  \bibfield  {author} {\bibinfo {author} {\bibfnamefont {D.~T.}\ \bibnamefont
  {Limmer}},\ }\bibfield  {title} {\emph {\bibinfo {title} {The Length and Time
  Scales of Water's Glass Transitions}},\ }\href@noop {} {\bibfield  {journal}
  {\bibinfo  {journal} {J. Chem. Phys.}\ }\textbf {\bibinfo {volume} {140}},\
  \bibinfo {pages} {214509} (\bibinfo {year} {2014})}\BibitemShut {NoStop}%
\bibitem [{\citenamefont {Rammal}\ and\ \citenamefont
  {Toulouse}(1983)}]{Rammal1983}%
  \BibitemOpen
  \bibfield  {author} {\bibinfo {author} {\bibfnamefont {R.}~\bibnamefont
  {Rammal}}\ and\ \bibinfo {author} {\bibfnamefont {G.}~\bibnamefont
  {Toulouse}},\ }\bibfield  {title} {\emph {\bibinfo {title} {Random Walks on
  Fractal Structures and Percolation Clusters}},\ }\href
  {https://doi.org/10.1051/jphyslet:0198300440101300} {\bibfield  {journal}
  {\bibinfo  {journal} {J. Phys. Lett. (Paris).}\ }\textbf {\bibinfo {volume}
  {44}},\ \bibinfo {pages} {13} (\bibinfo {year} {1983})}\BibitemShut {NoStop}%
\bibitem [{\citenamefont {Strichartz}(2006)}]{Strichartz2006}%
  \BibitemOpen
  \bibfield  {author} {\bibinfo {author} {\bibfnamefont {R.~S.}\ \bibnamefont
  {Strichartz}},\ }\href {https://doi.org/10.2307/j.ctv346nvv} {\emph {\bibinfo
  {title} {Differential Equations on Fractals: A Tutorial}}}\ (\bibinfo
  {publisher} {Princeton University Press},\ \bibinfo {year}
  {2006})\BibitemShut {NoStop}%
\bibitem [{\citenamefont {Berthier}\ and\ \citenamefont
  {Garrahan}(2005)}]{Berthier2005}%
  \BibitemOpen
  \bibfield  {author} {\bibinfo {author} {\bibfnamefont {L.}~\bibnamefont
  {Berthier}}\ and\ \bibinfo {author} {\bibfnamefont {J.~P.}\ \bibnamefont
  {Garrahan}},\ }\bibfield  {title} {\emph {\bibinfo {title} {Numerical Study
  of a Fragile Three-Dimensional Kinetically Constrained Model}},\ }\href
  {https://doi.org/10.1021/jp045491e} {\bibfield  {journal} {\bibinfo
  {journal} {J. Phys. Chem. B}\ }\textbf {\bibinfo {volume} {109}},\ \bibinfo
  {pages} {3578} (\bibinfo {year} {2005})}\BibitemShut {NoStop}%
\bibitem [{\citenamefont {Van~Kampen}(2007)}]{van2007stochastic}%
  \BibitemOpen
  \bibfield  {author} {\bibinfo {author} {\bibfnamefont {N.~G.}\ \bibnamefont
  {Van~Kampen}},\ }\href
  {https://doi.org/https://doi.org/10.1016/B978-0-444-52965-7.X5000-4} {\emph
  {\bibinfo {title} {Stochastic Processes in Physics and Chemistry}}}\
  (\bibinfo  {publisher} {North Holland},\ \bibinfo {year} {2007})\BibitemShut
  {NoStop}%
\bibitem [{\citenamefont {Cahn}(1995)}]{cahn1995time}%
  \BibitemOpen
  \bibfield  {author} {\bibinfo {author} {\bibfnamefont {J.~W.}\ \bibnamefont
  {Cahn}},\ }\bibfield  {title} {\emph {\bibinfo {title} {The Time Cone Method
  for Nucleation and Growth Kinetics on a Finite Domain}},\ }\href
  {https://doi.org/10.1557/PROC-398-425} {\bibfield  {journal} {\bibinfo
  {journal} {Mat. Res. Soc. Symp. Proc.}\ }\textbf {\bibinfo {volume} {398}},\
  \bibinfo {pages} {427} (\bibinfo {year} {1995})}\BibitemShut {NoStop}%
\bibitem [{\citenamefont {Ohta}\ \emph {et~al.}(1987)\citenamefont {Ohta},
  \citenamefont {Ohta},\ and\ \citenamefont {Kawasaki}}]{ohta1987domain}%
  \BibitemOpen
  \bibfield  {author} {\bibinfo {author} {\bibfnamefont {S.}~\bibnamefont
  {Ohta}}, \bibinfo {author} {\bibfnamefont {T.}~\bibnamefont {Ohta}},\ and\
  \bibinfo {author} {\bibfnamefont {K.}~\bibnamefont {Kawasaki}},\ }\bibfield
  {title} {\emph {\bibinfo {title} {Domain Growth in Systems with
  Multiple-Degenerate Ground States}},\ }\href
  {https://doi.org/https://doi.org/10.1016/0378-4371(87)90077-X} {\bibfield
  {journal} {\bibinfo  {journal} {Physica A}\ }\textbf {\bibinfo {volume}
  {140}},\ \bibinfo {pages} {478 } (\bibinfo {year} {1987})}\BibitemShut
  {NoStop}%
\bibitem [{\citenamefont {Sekimoto}(1986)}]{sekimoto1986evolution}%
  \BibitemOpen
  \bibfield  {author} {\bibinfo {author} {\bibfnamefont {K.}~\bibnamefont
  {Sekimoto}},\ }\bibfield  {title} {\emph {\bibinfo {title} {Evolution of the
  Domain Structure During the Nucleation-and-Growth Process With Non-Conserved
  Order Parameter}},\ }\href
  {https://doi.org/https://doi.org/10.1016/0378-4371(86)90146-9} {\bibfield
  {journal} {\bibinfo  {journal} {Physica A}\ }\textbf {\bibinfo {volume}
  {135}},\ \bibinfo {pages} {328 } (\bibinfo {year} {1986})}\BibitemShut
  {NoStop}%
\bibitem [{\citenamefont {Onsager}(1944)}]{Onsager1944}%
  \BibitemOpen
  \bibfield  {author} {\bibinfo {author} {\bibfnamefont {L.}~\bibnamefont
  {Onsager}},\ }\bibfield  {title} {\emph {\bibinfo {title} {Crystal
  Statistics. I. A Two-Dimensional Model with an Order-Disorder Transition}},\
  }\href {https://doi.org/10.1103/PhysRev.65.117} {\bibfield  {journal}
  {\bibinfo  {journal} {Phy. Rev.}\ }\textbf {\bibinfo {volume} {65}},\
  \bibinfo {pages} {117} (\bibinfo {year} {1944})}\BibitemShut {NoStop}%
\bibitem [{\citenamefont {Kob}\ and\ \citenamefont {Andersen}(1994)}]{Kob1994}%
  \BibitemOpen
  \bibfield  {author} {\bibinfo {author} {\bibfnamefont {W.}~\bibnamefont
  {Kob}}\ and\ \bibinfo {author} {\bibfnamefont {H.~C.}\ \bibnamefont
  {Andersen}},\ }\bibfield  {title} {\emph {\bibinfo {title} {Scaling Behavior
  in the $\ensuremath{\beta}$-Relaxation Regime of a Supercooled Lennard-Jones
  Mixture}},\ }\href {https://doi.org/10.1103/PhysRevLett.73.1376} {\bibfield
  {journal} {\bibinfo  {journal} {Phys. Rev. Lett.}\ }\textbf {\bibinfo
  {volume} {73}},\ \bibinfo {pages} {1376} (\bibinfo {year}
  {1994})}\BibitemShut {NoStop}%
\bibitem [{\citenamefont {Ingebrigtsen}\ \emph {et~al.}(2019)\citenamefont
  {Ingebrigtsen}, \citenamefont {Dyre}, \citenamefont {Schr{\o}der},\ and\
  \citenamefont {Royall}}]{Ingebrigtsen2018}%
  \BibitemOpen
  \bibfield  {author} {\bibinfo {author} {\bibfnamefont {T.~S.}\ \bibnamefont
  {Ingebrigtsen}}, \bibinfo {author} {\bibfnamefont {J.~C.}\ \bibnamefont
  {Dyre}}, \bibinfo {author} {\bibfnamefont {T.~B.}\ \bibnamefont
  {Schr{\o}der}},\ and\ \bibinfo {author} {\bibfnamefont {C.~P.}\ \bibnamefont
  {Royall}},\ }\bibfield  {title} {\emph {\bibinfo {title} {Crystallization
  Instability in Glass-Forming Mixtures}},\ }\href@noop {} {\bibfield
  {journal} {\bibinfo  {journal} {Phs. Rev. X}\ }\textbf {\bibinfo {volume}
  {9}},\ \bibinfo {pages} {031016} (\bibinfo {year} {2019})}\BibitemShut
  {NoStop}%
\bibitem [{\citenamefont {Zhang}\ and\ \citenamefont {Han}(2018)}]{Zhang2018}%
  \BibitemOpen
  \bibfield  {author} {\bibinfo {author} {\bibfnamefont {H.}~\bibnamefont
  {Zhang}}\ and\ \bibinfo {author} {\bibfnamefont {Y.}~\bibnamefont {Han}},\
  }\bibfield  {title} {\emph {\bibinfo {title} {{Compression-Induced
  Polycrystal-Glass Transition in Binary Crystals}}},\ }\href
  {https://doi.org/10.1103/PhysRevX.8.041023} {\bibfield  {journal} {\bibinfo
  {journal} {Phys. Rev. X}\ }\textbf {\bibinfo {volume} {8}},\ \bibinfo {pages}
  {041023} (\bibinfo {year} {2018})}\BibitemShut {NoStop}%
\bibitem [{\citenamefont {Debenedetti}(1996)}]{Debenedetti1996}%
  \BibitemOpen
  \bibfield  {author} {\bibinfo {author} {\bibfnamefont {P.~G.}\ \bibnamefont
  {Debenedetti}},\ }\href@noop {} {\emph {\bibinfo {title} {Metastable Liquids:
  Concepts and Principles}}}\ (\bibinfo  {publisher} {Princeton University
  Press},\ \bibinfo {year} {1996})\BibitemShut {NoStop}%
\bibitem [{\citenamefont {Tan}\ \emph {et~al.}(2014)\citenamefont {Tan},
  \citenamefont {Xu},\ and\ \citenamefont {Xu}}]{Tan2014}%
  \BibitemOpen
  \bibfield  {author} {\bibinfo {author} {\bibfnamefont {P.}~\bibnamefont
  {Tan}}, \bibinfo {author} {\bibfnamefont {N.}~\bibnamefont {Xu}},\ and\
  \bibinfo {author} {\bibfnamefont {L.}~\bibnamefont {Xu}},\ }\bibfield
  {title} {\emph {\bibinfo {title} {Visualizing Kinetic Pathways of Homogeneous
  Nucleation in Colloidal Crystallization}},\ }\href
  {https://doi.org/10.1038/nphys2817} {\bibfield  {journal} {\bibinfo
  {journal} {Nat. Phys.}\ }\textbf {\bibinfo {volume} {10}},\ \bibinfo {pages}
  {73} (\bibinfo {year} {2014})}\BibitemShut {NoStop}%
\end{thebibliography}%

\end{document}